\documentclass[3p,times,procedia]{elsarticle}

\usepackage{amsmath,amssymb,amsfonts}
\usepackage{amsthm}
\usepackage{xcolor}
\usepackage{dsfont}
\usepackage{tikz}
\usetikzlibrary{positioning}
\usetikzlibrary{calc}
\usepackage{floatrow}
\usepackage{graphicx}
\usepackage{subfig}
\usepackage{caption}
\usepackage[linesnumbered, ruled, vlined]{algorithm2e}
\usepackage{graphicx,lipsum}
\usepackage{url}
\usepackage{hyperref}
\usepackage[capitalize]{cleveref}
\usepackage{graphicx}

\renewcommand{\d}[1]{\textnormal{d}#1}

\usepackage{todonotes}
\newboolean{SHOWCOMMENTS}
\setboolean{SHOWCOMMENTS}{true}
\ifthenelse{\boolean{SHOWCOMMENTS}}{
\newcommand{\tdMK}[1]
{\todo[color=blue!25,inline]{\footnotesize{\bf Martin:} #1}}
}{
\newcommand{\tdMK}[1]{}
}
\newcommand{\MK}[1]{{\color{blue}#1}}

\ifthenelse{\boolean{SHOWCOMMENTS}}{
\newcommand{\tdJB}[1]
{\todo[color=green!80!gray!25,inline]{\footnotesize{\bf Julia:} #1}}
}{
\newcommand{\tdJB}[1]{}
}

\ifthenelse{\boolean{SHOWCOMMENTS}}{
\newcommand{\RS}[1]
{\todo[color=teal!20,inline]{\footnotesize{\bf René:} #1}}
}{
\newcommand{\RS}[1]{}
}

\newcommand{\flow}{{\widehat{f}}}
\journal{Infectious Disease Modelling}

\begin{document}

\begin{frontmatter}



\title{Hybrid metapopulation agent-based epidemiological models for efficient insight on the individual scale: a contribution to green computing}


\author[label1]{Julia Bicker}
\author[label2]{Ren\'{e} Schmieding}
\author[label2]{Michael Meyer-Hermann}
\author[label1,label3]{Martin~J.~Kühn}

\affiliation[label1]{organization={Institute of Software Technology, Department of High-Performance Computing, German Aerospace Center},%Department and Organization, 
            city={Cologne},
            country={Germany}}
\affiliation[label2]{organization={Department of Systems Immunology and Braunschweig Integrated Centre of Systems Biology (BRICS), Helmholtz Centre for Infection Research},%Department and Organization, 
            city={Brunswick},
            country={Germany}}
\affiliation[label3]{organization={Life and Medical Sciences Institute, University of Bonn},%Department and Organization, 
            city={Bonn},
            country={Germany}}

\begin{abstract}
Emerging infectious diseases and climate change are two of the major challenges in 21st century. Although over the past decades, highly-resolved mathematical models have contributed in understanding dynamics of infectious diseases and are of great aid when it comes to finding suitable intervention measures, they may need substantial computational effort and produce significant $\text{CO}_2$ emissions. Two popular modeling approaches for mitigating infectious disease dynamics are agent-based and \MK{population}-based models. Agent-based models (ABMs) offer a \MK{microscopic view} and are thus able to capture heterogeneous human contact behavior and mobility patterns. However, insights on individual-level dynamics come with high computational effort that scales with the number of agents. On the other hand, \MK{population-based models using e.g. ordinary} differential equations (\MK{ODE}s) are computationally efficient even for large populations due to their complexity being independent of the population size. Yet, \MK{population}-based models are restricted in their granularity as they assume a (to some extent) homogeneous and well-mixed population. To manage the trade-off between computational complexity and level of detail, we propose spatial- and temporal-hybrid models that use ABMs only in an area or time frame of interest. To account for relevant influences to disease dynamics, e.g., from outside, due to commuting activities, we use \MK{population}-based models, only adding moderate computational costs. Our hybridization approach demonstrates significant reduction in computational effort by up to 98\% -- without losing the required depth in information in the focus frame. The hybrid models used in our numerical simulations are based on two recently proposed models, however, any suitable combination of ABM-EBM could be used, too. Concluding, hybrid epidemiological models can provide insights on the individual scale where necessary, using aggregated models where possible, thereby making a contribution to green computing.
%Test \citep{kuhn_assessment_2021}

\end{abstract}




\end{frontmatter}



\section{Introduction}\label{sec::Introduction}

Two of the major 21st century's challenges are accelerating climate change~\cite{callaghan_new_2010} and emerging infectious diseases with an expected increase in frequency of epidemics and pandemics~\cite{daszak_workshop_2020}. As these challenges are driven by the same underlying causes, sets of proposed counteractions might go hand-in-hand~\cite{daszak_workshop_2020,epstein_climate_2001}. In order to assess counteractions and support decision-makers, mathematical models are viable tools that have been used extensively in both domains. Models can help to understand the situation at hand and to analyze potential future developments, which often cannot be conducted in classical experiments. The recent COVID-19 pandemic has demonstrated the benefit of mathematical models to analyze ongoing infectious disease dynamics and to evaluate potential scenarios, see, e.g.,~\cite{reiner_modeling_2020,kerr_covasim_2021,koslow_appropriate_2022}. Extensive testing strategies maintaining mobility as a social good, have, for instance, been studied in~\cite{kuhn_regional_2022}. The set of mathematical models to \MK{address} epidemiological emergencies \MK{to provide guidance to decision makers} is diverse. It ranges from classical statistical models such as generalized regression models, through population-based~\cite{reiner_modeling_2020,medlock_integro-differential-equation_2004}, metapopulation~\cite{liu_modelling_2022,kuhn_assessment_2021}, and agent-based~\cite{bershteyn_implementation_2018,kerr_covasim_2021,muller_predicting_2021,bracher_pre-registered_2021,KKN24} models to machine and deep learning models~\cite{garnett_mathematical_2011,wynants_prediction_2020,SZHK24}. Population-based models using sets of ordinary~\cite{reiner_modeling_2020,kuhn_assessment_2021} or integro-differential-equation-based models~\cite{medlock_integro-differential-equation_2004,Wendler2024IDE} are well established methods that have been used by hundreds of work groups worldwide to analyze infectious disease dynamics. Their advantages are the ease of analysis, their well-understood character and the low computational requirements which do not depend on the number of individuals in the population but on the number of different subpopulations. However, their insight on the individual-level dynamics or on (singular) stochastic effects is limited and they also fail in the beginning of an outbreak when the number of infected is small.
Agent- or individual-based models (ABMs or IBMs), on the other hand, allow the stochastic simulation of individual behavior and reaction to diseases, in a social and immunological dimension, see, e.g.,~\cite{kerr_covasim_2021,muller_explicit_2023}. Though, this comes at the cost of superlinear complexity. While the energy consumption corresponding to computational cost can be reduced through, e.g., optimizations on the cache level~\cite{willem_optimizing_2015} or the optimal use of heterogeneous hardware structures such as SIMD registers~\cite{kuhn2023simd}, we here propose a complementary approach to directly avoid less impactful computations through the development of well-designed hybridization concepts.

The authors of~\cite{muller_predicting_2021} demonstrated that the inclusion of disease import, i.e., infections caused by travelers and commuters, is an important factor to explain disease dynamics. Via the metapopulation approach in~\cite{kuhn_regional_2022}, it was shown how large the effects of disease import can become if testing is not done properly. Using simple statistics for disease import can be a suitable first step in improving model precision in a considered focus region. However, if disease import is subject to change through interventions, disease dynamics in coupled regions are advised to also be modeled dynamically.

In this work, we will present a framework for spatial- and temporal-hybrid epidemiological modeling, combining any pair of suitable fine-granular agent-based and coarse-granular \MK{population}-based models. We will then suggest a particular combination of two suitable models introduced recently in~\cite{winkelmann_mathematical_2021} and present results for three \MK{simplified} examples thereby demonstrating potential areas of application and the benefit of the hybrid models. In this context, we will simulate a COVID-19-like disease with parameters motivated by~\cite{kuhn_assessment_2021}. However, as the scope of the current paper is the disease-agnostic introduction of novel hybrid models, no further parameter refinement or real-world application for COVID-19 has been performed. Let us note that any other disease model can be put in our framework. We will show that our hybrid modeling approach can account for individual developing disease dynamics, otherwise modeled with pure ABMs, while drastically reducing computational effort. We will provide discussions, limitations, and directions for future research before we eventually draw a conclusion.

\section{Materials and methods}
\label{sec::Materials_and_Methods}
\MK{
In this section, we first present related work on hybrid models. Then we present an agent-based and a metapopulation model, and eventually we introduce our novel, hybrid modeling framework which is applied to the two models introduced before.

\subsection{Related work}
The development of hybrid models is often driven by the intention to combine the advantages of two different modeling approaches. In the context of infectious disease dynamics, several suggestions for hybrid modeling have been made. While the authors of~\cite{avegliano_equation-based_2023,angione_using_2022,robertson_bayesian_2024} used ABMs together with ordinary differential equation (ODE)-based and machine learning models, respectively, to improve parameter estimation, we focus on hybrid models where both model types are used in the simulation and prediction process. However, also in this context, the term \textit{hybrid model} is used for many different realizations.

In~\cite{yoneyama_hybrid_2012,kasereka_hybrid_2014,bradhurst_hybrid_2015,banos_importance_2015,marilleau_coupling_2018}, the authors used metapopulation or network-based models which are sometimes also denoted as hybrid models. These metapopulation models combined ODE-based models for different considered regions while the mobility between regions was modeled with an agent-based modeling principle. Similarly, although not coined hybrid, the PDMM of~\cite{winkelmann_mathematical_2021} used ODE-based formulations for single regions and a stochastic jump process to model mobility between the regions. In~\cite{kuhn_assessment_2021,zunker_novel_2024}, other examples of network-based metapopulation models were presented. There, ODE-based models were used for particular regions representing the nodes of a graph and deterministic mobility was realized through the edges of the graph. While also using a metapopulation model combined with an ABM in our hybridization framework, our understanding of hybrid goes beyond the aforementioned.

In~\cite{bobashev_hybrid_2007,quang_toward_2016,hunter_hybrid_2020}, hybrid models were implemented that used different model types, namely agent-based and \MK{population}-based models, in different temporal regimes. The authors used ABMs for low-incidence regimes and upon passing a threshold on the number of infected individuals, a switch to an ODE- or difference equation-based model was executed. We call this temporal coupling of model types \textit{temporal hybridization}.

The authors of~\cite{sewall_interactive_2011} implemented a hybrid model for simulating large-scale traffic. They coupled an ABM and an \MK{population}-based model and provided rules for spatial exchange when one region was modeled with the ABM and a neighboring region with the \MK{population}-based model. We call this method \textit{spatial hybridization}. Similar to the temporal hybridization, they also implemented the conversion of a whole region from one model to another, here not dependent on time but on user input.

In this paper, we will introduce novel \textit{spatial-hybrid models} for infectious disease dynamics which use an ABM in a focus region and macroscopic ODE-based models in all connected regions. Furthermore, we will extend the application of \textit{temporal-hybrid models} where switching takes place upon exceeding a threshold.

\subsection{High-performance computing, carbon footprint, and green computing}

High-performance computing (HPC) can enable breakthrough research results for ABMs~\cite{collier_parallel_2013} but require great computing capability which naturally comes with high energy demands. The regularly updated Top500 list\footnote{https://top500.org/lists/top500/list/2024/11/} of the 500 most powerful commercially available computer systems known to the providers, lists the (peak) power consumption for 199 out of the 500 supercomputers in November 2024. The total power consumptions for these 199 system sums up to 431,222.57 kW. Using an optimistic emission of 380g $CO_2$ for the production of 1 kWh power, provided by the German Federal Environment Agency for the 2023 German energy mix with 51.9\% of renewable energy~\cite{icha_entwicklung_2024}, these systems alone have a daily emission of 3932 tons of $CO_2$. The author of~\cite{portegies_zwart_ecological_2020} situated the carbon footprint for the use of one million compute cores between an 8-hour air travel and a launch of a Falcon 9 rocket into space. On a broader scale, green computing considers the entire product life cycle of technological devices from production through operation and recycling, resulting in several challenges as highlighted by~\cite{wang_meeting_2008}. While substantial progress has been made on the hardware side to increase energy efficiency on the operational level, in particular through graphic processing units~ \cite{portegies_zwart_ecological_2020,rofouei_energy-aware_2008}, carbon emission of HPC systems cannot be neglected. The manufacturing or recycling processes targeted by green computing have to be considered on another level, nevertheless, contributions to green computing can be made by scientists when designing large-scale (scientific) software with energy-awareness. Aside from classical performance engineering and code optimization to speed-up algorithms or achieve more efficient hardware usage, hybrid, multiscale, or adaptive frameworks can achieve substantial performance gains or energy consumption reductions. Through the principle of computing on a finest level where necessary and aggregating information where possible,  green computing advancements can be made by scientists and software developers.

\subsection{Explicit agent- and metapopulation-based models}\label{sec:explicit_models}

In this section, we will introduce an ODE-based metapopulation model and an ABM, originally introduced in~\cite{winkelmann_mathematical_2021}. These models will be used in the applications of our hybrid modeling framework. We will extend the set of \textit{infection states} to allow for a- and presymptomatic transmission. The description of the transmission and course of the disease model is provided in~\cref{sec:cods}.  
}

We start with the ABM, whose change over time is described by a so called master equation motivated by chemical reaction systems, cf., e.g.,~\cite{gillespie_exact_1977}. While this does impose some restriction on what the agents' (inter)actions may be, it allows us to derive a piecewise deterministic metapopulation model (PDMM) that shares parameters and behavior with the ABM. The PDMM can be obtained by a model reduction of the presented ABM, see~\cite{winkelmann_mathematical_2021}. All models are implemented \MK{in C++, which is among the fastest and most efficient programming languages~\cite{pereira_energy_2017,portegies_zwart_ecological_2020}}, and within the MEmilio software~\cite{kuhn_memilio_2023} for modular epidemics simulations.

\subsubsection{A diffusion- and drift-based ABM}\label{sec:abmwinkelmann}
The nature of agent-based modeling is to capture more intuitively and to model in a direct manner individual (human) properties and interactions. Corresponding models have already been explored before the availability of modern computers; see, e.g.,~\cite{schelling_dynamic_1971}. The authors of~\cite{hunter_taxonomy_2017} have set up a so far nonexisting taxonomy for ABMs in infectious diseases and considered if particular ABMs included modules or submodels for society, transportation, and environment. 

\MK{According to our understanding, an} ABM consists of a finite number of \MK{unique} \emph{agents} $\alpha_1,\ldots,\alpha_{n_a}$ and an \emph{environment} in which and to which an agent reacts. Furthermore,
\begin{itemize}
    \item an agent is characterized by a finite number of \emph{features} that determine its \emph{state},
    \item an agent interacts with other agents and their joint environment according to \emph{interaction rules},
    \item the state of an agent or the environment changes through interactions or with time.
\end{itemize}

The features of the used ABM are infection states given by the overall infection dynamics and agents' positions determined by agents' movements. Hence, an agent is represented as a tuple $(x, z) \in \Omega \times \mathcal{Z}$, where $\Omega \subset \mathbb{R}^2$, the environment, is a compact domain and $\mathcal{Z} = (z_1, z_2, \dots , z_{n_I})$ a set of infection states. Considering $n_a$ agents, the system state is defined as a vector $Y := (X, Z) \in \mathbb{Y}:=\Omega^{n_a} \times \mathcal{Z}^{n_a}$, where an agent $\alpha$ (with its two features, position and infection state) is represented by the $\alpha$-th component of the system state $Y_{\alpha}=(x_{\alpha}, z_{\alpha}) = (X, Z)_{\alpha}$. The evolution of the system state over time is modeled as a continuous-time Markov process $(Y(t), t\geq 0)$. The evolution of the process is determined by the master equation, that describes the relevant probability measure by the change of its probability density
  \begin{align} \label{eq:MasterEquation}
    \partial_t p(X,Z;t) = G p(X,Z;t) + L p(X,Z;t) \, .
  \end{align}
The operator $G$ defines the infection state adoptions and only acts on $Z$, while $L$ defines location changes, only acting on $X$.

\MK{For the formal definition of $G$ and $L$ the reader is referred to~\cite{winkelmann_mathematical_2021}. Here, we will only describe the realization of the operators used for the later applications. Starting with the infection state dynamics,} let $i \rightarrow j$ be an infection state adoption, where $i\in \mathcal{Z}$ is the current infection state, and $j\in \mathcal{Z}\setminus\{i\}$ the infection state to adopt. For any particular agent, we assume that the likelihood of this adoption to happen in a given amount of time is determined by some rate $f_{i,j}^{(\alpha)}(Y(t))$ depending on the system state at the current time $t$. We call this rate \emph{(infection state) adoption rate}. A suitable method to realize this rate-dependent infection dynamic is using independent inhomogeneous Poisson processes $\mathcal{P}^{(\alpha)}_{i,j}(t) := \mathcal{P}_{f_{i,j}^{(\alpha)}(Y(t))}$ for infection state adoption $i \rightarrow j$ and agent $\alpha$. As a natural extension of $f_{i,j}^{(\alpha)}$, for $j=i$, we define $f_{i,i}^{(\alpha)}\equiv 0$. These rates should be interpreted as rates to change the current state.

For a given infection state adoption, various compartments can influence 
the adoption rate. The influence of an adoption $i \rightarrow j$ is defined as an index set $\Psi_{i,j} \subset \mathcal{Z}\setminus \{i\}$. We use these influences to differentiate between two types of adoptions:
\begin{enumerate}
    \item First-order adoption: An adoption event that does not require interactions with other agents, i.e., spontaneous infection state change, like recovery or death from the disease. Its adoption rate only depends on the compartment the agent is currently in, thus $\Psi_{i,j} = \emptyset$.
    \item Second-order adoption: An adoption based on pairwise interactions, i.e., infection via a contact with an infectious agent. The rate depends on the compartment of origin of an agent as well as all influences in $\Psi_{i,j} \neq \emptyset$.
\end{enumerate}
Therefore, if there are no influences, the adoption event is of first-order, and of second-order if there are any.

The adoption rates are given by adoption rate functions $f_{i,j} : \mathbb{Y} \to [0, \infty)^{n_a}$, which in general depend on the whole system state. For each agent $\alpha$, the $\alpha$-th component of $f_{i,j}$ is given by the function $f_{i,j}^{(\alpha)} : \mathbb{Y} \to [0, \infty)$, which depending on the order of the adoption $i \rightarrow j$ is defined as
\begin{align}
    f_{i,j}^{(\alpha)} (X, Z) & = \delta_i(z_\alpha) \gamma_{i,j}(x_\alpha)\\
    \intertext{for first-order adoption events, or}
    f_{i,j}^{(\alpha)} (X, Z) & = \delta_i(z_\alpha) \sum_{\tau \in \Psi_{i,j}} \sum_{\beta = 1}^{n_a} \delta_\tau(z_\beta) \gamma_{i,j,\tau}(x_\alpha, x_\beta) \label{eq:ABMSecondOrderAdoptionF}
\end{align}
for second-order adoptions. Here $\delta_i(z_\alpha)$ is the Kronecker-delta, being one when agent $\alpha$ has infection state $i$ and zero otherwise. The functions $\gamma_{i,j}: \Omega \to [0, \infty)$ and $\gamma_{i,j,\tau}: \Omega^2 \to [0, \infty)$ give the magnitude of the adoption rate, only depending on an agent's position. We choose
\begin{align} \label{eq:ABMGammaExample}
    \gamma_{i,j}(x) = c_{i,j} \text{ and } \gamma_{i,j,\tau}(x, y) = c_{i,j,\tau} \mathds{1}_{\| x - y \|_\Omega \leq r}\ ,
\end{align}
with a cutoff for second-order adoptions beyond the interaction radius $r>0$. The radius determines how close two agents have to be for being considered contacts, and we use rate constants $c_{i,j} \geq 0$, $c_{i,j,\tau} \geq 0$. We always set $c_{i,i} = c_{i,i,\tau} = 0$ as adoptions from one state to itself are not possible. Note that rate constants $c_{i,j}$ and $c_{i,j,\tau}$ could also be time-varying $c_{i,j}(t)$ and $c_{i,j,\tau}(t)$.

Agents' movement trajectories $y$ (omitting a dependency \MK{on infection state} $i$) are determined by independent diffusion processes. They are given by stochastic differential equations of the form
\begin{align} \label{eq:SDE}
    \frac{\d x(t)}{\d t} = b(t, x(t)) + \sigma(t, x(t))\xi(t)\,.
\end{align}
\MK{Here, $b:[0,T]\times \mathbb{R}^2 \to \mathbb{R}^2$ is called drift coefficient or potential on the domain,} and $\sigma:[0,T]\times \mathbb{R}^2 \to \mathbb{R}^{2 \times m}$ is called diffusion coefficient or noise. The potential can be given by \MK{functions $F(t): \Omega \to \mathbb{R}$} in which case $b\MK{(t)}:=-\nabla F\MK{(t)}$. As \MK{$b$} is assumed to be deterministic, all random behavior is solely caused by the noise. The magnitude of the noise determines the influence of the white noise process $\xi : [0, T] \to \mathbb{R}^{m}$ on the diffusion. Formally, it is defined as $\xi = \frac{\d{}W}{\d{t}}$, the derivative of the Brownian motion in $\mathbb{R}^m$. The parameter $m$ can be used to stratify agents' movement, for example with respect to infection state or region.

In practice, we use an agent's current position as initial value for \eqref{eq:SDE}, and only integrate a small time step\MK{, using the Euler-Maruyama method,} to get its new position.

\subsubsection{The corresponding piecewise deterministic metapopulation model}\label{sec:pdmm}
The corresponding piecewise deterministic metapopulation model (PDMM) simplifies the ABM by using two approximations. In the following, we will shortly describe these approximations. The full derivation of the PDMM from the ABM introduced above can be found in~\cite{winkelmann_mathematical_2021}. 

First, computational effort is reduced by discretizing the domain. The domain is split into subregions $\Omega_1,...,\Omega_{n_R}$ with $\Omega=\bigcup_{k=1}^{n_R} \Omega_k$ and agents are aggregated to subpopulations in the subregions. Hence, an agent's position is now defined only by its subregion index $k \in \{1,...,n_R\}$ instead of an exact position $x \in \Omega$. Consequently, all agents in one subregion have the same position. The system state is defined as the matrix $N = {(N^{(k)}_{i})}_{i \in \mathcal{Z}, k=1,\dots, n_R} \in \mathbb{R}_{\geq 0}^{n_I\times n_R}$, with $N^{(k)}_{i}$ the number of agents in region $\Omega_k$ and \MK{infection state} $i$. Let $e_i^{(k)}$ be the matrix where the entry for $(i,k)$ is one and all others are zero. The movement between subpopulations is modeled by Poisson processes $\mathcal{L}_i^{\left(k,l\right)}$ with rates $\lambda_i^{(k,l)}$ for spatial transitions from subregion $k$ to $l$, $k,l\in\{1,\ldots,n_R\}$. These rates depend on the transitioning agent's infection state $i$ and are of the form $N \rightarrow N + e^{(l)}_i - e^{(k)}_i$. Formally, the spatial transition rates are given by the location change operator $L$ through 
\begin{align} \label{eq:PdmmAdoptionRate}
    \lambda_i^{(k,l)} := \frac{\int_\Omega \delta_{\Omega_l}(x) L_i[\delta_{\Omega_k}(x)] \d{x}}{\int_\Omega \delta_{\Omega_k}(x) \d{x}}\,,
\end{align}
but in practice they may be easier to sample than to calculate analytically.
The space of all possible system states is given by
\begin{align}
    \mathbb{M}_{n_a} = \left\{ N \middle| \sum_{i=1}^{n_I} \sum_{k=1}^{n_R} N^{(k)}_i = n_a \right\},
\end{align}
where $n_a$ is again the total number of agents in the model. 

Secondly, model and computational cost is reduced by approximating the infection state adoption processes by a deterministic system of equations. The underlying assumption is that the populations are large such that infection state adoptions are relatively rapid and spatial transitions are comparatively rare. Using region-dependent adoption rate functions $\flow_{i,j}^{(k)} : \mathbb{R}^{n_I}_{\geq0} \rightarrow \mathbb{R}_{\geq0}$, the adoption dynamics are given by ODEs of the form
\begin{align} \label{eq:PDMMBasic}
    \frac{d}{dt} N_i^{(k)}(t) = \sum_{\substack{j \in \mathcal{Z} \\ i\neq j}}(\flow_{j,i}^{(k)}(N^{(k)}(t)) - \flow_{i,j}^{(k)}(N^{(k)}(t)))\ ,
\end{align}
without accounting for spatial transitions. The evolution of the system state $N(t)$ over time is described by a continuous-time Markov jump process. Writing $\left(N(t)\right)_{t\geq 0}$ as a path, we obtain
\begin{align} \label{eq:PDMMState}
\begin{aligned}
N(t) = N(0) & + \underset{k\neq l}{\sum_{k,l=1}^{n_R}} \sum_{i=1}^{n_I} \mathcal{L}_{i}^{(k,l)}\left( \int_0^t\lambda_i^{(k,l)} N_i^{(k)}(s) \text{d}s \right)(e_i^{(l)} - e_i^{(k)}) \\
& + \sum_{k=1}^{n_R} \sum_{i,j=1}^{n_I} \int_0^t\flow_{i,j}^{(k)}(N(s)) \text{d}s (e_j^{(k)} - e_i^{(k)})\ .
\end{aligned}
\end{align}
For the full derivation from the ABM through a stochastic metapopulation model to the PDMM using convergence results for Markov processes, see again~\cite{winkelmann_mathematical_2021}.

\subsubsection{Transmission and course of the disease}
\label{sec:cods}
As course of the disease model, we use a Susceptible-Exposed-Carrier-Infected-Recovered-Dead model, thus $\mathcal{Z}=\left(S, E, C, I, R, D\right)$. Individuals in the \textit{Carrier} compartment may be either pre- or asymptomatic, hence they do not show symptoms while the \textit{Infected} compartment encompasses all individuals showing symptoms ranging from mild to severe. Both compartments are infectious to \textit{Susceptibles}. There is only one second-order adoption (see ~\cref{fig:transmission_model}) which is $S \rightarrow E$ with influences $\Psi_{SE}=\{C,I\}$. As there is no outflow from the $R$ compartment, this model does not consider the possibility of reinfection. 

\begin{figure}
\tikzstyle{box}=[draw, rounded corners=2pt, minimum size=0.7cm, align = center, minimum height = 0.7cm, line width = 0.5pt]
\tikzstyle{arrow}=[->, black, thick, text = black]
    \centering
    \begin{tikzpicture}
        \node[box] at (0,0) (S) {S};
        \node[box, right = 2cm of S] (E) {E};
        \node[box, right = 2cm of E] (C) {C};
        \node[box, right = 2cm of C] (I) {I};
        \node[box, below left = 1.5cm and 0.7cm of I] (R) {R};
        \node[box, below right = 1.5cm and 0.7cm of I] (D) {D};
        \draw[arrow] (S) -- node [above] {} (E);
        \draw[arrow] (E) -- node [above] {} (C);
        \draw[arrow] (C) -- node [above] {} (I);
        \draw[arrow] (C) |- node [above left=0.7cm and 0.1cm] {} (R);
        \draw[arrow] (I) |- node [above left=0.7cm and 0.1cm] {} (R);
        \draw[arrow] (I.east) -| node [below right=0.7cm and 0.1cm] {} (D);
        \draw[thick,dotted] ($(C.north west)+(-0.2,0.2)$)  rectangle ($(I.south east)+(0.2,-0.2)$);
        \draw[thick,dotted] (6, -1) -- (6,-0.7);
        \draw[arrow, thick,dotted] (6, -1) -| (1.8,0);
    \end{tikzpicture}
    \caption{\textbf{Flow chart of the \MK{infection state} adoption model.} \MK{The infection states are Susceptible (S), Exposed (E), Carrier (C), Infected (I), Recovered (R) and Dead (D).} The dotted lines shows the influences for the second-order adoption.}
    \label{fig:transmission_model}
\end{figure}

Given the courses of the disease defined here, the local equations giving the \MK{infection state} dynamics in \MK{region $\Omega_k$ of the PDMM from~}\cref{sec:pdmm} write
\begin{align}
\begin{aligned}
    \frac{d}{dt} N_S^{(k)} &= -\frac{\rho^{(k)}\left(\xi_C^{(k)} N_C^{(k)} +\xi_I^{(k)} N_I^{(k)}\right)}{n_a^k} N_S^{(k)},\\
    \frac{d}{dt} N_E^{(k)} &= \frac{\rho^{(k)}\left(\xi_C^{(k)} N_C^{(k)} +\xi_I^{(k)} N_I^{(k)}\right)}{n_a^k} N_S^{(k)} - \gamma_{E,C}^{(k)}N_E^{(k)},\\
    \frac{d}{dt} N_C^{(k)} &= \gamma_{E, C}^{(k)}N_E^{(k)} - (\gamma_{C, R}^{(k)}+\gamma_{C, I}^{(k)})N_C^{(k)},\\
    \frac{d}{dt} N_I^{(k)} &= \gamma_{C, I}^{(k)}N_C^{(k)} - (\gamma_{I, R}^{(k)}+\gamma_{I, D}^{(k)})N_I^{(k)},\\
    \frac{d}{dt} N_R^{(k)} &= \gamma_{C, R}^{(k)}N_C^{(k)}+\gamma_{I, R}^{(k)}N_I^{(k)},\\
    \frac{d}{dt} N_D^{(k)} &= \gamma_{I, D}^{(k)}N_I^{(k)}.
\end{aligned}
\end{align}

\subsection{A framework of hybrid modeling \MK{and its application}}
\label{subsec: Hybrid_modeling}

\begin{figure}
    \centering
    \sidesubfloat[]{\includegraphics[width=0.24\textwidth]{"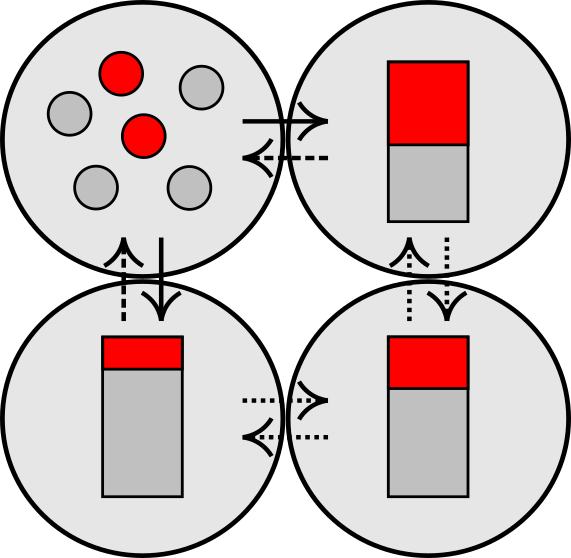"}\hspace{0.5cm}\label{fig:spatial_hybrid}}%
    \sidesubfloat[]{\includegraphics[width=0.68\textwidth]{"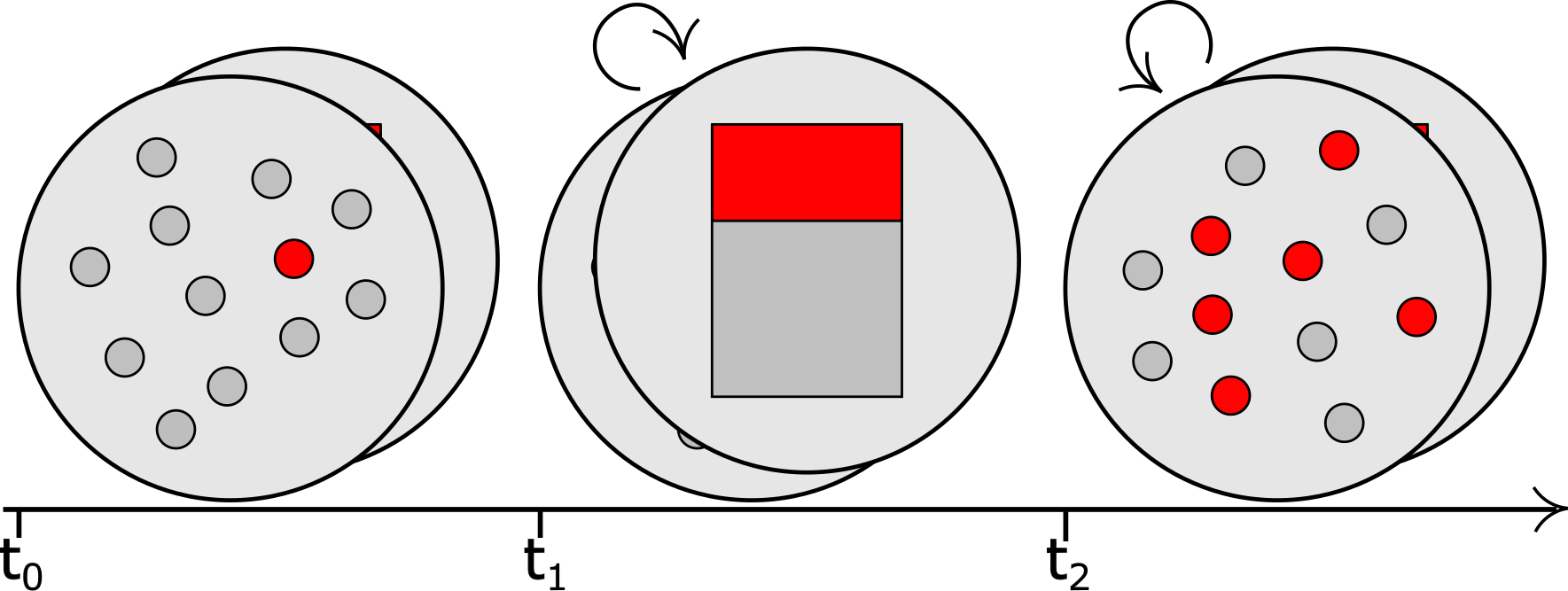"}\label{fig:temporal_hybrid}}%
    \caption{\textbf{Design sketch of a spatial-hybrid model (a) and a temporal-hybrid model (b).} Large gray disks represents (sub)regions, with possible population exchanges indicated by arrows. The population is represented by small disks for agents and bar charts for ODE compartments. Red indicates infected agents or population shares, gray indicates noninfected agents or population shares. The exchange in (a) only moves small parts of a local population to another region, the exchange in (b) moves the entire population to the other model.}
    \label{fig:hybrid_models}
\end{figure}

\MK{In this section, w}e introduce \MK{the general concept of} two different types of hybrid epidemiological models: spatial-hybrid and temporal-hybrid models. These approaches do not exclude each other so that also spatio-temporal-hybrid models can be realized. Both hybridization approaches are shown in~\cref{fig:hybrid_models}. \MK{For both approaches, we will provide explicit \MK{applications} with the models presented in the previous section.}

The motivation to use \MK{population}-based models combined with ABMs for disease dynamics is to reduce the computational effort of \MK{individual-level} models which scale with the number of considered agents or individuals. In particular, we here provide the computational cost for the two explicit models introduced before. For both models, we need to consider computational costs for spatial transitions or movements and infection state adoption dynamics. The movement of agents in the ABM is given by a diffusion process that has to be evaluated for every agent. Assuming a constant time for the evaluation, the cost to calculate movement dynamics in the ABM lies in $\mathcal{O}(n_a)$. Furthermore, the infection state adoption rates have to be evaluated in every iteration. As first order adoption rates are computed per agent, the complexity as well is in $\mathcal{O}(n_a)$. For second-order adoptions, pairwise comparison of agents is necessary, hence, we obtain a superlinear complexity and in the worst case one iteration has computational cost in $\mathcal{O}(n_a^2)$. For the PDMM the cost for infection state adoption dynamics in one iteration is given by integrating the system of differential equations, thus it is independent of $n_a$. Technically, the cost for integration scales with the number of equations meaning the number of regions and compartments, but their size is negligibly small compared to $n_a$. Therefore, the relevant complexity and runtime factor in the PDMM is given by the spatial transition events. The frequency of the spatial transitions depends on the magnitude of the rates. As the transition rates are multiplied with the corresponding subpopulation, see~\eqref{eq:PDMMState}, the complexity of the spatial dynamics lies in $\mathcal{O}(n_a)$. A summary of the model characteristics is given in~\cref{tab:Model comparison}.

\begin{table}[H]
    \centering
    \begin{tabular}{l|c|c}
         & \textbf{ABM} & \textbf{PDMM}  \\
         \hline\hline
         \textit{Complexity} & superlinear, up to $\mathcal{O}(n_a^2)$ &  $\mathcal{O}(n_a)^*$ \\
         \hline
         \textit{Spatial domain} & continuous & discrete\\
         \hline
         \textit{Adoption dynamics} & stochastic & deterministic
    \end{tabular}
    \caption{\textbf{Model comparison of ABM and PDMM.} The term $\mathcal{O}(n_a)^*$ describes the runtime scaling due to \MK{spatial} transition rates and can be replaced by \MK{$\mathcal{O}(n_rn_I)$, with $n_rn_I \ll n_a$}, under the assumption that spatial transitions are several orders of magnitude lower than infection state adoptions.}
    \label{tab:Model comparison}
\end{table}

A requirement to the two models used is that we can define a mapping between them. In its simplest form,  the coarse-granular model may be obtained by projection from the fine model. In a more elaborated setting, we need to map fine-granular, continuous features like viral load to a discrete, coarse-granular definition. Naturally, going from fine-to-coarse is easier as information gets reduced. Going from coarse-to-fine either needs sampling or the replication of retained information to realize a realistic model over the whole simulation horizon (spatial or temporal). \MK{While in the current paper we only needed to sample a corresponding position in a 2D plane, more sophisticated sampling (based on retained information on travelers) will be necessary for more realistic ABMs. The corresponding sampling then could include information on households or workplaces and agents' immune histories.}
\MK{In general, the coarse (and hybrid) results should approximate the fine result. One condition for this would be that the models' mean results converge. In addition, both models should match in the sense that there is a projection from the fine to the coarse model population, with a suitable mapping from coarse to fine such that the information lost when retrieving the fine model population from the coarse model, does not exceed the approximation error. Nevertheless, to determine whether the results of a particular hybrid model will be accepted depends on individual evaluation of the trade-off between approximation error and speed-up.}

\subsubsection{Spatial hybridization}
\label{sec: spatial-hybrid}

The idea of a \emph{spatial-hybrid model} originates from research questions on fine-granular infection spread, like household or workplace transmission in a particular region. To consider only this (focus) region would neglect the influence of connected regions on infection dynamics through, e.g., commuting activities. However, the use of fine-granular models for all connected regions might be prohibitively expensive or individual-level data is either exclusively available in the focus region, or substantially costly to collect for outside regions.

\MK{Consequently, we model only the} focus region by a fine-granular model and use coarse-granular models in any connected region. With this approach we get results with the desired resolution in the focus region while considering the dynamic, time-dependent influence of the connected regions in a runtime efficient manner. In the following, we will denote this approach as \emph{spatial hybridization} and will refer to hybrid models with higher granularity in a focus region as \emph{spatial-hybrid models}. 

For the spatial hybridization, we require a disjoint partition of the modeled domain into at least two subregions. The exchange of populations between the different regions can be realized with different mobility models, however, sending agents to a coarser model does cause loss of information (e.g. sending an agent to a simple ODE-based model will only retain its infection state and lose all spatial information). To the best of our knowledge, there is only little theory on this exchange problem and the exchange is highly dependent on the concrete models used. Some discussions and suggested models can be found in~\cite{sewall_interactive_2011, bradhurst_hybrid_2015,hunter_hybrid_2020}.
\MK{While a particular goal is that the hybridization approaches the ABM outcome, for instance,} ODE-based models are not able to let a virus go completely extinct\MK{ while} ABMs can simulate this extinction. This does not mean that a spatial-hybrid of ODE-ABM cannot be used if infection numbers are small in a region \MK{where the potential of extinction is real. This only means} that the results may differ from the results of a fully fine-resolved (stochastic) model and hence that this \MK{limitation and its impact} needs to be considered when evaluating simulation results. For further discussion on this effect, see e.g.~\cite{schaller_continuum_2006,shnerb_importance_2000}. More suitable combinations could use stochastic differential equation-based models or a mixed spatial-temporal-hybrid (in combination with the next section). 

Two ways to obtain compatible models are to fit the two models against each other, or to derive the coarse model (parameters) from the fine one. However, in both cases it is important to understand the model behavior very well to be aware of inter-dependencies. For instance, location and infection state changes may depend on each other and cannot always be fitted separately. An interesting discussion can be found in~\cite{NIEMANN2024102242}.

\MK{In the following}, we will provide some further details on the precise realization \MK{of the spatial hybridization for the ABM and PDMM introduced in~\cref{sec:explicit_models}. Let} $\Omega_1$ be the focus region. 
For the precise implementation, ABM and PDMM are set up for the whole domain while both models have no movement or infection state adoptions in the non-relevant regions meaning the focus region for the PDMM and all regions outside the focus regions for the ABM, i.e., the corresponding rates are zero. Then, both models are run independently from each other as long as no agents are leaving or entering the focus region. On time points for population exchange, models will be synchronized. As mentioned before, the exchange of agents and populations between the models is a crucial aspect of the hybridization. As the mobility process of the PDMM is also based on agents and not on shares of the whole population \MK{and as we set the adoption rates in the PDMM in the focus region to zero, it holds $N_i^{(1)}\in \mathbb{N}$ for $i\in\mathcal{Z}$}. The exchange between \MK{the models is done as follows}:
\begin{itemize}
    \item ABM $\rightarrow$ PDMM: \MK{Let $t$ be the time point of the last exchange and $Y^{-}(t, t') \subseteq Y(t')$ be the set of agents that need to be exchanged from ABM to PDMM, i.e., agents with positions $x \notin \Omega_1$, at time point $t'>t$.} As we switch from fine-granular to the corresponding coarse-granular model, the exchange can be done by a trivial projection. \MK{Agents in $Y^{-}(t, t')$} are just added to the subpopulation $N_i^{(\MK{\phi(x)})}(t')$ in the PDMM according to their infection state $i\in\mathcal{Z}$ \MK{and a projection $\phi: \Omega \rightarrow \{2,...,n_R\}$ with $\phi(x)=k$ for $x\in\Omega_k$}. Subsequently, this agent is removed from the ABM.
    \item PDMM $\rightarrow$ ABM: The exchange from coarse-granular to fine-granular \MK{model} is not as trivial\MK{, since} the PDMM only has \MK{spatial} information aggregated to \MK{subregions} and\MK{, hence,} misses concrete positions inside the subregions. For the \MK{exchange at time point $t'$, $\sum_{i \in \mathcal{Z}}N_i^{(1)}(t')$ new agent tuples are created and added to the ABM system state $Y(t')$. The infection state $z\in\mathcal{Z}$ of a newly created agent is directly given by the PDMM subpopulation index. Its position $x\in\Omega_1$ is sampled from a distribution $\mathcal{P}$ which depends in general on the ABM movement operator $L$, see~\cref{eq:MasterEquation}. As we use a diffusion process in our applications, $\mathcal{P}$ depends on the potential $b$, see~\cref{eq:SDE}, and will be precisely defined in~\cref{sec::spatial_hybrid_results} for the two examples provided there.}
\end{itemize}

To get the initial populations for the spatial-hybrid model, we first only add agents to the ABM and then make one exchange step to set the initial system state $N(0)$ in the PDMM. With this, all agents outside $\Omega_{1}$ are removed from the ABM, and in the PDMM the population in $\Omega_{1}$ is set to $0$.

\MK{The implementation of the spatial hybridization is shown in Algorithm~\ref{alg:Spatial_hybridization}.}

\begin{algorithm}[t]
\MK{
    Create ABM and PDMM for $\Omega=\bigcup_{k=1}^{n_R} \Omega_k$ and $t=t_0$\;
    Set and restrict rates and populations, see~\cref{sec: spatial-hybrid}\;
    \hspace{0.5cm}ABM for $\Omega_1$\;
    \hspace{0.5cm}PDMM for $\Omega_2,\ldots,\Omega_{n_R}$\;
   \hspace{0.0cm}\textbf{While} $t\in[t_0, t_{max}]$ \textbf{do}\;
        \hspace{0.5cm}Define next synchronization point $t'$\;
        \hspace{0.5cm}Advance ABM and PDMM from $t$ to $t'$\;
        \hspace{0.5cm}\textcolor{green!50!black!90}{/$\ast$ Exchange agents between models $\ast$/}\\
        \hspace{0.5cm}\textbf{Foreach} $\alpha \in Y^{-}(t, t')$\;
            \hspace{1cm} Set $N_{z_\alpha}^{(\phi(x_\alpha))}(t')\mathrel{+}=1$\;
            \hspace{1cm} Remove $\alpha$ from $Y(t')$\;
        \hspace{0.5cm}\textbf{Foreach} $i \in \mathcal{Z}$\;
             \hspace{1.0cm}\textbf{While} $N_i^{(1)}(t')>0$ \textbf{do}\;
                \hspace{1.5cm} Sample position $x\sim \mathcal{P}$\;
                \hspace{1.5cm} Add agent $(x, i)$ to $Y(t')$\;
                \hspace{1.5cm} Set $N_i^{(1)}(t')\mathrel{-}=1$\;
        \hspace{0.5cm}Set $t=t'$\;
    }
\caption{Spatial hybridization}
\label{alg:Spatial_hybridization}
\end{algorithm}

\subsubsection{Temporal hybridization}
\label{subsubsec: temporal-hybrid}
The idea of \emph{temporal-hybrid models} is based on the assumption that the lower the case numbers, the higher is the impact of stochastic events to the (simulation) outcome. On the other hand, if case numbers are high, individual behavior and stochastic events become less influential. This means that single simulation results of stochastic models come closer to averaged outcomes \MK{and hence mean-field models deliver a suitable approximation also for \MK{individual} simulations.} 

The temporal hybridization is based on the same requirements as the spatial-hybrid model, except that we only use one model for the whole domain and switch between coarse- and fine-granular model during the course of the simulation. Then, instead of exchanging agents or populations on spatial transitions, we exchange the entire model at certain time points. These time points depend on the current system state and are thus selected dynamically. As a consequence, we need fine granular data for the whole domain.

In~\cref{sec: results_temporal_hybrid}, we use an ABM to model disease spread when the number of infected individuals is below a predefined threshold and switch to a coarser, hence ODE-based or metapopulation model, as soon as the number of infected individuals exceeds the threshold. \MK{We denote hybrid models that switch between fine and coarse model using a predefined, time-dependent criterion as} \emph{temporal-hybrid models}. In literature this kind of hybrid model is also referred to as stage-based hybrid model, see e.g. \citep{bradhurst_hybrid_2015} and \citep{bobashev_hybrid_2007}.
In case of the virus-extinction example from above, we could change from an ODE-based or metapopulation model to an ABM whenever the number of virus carriers drops below a threshold that would realistically allow an extinction to occur and switch back whenever the virus will, with some certainty or probability, spread to a larger population. Note that this approach needs careful evaluation of the switching condition. 

Like for the spatial-hybrid model the exchange of information between models is crucial and dependent on the concrete models used. A benefit of a temporal-hybrid model is that it can effectively save computational cost in periods of diffusive spread while maintaining the required level of detail in highly stochastic periods.

The setup for the temporal-hybrid model is similar to the implementation of the spatial hybridization described in~\cref{sec: spatial-hybrid}, in so far as both models are setup with matching parameters. \MK{Instead of restricting the models to spatial regions, both are} used on the whole domain. However, only one model runs at a time. To decide which model that is, we define a function $\Gamma$ that returns, based on time, the system state at $t$ and 
predefined criteria, a model $\{\textnormal{ABM},\textnormal{PDMM}\}$, i.e., $\Gamma$ is a function that determines with which model to continue from a given time point. In the application presented in~\cref{sec: results_temporal_hybrid}, the model choice depends on the number of infectious agents, which makes most sense as infectious disease dynamics are driven by the stochasticity in the behavior of infectious agents. The procedure \MK{of the temporal hybridization} is shown in ~\cref{alg:Temporal_hybridization}.
\begin{algorithm}[t]
    Create ABM and PDMM for $\Omega$ at $t=t_0$\;
    Select one model as $m_{current}\in\{\textnormal{ABM},\textnormal{PDMM}\}$\;
    Initialize population only for $m_{current}$\;
    Define time step $\Delta t$\;    
    \hspace{0.0cm}\textbf{While} $t\in[t_0, t_{max}]$ \textbf{do}\;   
    \hspace{0.5cm}$m_{next} =  \Gamma(m_{current}, t)$\;
        \hspace{0.5cm}\textbf{If} $m_{current} \neq m_{next}$\;
            \hspace{1.0cm}Move the population from $m_{current}$ to $m_{next}$\;
            \hspace{1.0cm}$m_{current}$ = $m_{next}$\;
        
        \hspace{0.5cm}Set $t=t+\Delta t$\;
        \hspace{0.5cm}Advance $m_{current}$ to $t$\;
\caption{Temporal hybridization}
\label{alg:Temporal_hybridization}    
\end{algorithm}

Moving the population between \MK{both} models follows the same principle as the agent exchange described in~\cref{sec: spatial-hybrid}, applied to all agents in all regions at once\MK{, i.e., for changing from ABM to PDMM agents are just aggregated to subpopulations according to their position and infection state and for the change from PDMM to ABM, agents' concrete positions on the domain are sampled from a distribution $\mathcal{P}$ while their infection state is given by the subpopulation index. The distribution $\mathcal{P}$ we used for our examples will be defined in~\cref{sec: results_temporal_hybrid}.}

\section{Results}
\label{sec::Results}

In this section, we provide simulation results for the agent-based and the metapopulation model presented in~\cref{sec:explicit_models}. We will compare these results against our spatial- and temporal-hybrid models.
 \MK{The parameters related to infection dynamics are motivated by wild-type COVID-19 and based on~\cite{kuhn_assessment_2021} for all examples (see~\cref{tab:joint_transmission_params}) -- apart from the transmission rate $\rho^{(k)}$ from Susceptible to Exposed, which varies between the examples and also between regions of one example. The chosen values for the transmission rate $\rho^{(k)}$ are within the range presented in~\cite{kuhn_assessment_2021} and lead to initial reproduction numbers $R_0$ between $0.8$ and $5$, roughly corresponding to estimates for the effective reproduction number for wild-type COVID-19 in different German Federal states between March and August 2020, see~\cite{khailaie_development_2021}. However, the examples presented use simplified models as the scope of the current paper is the disease-agnostic introduction of novel hybrid models. Therefore no further parameter refinement for COVID-19 and comparison to real-world data has been done.} All models are simulated using a temporal Gillespie algorithm, allowing us to obtain stochastically exact results without sampling the distribution~\cite{djurdjevac_conrad_human_2018,vestergaard_temporal_2015}. We will highlight differences and similarities in the particular model outcomes and compare model runtimes as a proxy indicator for computational effort. All simulations were conducted on a Intel Xeon "Skylake" Gold 6132 (2.60 GHz) with four nodes with 14 CPU cores each and 384GB DDR4 memory, using separate cores to conduct Monte Carlo runs in parallel.

\subsection{Spatial Hybridization}
\label{sec::spatial_hybrid_results}
In the following, we present the results for two applications of the spatial hybridization. The first application naturally extends the theoretical setting of~\cite{winkelmann_mathematical_2021} by using four instead of two potential wells (i.e., local minima of the potential) and significantly more agents. In the second application, we consider \MK{a potential depicting} the city of Munich and its surrounding counties. \MK{For both applications, we performed a sensitivity analysis to determine the most influential parameters, see~\ref{sec:sensitivity_analysis}. Additionally, the implicit influence of parameters on the runtime of ABM, PDMM, and spatial-hybrid model was analyzed in~\ref{sec:Runtime analysis}}.

\subsubsection{Application 1 - A mathematical quadwell example}
\label{sec:qw_example}
\begin{figure}[b]
    \centering
    \sidesubfloat[]{\includegraphics[width=0.35\textwidth]{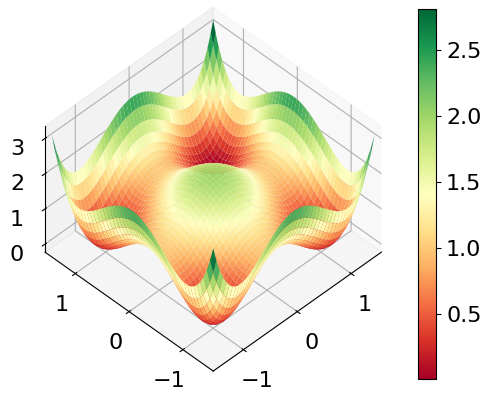}\label{fig:qw_a}}\hspace{0.5cm}
    \sidesubfloat[]{\includegraphics[width=0.4\textwidth]{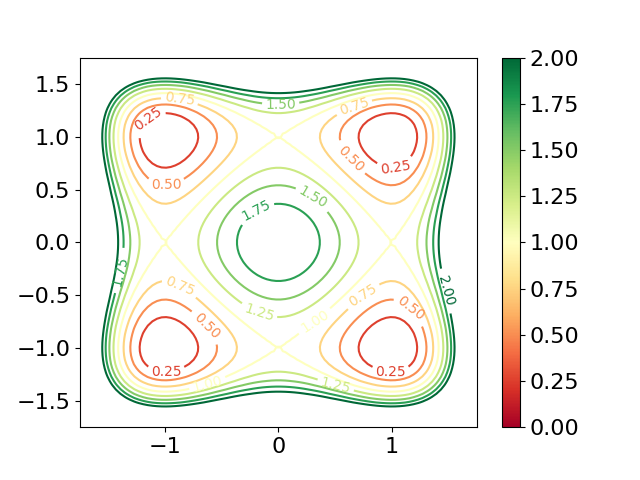}\label{fig:qw_b}}
    \caption{\textbf{Quadwell potential~\eqref{eq:quadwell} used for the diffusion process in the ABM.} The four metaregions are separated by the axes $x=0$ and $y=0$. The figure shows an (a) isometric view and (b) a contour plot of the potential.}
    \label{fig:quadwell potential}
\end{figure}

\begin{figure}
    \centering
    \includegraphics[width=0.6\textwidth]{"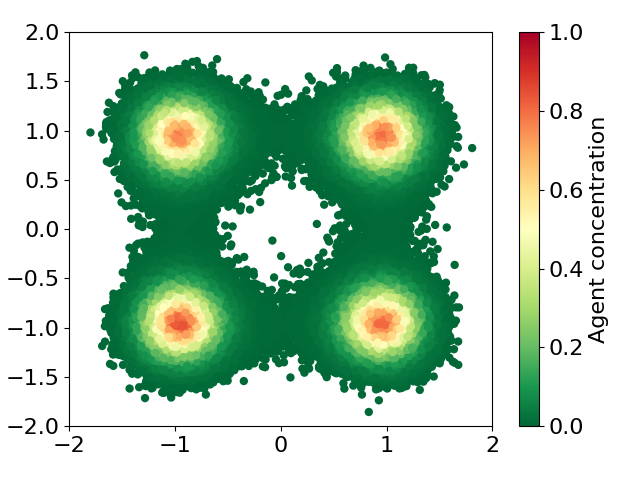"}
    \caption{\textbf{Position distribution for a simulation of $800$ agents for $50$ days in the quadwell scenario for noise term $\sigma=0.55$.} Agents are initialized with positions having $0.3$ distance from the axes \MK{$x=0,\pm 2$ and $y=0, \pm 2$.}}
    \label{fig:sigma_0.55}
\end{figure}

\MK{In our first application, we use a function with four minima - a quadwell potential - and a  constant noise term $\sigma \in \mathbb{R}$ for the diffusion process modeling agent movement in the ABM, see~\cref{eq:SDE}.} The quadwell potential is \MK{constant in time and} given by
\begin{align}\label{eq:quadwell}
    F(x,y)=(x^2-1)^2+(y^2-1)^2;
\end{align}
see~\cref{fig:quadwell potential}.
\MK{This function naturally defines four subregions for the PDMM that are given by the potential wells as $\Omega_1=(-\infty,0)\times(0,\infty), \Omega_2=(0,\infty)\times(0,\infty), \Omega_3=(-\infty,0)\times(-\infty,0)$ and $\Omega_4=(0,\infty)\times(-\infty,0)$.} This corresponds to numbering the wells in~\cref{fig:qw_b} from 1 to 4 starting with the upper left and ending with the lower right well, numbered line by line. The magnitude of the noise term determines the number of transitions between metaregions, see~\cref{fig:sigma_movement}. \MK{As the infection dynamics in the focus region should be influenced by the dynamics in the other regions, we do want spatial transitions between the metaregions, which do not occur for $\sigma=0.3$ (\cref{fig:sigma_0.3}) and only very few for $\sigma=0.5$ (\cref{fig:sigma_0.5}). However, the spatial transitions should not happen as frequent as for $\sigma=0.6$ (\cref{fig:sigma_0.6}) as the assumption for the runtime advantage of the PDMM is that spatial transitions are rare compared to infection state adoptions (see~\cref{tab:Model comparison}). Therefore, we chose $\sigma=0.55$ for our application, see~\cref{fig:sigma_0.55}.}
\MK{Given the ABM noise term $\sigma$, the spatial transition rates for the PDMM can be sampled by calculating the average number of transitions in a unit-time step relative to the population in the subregion.}
We chose $\rho^{(k)}=0.1$ for $k=1,3,4$ corresponding to $R_0=0.8$ in region $\Omega_1, \Omega_3$ and $\Omega_4$ and $\rho^{(2)}=0.3$ corresponding to $R_0=2.4$ in region $\Omega_2$. The parameters used for the quadwell potential are summarized in~\cref{tab:joint_transmission_params} and~\cref{tab:params_qw}. The system state at four different time points for one ABM realization with $1000$ agents in~\cref{fig:qw_trajectory} shows that, due to the higher transmission rate, the infection spreads much faster in $\Omega_2$ compared to the other regions. We define $\Omega_1$ as the focus region for the spatial-hybrid model and simulate the ABM, the PDMM and the spatial-hybrid model for $t_{max}=150$ days.
\MK{When agents transition to the focus region in the spatial-hybrid model, their exact position has to be sampled from a distribution $\mathcal{P}$, see~\cref{alg:Spatial_hybridization}. For this application we use a normal distribution whose parameters, i.e., mean and variance depend on the region an agent transitions from. The mean and variance were calculated from the transitions of $2850$ agents shown in~\cref{fig:qw_trans_dist}. Denoting the region the transitioning agent came from as $\Omega_\alpha^{(old)}$, this yields for the new position $x_\alpha^{(new)}\in \Omega_1$ of agent $\alpha$, $x_\alpha^{(new)} \sim \mathcal{P}_\alpha$ with
\begin{align}
    \mathcal{P}_\alpha = \begin{cases}
        \left(\mathcal{N}(-0.1, 0.03^2), \mathcal{N}(0.94, 0.19^2)\right), &\text{if } \Omega_\alpha^{(old)} = \Omega_2\\[7pt]
        \left(\mathcal{N}(-0.94, 0.19^2), \mathcal{N}(0.1, 0.03^2)\right), &\text{if } \Omega_\alpha^{(old)} = \Omega_3
        \\[7pt]
        \left(\mathcal{N}(-0.1, 0.03^2), \mathcal{N}(0.1, 0.03^2)\right), &\text{if } \Omega_\alpha^{(old)} = \Omega_4
    \end{cases}
\end{align}}

\begin{figure}
    \centering
    \sidesubfloat[]{\includegraphics[width=0.33\textwidth]{"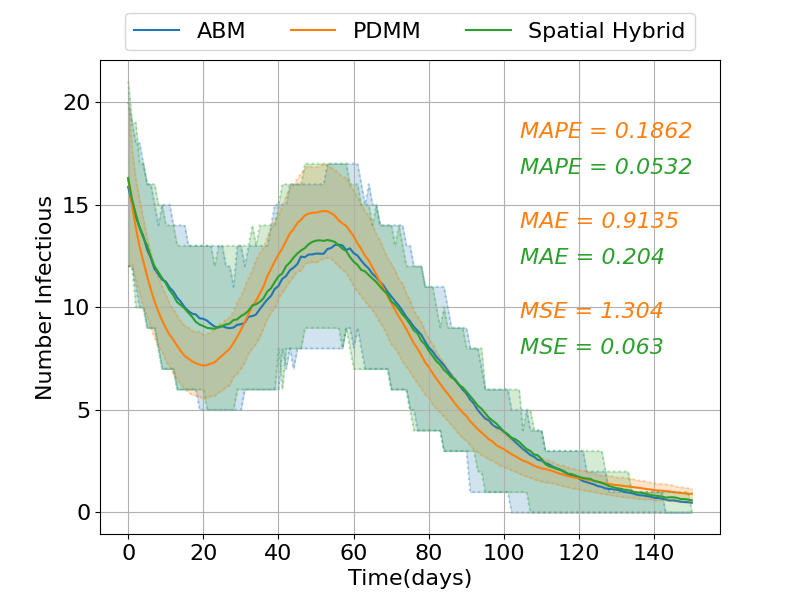"}\label{fig:qw_focus}}%
    \sidesubfloat[]{\includegraphics[width=0.33\textwidth]{"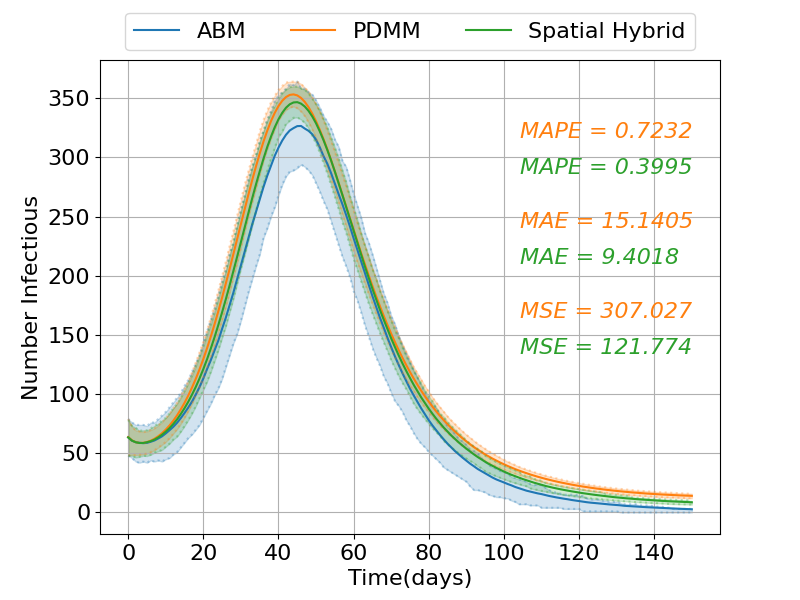"}\label{fig:qw_all}}
    \sidesubfloat[]{\includegraphics[width=0.33\textwidth]{"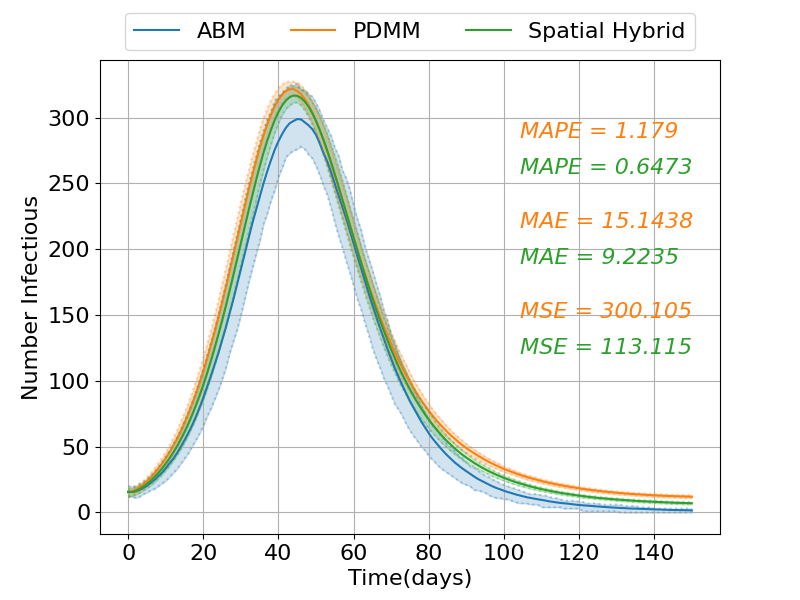"}\label{fig:qw_r2}}%
    \caption{\textbf{Spatial hybridization for the quadwell potential: Focus region, sum of all regions and Region 2.} Number of infectious agents (compartments $C$ and $I$) for (a) the focus region $\Omega_1$ (b) the sum of all regions and (c) Region 2. The figures show the mean outcomes in solid lines with a partially transparent face between the p25 and p75 percentiles from 500 runs. \MK{Additionally, MAPE, MAE, and MSE between the ABM mean and the mean of PDMM and spatial-hybrid model are displayed.}}
    \label{fig:results_qw}
\end{figure}

 We chose $n_a=8000$ agents with $1\%$ of the population being initially infected in every well ($0.2\%$ Exposed, $0.3\%$ Carrier and $0.5\%$ Infected). The results of $500$ simulations are shown in~\cref{fig:results_qw}. \MK{We consider the ABM - the detailed model - as ground truth and calculate the \textit{Mean Absolute Percentage Error (MAPE)}, the \textit{Mean Absolute Error (MAE)} and the \textit{Mean Squared Error (MSE)} between the ABM mean and the mean of PDMM and spatial-hybrid model.} Comparing \cref{fig:qw_all} and \cref{fig:qw_r2}, it can be seen that region 2 is the main driver of the infection dynamics \MK{i.e., most infectious agents are in $\Omega_2$. \cref{fig:qw_focus} shows that in the focus region the spatial-hybrid model approximates the ABM outcomes much better than the PDMM as the percentiles of the hybrid model - in contrast to the PDMM percentiles - almost completely overlap the ABM percentiles. Furthermore, all error metrics calculated on the mean of simulation outcomes are smaller for the spatial-hybrid model.} As $\Omega_2$ is not the focus region, this region is modeled with the PDMM in the spatial-hybrid model. Therefore the PDMM and hybrid model curves \MK{for $\Omega_2$, especially the percentiles, have a similar shape, however the hybrid model has smaller errors than the PDMM, see~\cref{fig:qw_r2}}. For the sake of completeness, the results for region 3 and region 4 are shown in~\cref{fig:results_qw_r3_r4}.

\begin{figure}
    \centering
    \sidesubfloat[]{\includegraphics[width=0.4\textwidth]{"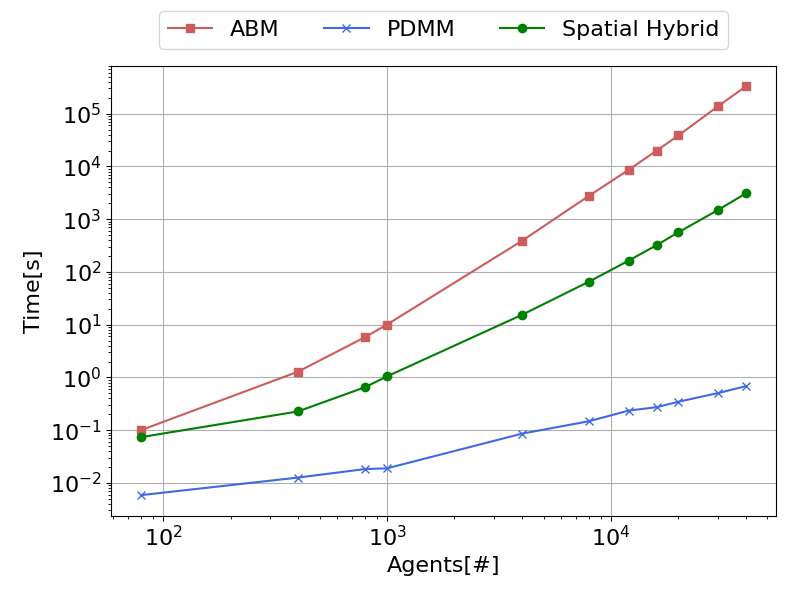"}\label{fig:qw_runtime}}%
    \sidesubfloat[]{\includegraphics[width=0.4\textwidth]{"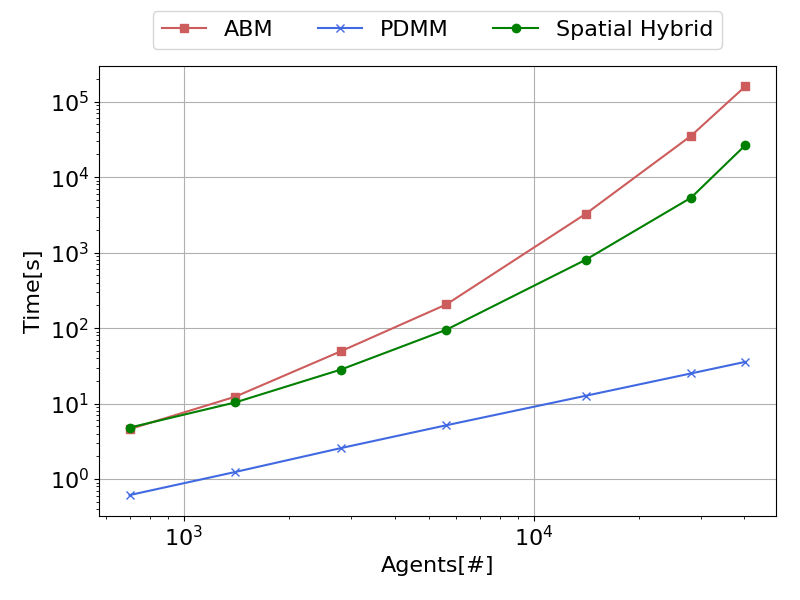"}\label{fig:munich_runtime}}%
    \caption{\textbf{Log-scaled runtime (in seconds) for ABM, PDMM and spatial-hybrid model.} \MK{Shown is the mean runtime of $56$ runs (one compute node)} for (a) the quadwell example with the setup according to~\cref{tab:joint_transmission_params}, ~\cref{tab:params_qw} and (b) the Munich example with the setup according to~\cref{tab:joint_transmission_params}, ~\cref{tab:params_munich}.}
    \label{fig:runtime}
\end{figure}

The runtime of the spatial-hybrid model \MK{for the chosen setup} is dominated by the ABM, but is still one order of magnitude smaller than the ABM runtime for $400$ agents and even two orders of magnitude for $16,000$ agents, see~\cref{fig:qw_runtime}. For $40,000$ agents, the runtime for the described scenario can be reduced by $98\%$ using the spatial-hybrid model.

\begin{figure}
    \centering
    \sidesubfloat[]{\includegraphics[width=\textwidth]{"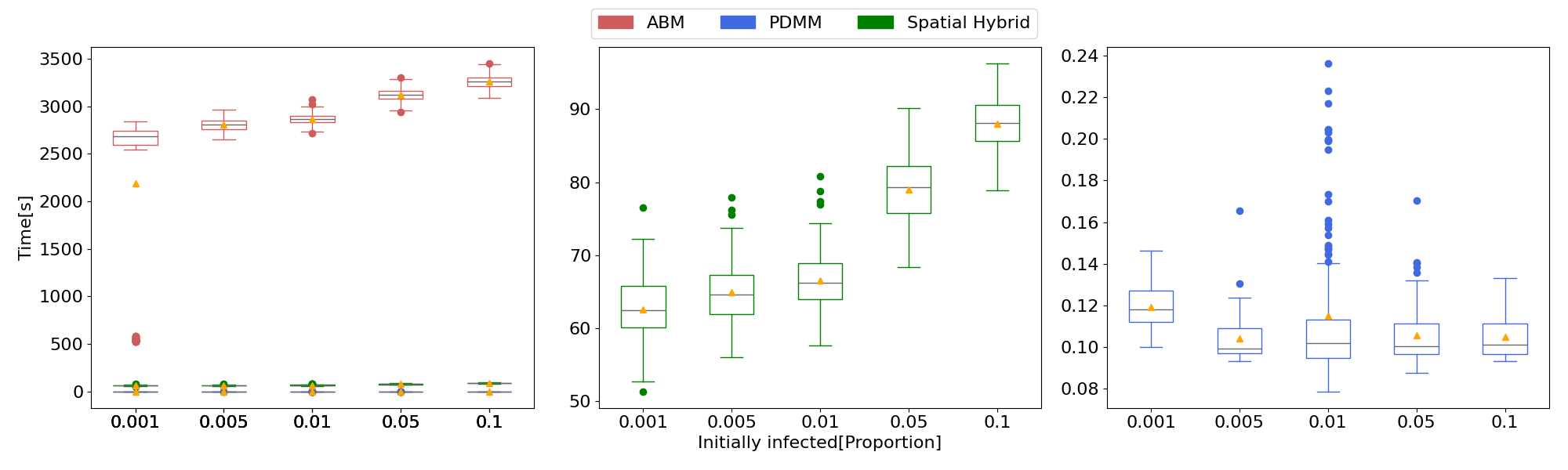"}\label{fig:qw_initially_infected}}\\
    \sidesubfloat[]{\includegraphics[width=\textwidth]{"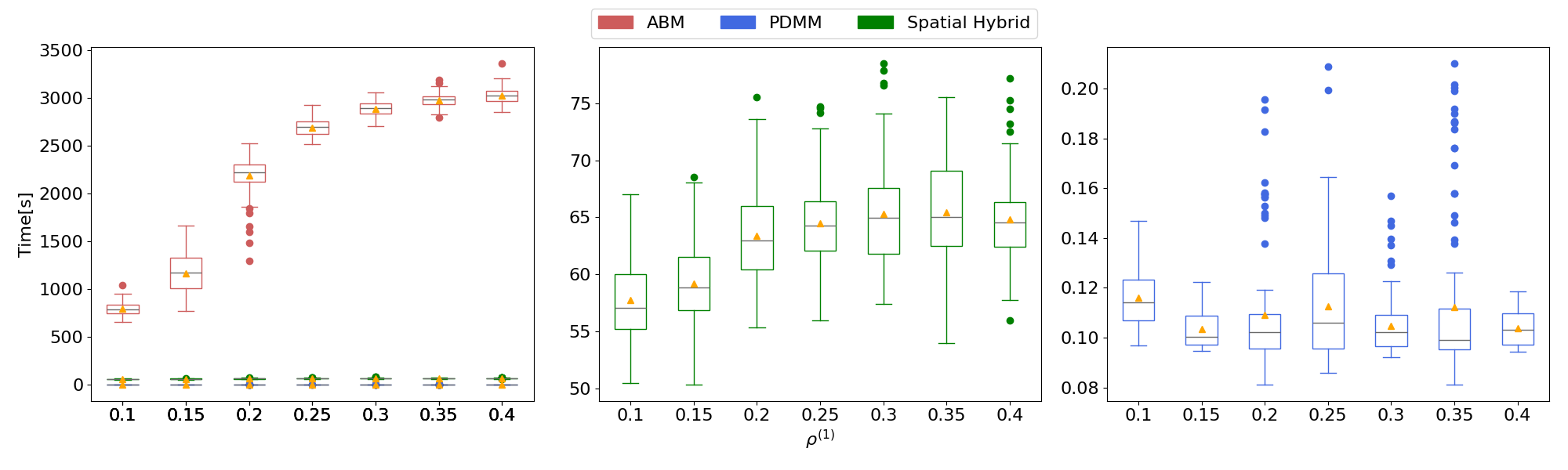"}\label{fig:qw_rho}}%
    \caption{\MK{\textbf{Runtime distribution for ABM, PDMM and spatial-hybrid model for a varying proportion of initially infected and varying values of $\rho^{(2)}$.}  Shown are the runtimes of $112$ runs for (a) five different values for the proportion of initially infected with the corresponding proportion distributed equally to compartments $E$, $C$ and $I$ and (b) seven different values for the transmission rate in $\Omega_2$. In (a) $\rho^{(2)}=0.3$ for all runs and in (b) $1\%$ of the population ($0.2\%$ Exposed, $0.3\%$ Carrier and $0.5\%$ Infected) is initially infected in all runs.}}
    \label{fig:runtime_analysis}
\end{figure}

\MK{In addition to the scaling behavior dependent on the number of agents, we consider the runtime scaling with respect to the number of initially infected and the value of the transmission rate $\rho^{(2)}$ in region 2 with, otherwise, the same setup as above. The runtime of ABM and spatial-hybrid model increases with an increasing proportion of initially infected while the runtime of the PDMM is independent of the proportion of initially infected, see~\cref{fig:qw_initially_infected}. The mean time of the ABM increases by $30\%$ when increasing the initially infected from $0.1\%$ to $1\%$ and by nearly $50\%$ when increasing from $0.1\%$ to $10\%$ initially infected. The mean time of the spatial-hybrid model increases by $7\%$ and $42\%$ when increasing the initial number of infected from $0.1\%$ to $1$ and $10\%$, respectively. For an increasing value of $\rho^{(2)}$, a similar behavior can be observed; see~\cref{fig:qw_rho}. The mean runtime of the ABM rises by $175\%$ when increasing $\rho^{(2)}$ from $0.1$ to $0.2$ and by $38\%$ when increasing $\rho^{(2)}$ from $0.2$ to $0.4$. The decreasing gradient can be explained by the increasing number of recovered and dead agents when choosing $\rho^{(2)}=0.4$. For recovered and dead agents, no infection state adoptions - needing computations and thus runtime - happen any more. The same behavior can be observed for the spatial-hybrid model with an even earlier saturation. The PDMM runtime is again not influenced by a varying value of $\rho^{(2)}$.}

\subsubsection{Application 2 - A potential for Munich and its surrounding counties}
\label{sec:Munich_example}
Secondly, we \MK{consider a less theoretical example}, in particular a potential for the city of Munich, defined as our focus region $\Omega_1$, and its surrounding counties. 

\begin{figure}
    \centering
    \includegraphics[width=0.6\textwidth]{"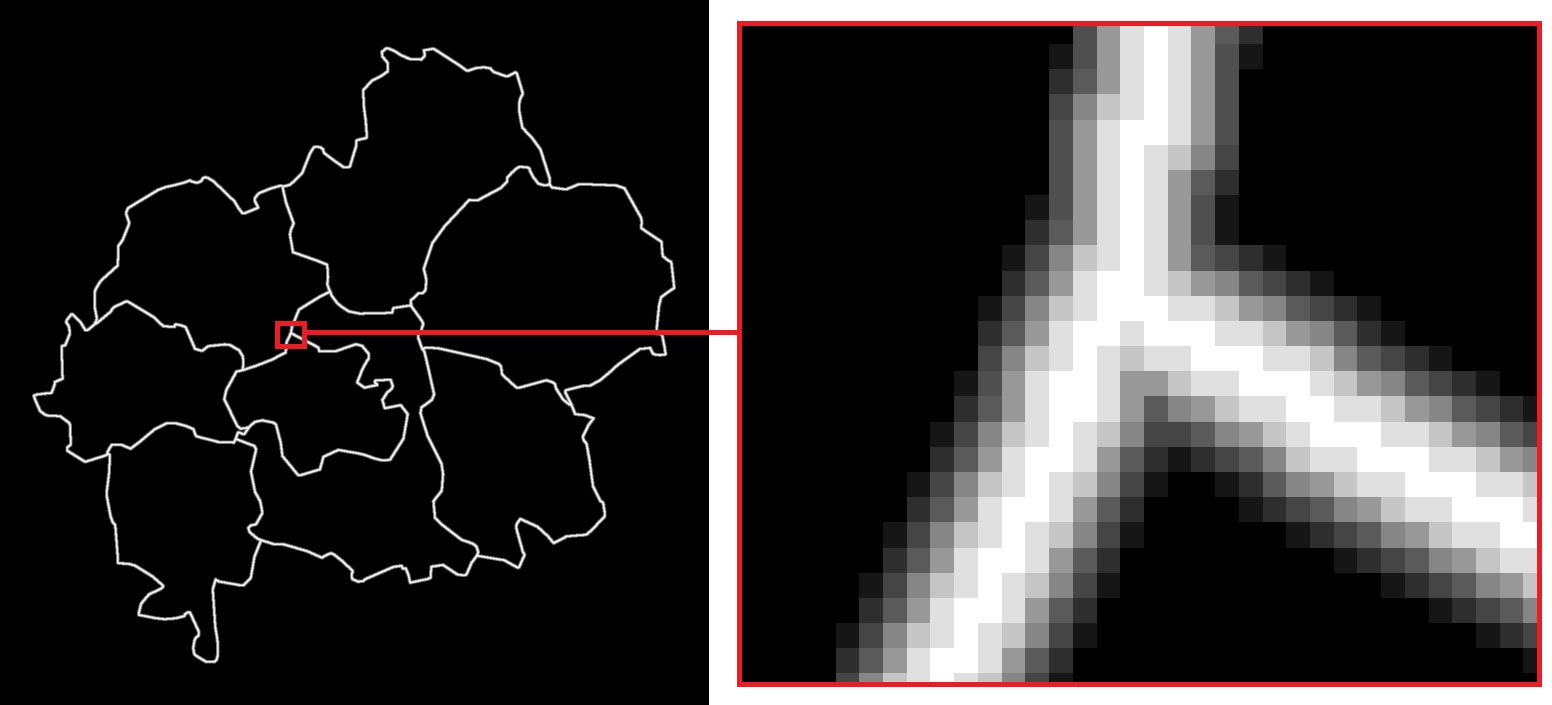"}
    \caption{\textbf{Potential defined for the ABM in and around Munich.} Black areas have value zero and white parts value one. We obtain the potential $F$ by discretizing the map, resulting in a matrix $P\in [0,1]^{p\times p}$ and then interpolating between matrix entries.}
    \label{fig:potential}
\end{figure}

\begin{figure}
    \centering
    \includegraphics[width=\textwidth]{"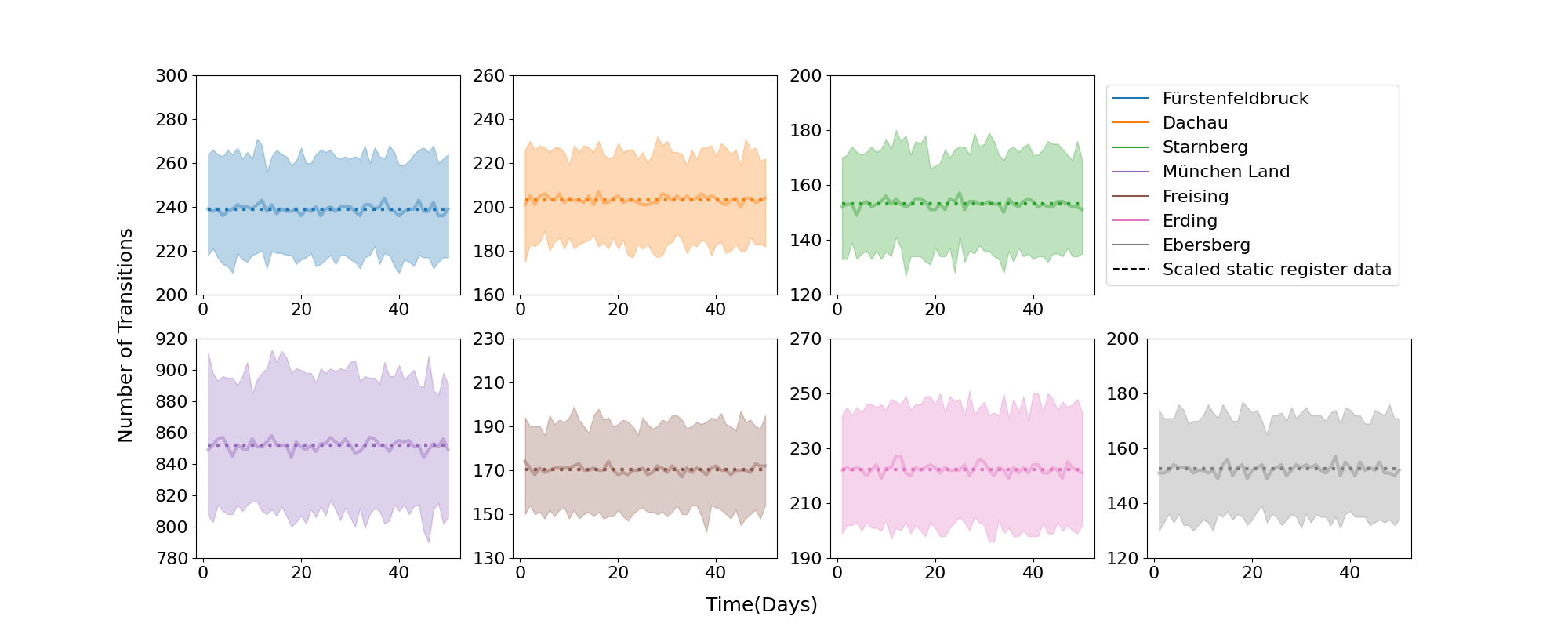"}
    \caption{\textbf{Daily commuting from and to Munich City for different surrounding counties.} $5\%$ and $95\%$ percentiles of daily transitions in simulations shown as colored area together with median as solid line. The static register data from Federal Agency of Work, scaled to number of simulated agents, is shown as dotted lines.}
    \label{fig:munich_transitions}
\end{figure}

To realize \MK{a difference between intercounty and intracounty} mobility patterns, we define a potential on the political map of the considered region. We apply a Gaussian curve on the discretized county borders (in pixels) to define a potential $F(x)$. More precisely, $F(x)=h$ if $x\in \mathbb{R}$ is on a border, $F(x) \in \left(0, h\right)$ modeled through a Gaussian curve if $x$ is in a given distance from the border and $F(x)=0$ otherwise, i.e., x inside a county. The resulting potential directly determines eight regions for the PDMM, aligning with the eight counties considered. \cref{fig:potential} shows the potential with a zoomed in snipped to the gradient on the border. 
The noise term $\sigma$ and the border height $h$ determine the number of agents crossing a border. 
A drawback of modeling agent movements with potential and diffusion process is the limited control on (intercounty) mobility and, in particular, that mostly agents next to the borders transition to neighboring regions in an appropriate time. However, in reality individuals that are further away from the borders should also transition to neighboring counties in a moderate time, e.g., on a daily basis caused by commuting activities. To that end, we modify the agents' movements. For short-distance travel, we use the diffusion process given by~\cref{eq:SDE}, where we set the border height $h$ for the potential $F$ to 10 and additionally prevent crossing borders through random walks. For long-distance travel, we introduce two normally distributed random variables $T_d^{\alpha}$ and $T_r^{\alpha}$ \MK{as additional attributes} for agent $\alpha$. Samples of these variables are only used when $\alpha$ is a commuter on the current day $\lfloor t\rfloor$. We use the random variable $T_d^{\alpha} \sim \mathcal{N}\left(\mu_d, \sigma_d^2  \right)$ as the time point an agent $\alpha$ leaves its home county on day $\lfloor t\rfloor$ and $T_r^{\alpha} \sim \mathcal{N}\left(\mu_r, \sigma_r^2  \right)$ as the time point an agent $\alpha$ returns on day $\lfloor t\rfloor$ -- ignoring and resampling values beyond three standard deviations to ensure that samples are in $[0,1]$. We introduce a commuting term $K(t)$, which is added onto $L$ from~\cref{eq:MasterEquation}, allowing us to better control the desired number of border crossings. Let $K_{\alpha}(t)$ be the $\alpha$-th component of $K(t)$. This entry is zero if agent $\alpha$ is not a commuter on day $\lfloor t\rfloor$. If $\alpha$ is a commuter, $K_{\alpha}(t)$ is given by
\begin{align}
K_{\alpha}(t) = \begin{cases}
    \pi(x_{\alpha})- x_{\alpha}, & \text{if } t_d^{\alpha} + \lfloor t\rfloor \in (t, t+\delta t] \vee t_r^{\alpha} + \lfloor t\rfloor \in (t, t+\delta t]\\
    0, & \text{else}
\end{cases}, \label{eq:K}   
\end{align}
where $\lfloor t\rfloor$ marks the beginning of the day and $t_d^{\alpha}$ and $t_r^{\alpha}$ are realizations of $T_d^{\alpha}$ and $T_r^{\alpha}$. Furthermore, $\pi: \Omega \rightarrow \Omega$ is a random variable on the domain using the probability function
\begin{align}
    \mathbb{P}\left(\pi(x_{\alpha}) = y\right) = \mathbb{P}\left(y|x_{\alpha}\in \Omega_k; y\in \Omega_l\right) \cdot \lambda^{\left(k, l\right)} = \begin{cases}
        \frac{\lambda^{\left(k, l\right)}}{\nu(\Omega_l)}, & l \neq k\\
        \lambda^{\left(k, k\right)}, & l=k \wedge x_{\alpha}=y\\
        0, & l=k \wedge x_{\alpha} \neq y
    \end{cases}\,, \label{eq:prob_K}
\end{align}
where $\lambda^{\left(k, l\right)}$ is the relative number of commuters from region $\Omega_k$ to region $\Omega_l$ and $\nu$ a counting measure on the discretized map (see~\cref{fig:potential}) ignoring border pixels. Hence, $\nu(\Omega_l)$ is the number of pixels inside region $\Omega_l$.
When using this adapted movement for the ABM, we need to adapt the exchange of agents from PDMM to ABM for the spatial-hybrid model as well. \MK{When we exchange an agent $\alpha$ from PDMM to ABM, we (uniformly) sample a new position $x_\alpha^{(new)}$ for it in the focus region $\Omega_1$, hence, $\mathcal{P}$ has the probability function
\begin{align}
    \mathbb{P}\left(x_\alpha^{(new)} = y\in \Omega_1\right)=\frac{1}{\nu(\Omega_1)}.
\end{align} 
In addition to sampling the new position in the focus region, we decide whether the agent has started its commute or is returning from it, depending on the current time $t$. In case of the former, i.e., if $t\leq \lfloor t\rfloor +\mu_d+z_{99.5}\cdot\sigma_d$ with $z_{99.5}=2.5758$ the $99.5\%$ quantile of the standard normal distribution, a return time $t_r^{\alpha}\sim \mathcal{N}\left(\mu_r, \sigma_r^2  \right)$ is drawn. 

We expect commuters to leave their home county on average at 9 am with the $99.5\%$ quantile between 5 am and 1 pm, resulting in $\mu_d=\frac{9}{24}$ and $\sigma_d=(\frac{13}{24}-\mu_d)/z_{99.5}=0.0647$. Further, we expect commuters to return on average at 6 pm with the $99.5\%$ quantile between 1 pm and 11 pm, hence $\mu_r=\frac{18}{24}$ and $\sigma_d=(\frac{23}{24}-\mu_r)/z_{99.5}=0.0809$. The spatial transition rates of the PDMM are equal to the commute weights $\lambda^{(k, l)}$, $k,l=1,\ldots,8$ used in~\cref{eq:prob_K}. Their values are based on~\cite{kuhn_assessment_2021} and are given in~\cref{eq:comm_matrix}.} \cref{fig:munich_transitions} shows that through the incorporation of the commuting term $K$, the number of daily transitions in Munich City matches the static daily commuting data \MK{from~\cite{bmas_pendlerverflechtungen_2020} used in~\cite{kuhn_assessment_2021}, and} scaled according to the number of agents $n_a$.

\begin{figure}
    \centering
    \sidesubfloat[]{\includegraphics[width=0.4\textwidth]{"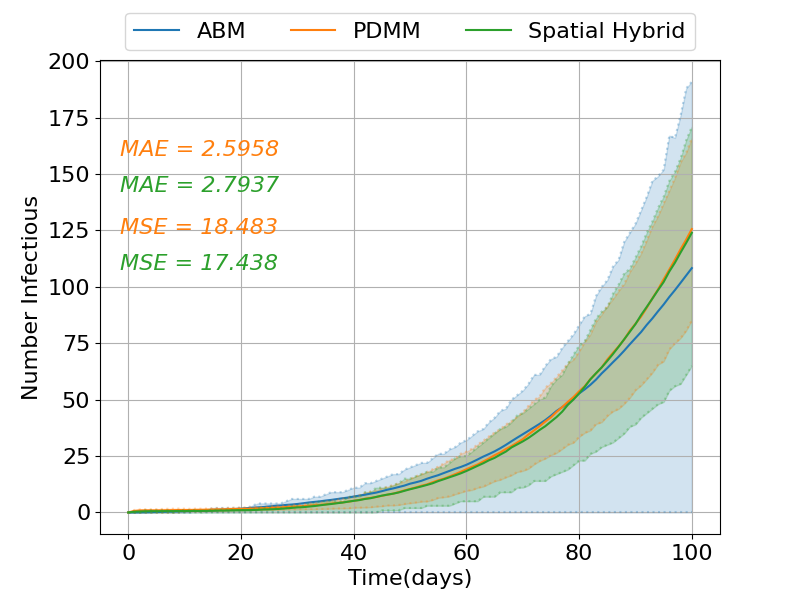"}\label{fig:munich_focus}}%
    \sidesubfloat[]{\includegraphics[width=0.4\textwidth]{"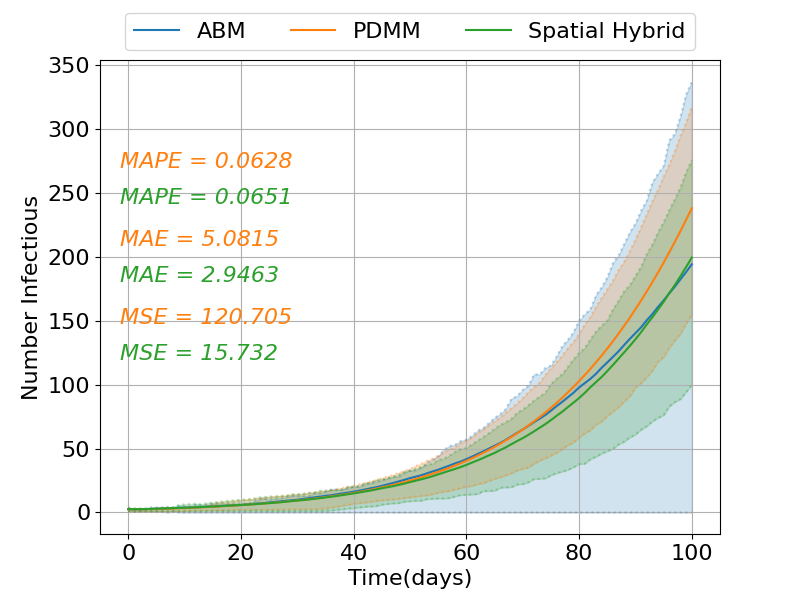"}\label{fig:munich_all}}
    \caption{\textbf{Spatial hybridization for the Munich potential: Focus region and sum of all regions.} Number of infectious agents (compartments $C$ and $I$) for (a) Munich City (focus region) and (b) the sum of all regions. The figures show the mean outcomes in solid lines with a partially transparent face between the p25 and p75 percentiles from 500 runs. \MK{If possible, MAPE, MAE, and MSE between the ABM mean and the mean of PDMM and spatial-hybrid model are displayed. Since the number of infected agents in the focus region is zero at the beginning, the MAPE cannot be calculated for the left figure.}}
    \label{fig:results_munich}
\end{figure}

\MK{For this application, we chose the same transmission rate for all regions with $\rho^{(k)}=0.2$ corresponding to $R_0=1.6$. All parameters used for the Munich potential are summarized in~\cref{tab:joint_transmission_params} and~\cref{tab:params_munich}. We chose $n_a=$14,064 agents with initially infected ($0.05\%$ Exposed, $0.05\%$ Carrier, and $0.1\%$ Infected) only in $\Omega_5$, see~\cref{fig:metaregions_munich}. In all other regions, all agents are susceptible at the beginning. We run $500$ simulations of ABM, PDMM and spatial-hybrid model for $t_{max}=100$ days, see~\cref{fig:results_munich}. We again consider MAPE, MAE, and MSE of the mean outcomes as error metrics. The MAPE is not defined for the focus region because the ABM mean is zero on the first few days, see~\cref{fig:results_munich}. In general, the spatial-hybrid model and PDMM have both low errors. In the focus region the MAE of the PDMM is even slightly lower than the one for the spatial-hybrid model. However, the MSE of the spatial-hybrid model is lower and the percentiles also fit the ABM percentiles better than the PDMM percentiles. Yet, the p25 percentile of the ABM displays that in at least $25\%$ of all runs, the virus dies out which is not captured by neither spatial-hybrid model nor PDMM. The reason for this is that ODE-based models are not able to let a virus go completely extinct and as the region with initially infected is not the focus region, it is modeled with the PDMM in the spatial-hybrid. To capture this behavior, a temporal-hybrid model could be used, see~\cref{sec: results_temporal_hybrid}.}

The runtime gain for the Munich potential is smaller than for the quadwell potential but the spatial-hybrid model is still 6-times faster than the ABM for \MK{40,000} agents, which is a runtime gain of \MK{more than} $80\%$; see~\cref{fig:munich_runtime}. The reason for this reduced effect is the very disadvantageous setup for the hybrid model with Munich City as focus region, whose population makes up approximately $75\%$ of all agents in the modeled counties. Consequently, in the spatial-hybrid model, the number of agents modeled with the ABM is roughly $75\%$ of $n_a$ while in the quadwell potential it is only $25\%$ of all agents. Secondly, the chosen initialization also has a relevant influence on the runtime gain. As we have \MK{an overall low} number of infectious agents in the simulation (\MK{on average} at most \MK{$1.5\%$} of the total population compared to approximately \MK{$4\%$} in the quadwell), there are \MK{fewer} infection state adoptions in the ABM per time step, which leads to less computational costs. One can observe that the runtime for the spatial-hybrid model for Munich is even slightly higher than the ABM runtime for $n_a<700$. Until that value, the exchange of agents from PDMM to ABM - which is slightly more complex for the Munich potential compared to the quadwell potential - produces overhead which only amortized for $n_a>700$.

\MK{We furthermore considered the scaling behavior with respect to the number of initially infected in region 5, Munich Land, and a changing value for $\rho^{(k)}$ with otherwise the same setup as above. The corresponding results are provided in~\cref{fig:runtime_munich} and align with the corresponding results from the quadwell potential.}

\subsection{Temporal hybridization}
\label{sec: results_temporal_hybrid}
\begin{figure}
    \centering
    \includegraphics[width=0.4\textwidth]{"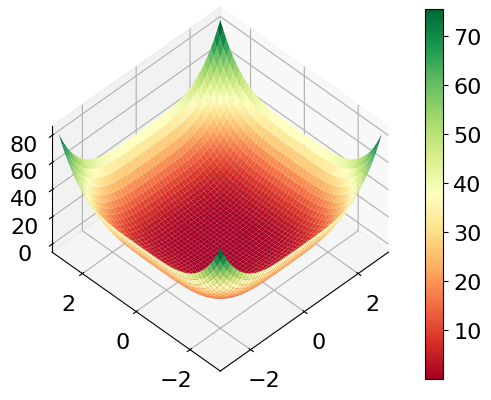"}
    \caption{\textbf{Single well potential~\eqref{eq:singlewell} used for the diffusion process in the ABM of the temporal-hybrid model.}}
    \label{fig:singlewell_potential}
\end{figure}

For the temporal hybridization we consider a single well potential given by
\begin{align} \label{eq:singlewell}
    F(x,y) = \frac{x^4+y^4}{2};
\end{align}
see \cref{fig:singlewell_potential}. By design we only have one region and the PDMM reduces to a classical ODE-based model without spatial \MK{resolution}. We use a constant noise $\sigma=0.5$ for the diffusion process and a relatively small contact radius $r=0.1$, thus $\sigma$ has a crucial influence on contacts between agents in the ABM and consequently on second-order adoptions. All parameters and their values for the single well application are listed \MK{in~\cref{tab:joint_transmission_params} and~\cref{tab:params_sw}.}

We chose a scenario with exactly one initially infected agent in compartment $I$ with an otherwise susceptible population. The transmission rate $\rho^{(1)}=0.6$ \MK{leading to $R_0=4.8$} is set such that in around $30\%$ of all simulations, the virus dies out. Otherwise, we get an epidemic outbreak and introduce a nonpharmaceutical intervention (NPI) which is implemented through a reduction of $\rho^{(1)}$ by \MK{$\kappa=80\%$} after $t_{NPI}=0.5(t_0 + t_{max})$. This NPI then leads to a slow decline of infections over the remaining \MK{simulation time}.

Since we are considering only one region, the derived PDMM is a deterministic ODE-model in which the virus never dies out. Obviously, we cannot use this PDMM to meaningfully simulate this specific scenario in its entirety. However, we can use the PDMM to speed up calculation once an outbreak reaches a certain size or after the virus has become extinct. Therefore, the temporal-hybrid model always starts with the ABM, and then switches to the PDMM once an extinction or an outbreak reaches a certain size. \MK{Note that after extinction, we could even stop the simulation but we continue with the ODE-based model until $t_{\max}$. This is done for practical reasons of comparison and since the runtime of this model is negligible.} The \MK{switch} is defined by a threshold $s > 1$ to be compared against the sum of agents in infection states $E$, $C$ and $I$. If this number is larger than $s$, we \MK{switch from ABM to PDMM}, projecting the whole population of the ABM. A suitable value for the threshold $s$ depends on the specific setup, especially on the transmission rate $\rho$. It is not trivial to be determined, particularly for real-world scenarios.

A second threshold $0 \leq s' < s$ could be introduced to switch from PDMM to ABM as soon as the number of infected agents falls below $s'$. This second threshold indicates a number of infected, where stochasticity is (again) important to determine the development of the disease dynamics. This can be of particular importance if disease mitigation, e.g., through NPIs, had been conducted successfully. \MK{When switching from PDMM to ABM, agents' positions and infection states have to be sampled according to the current PDMM system state. The infection state of an agent can be sampled according to the distribution given by the PDMM subpopulations $N_i^{(k)}$. The position is sampled from a distribution $\mathcal{P}$ which could for the considered single well potential be a uniform distribution on $[-1,1]\times[-1,1]$, see ~\cref{fig:sw_trajectory}.} In the presented scenario, we set $s'$ to zero. 

\begin{figure}
    \centering
    \sidesubfloat[]{\includegraphics[width=0.33\textwidth]{"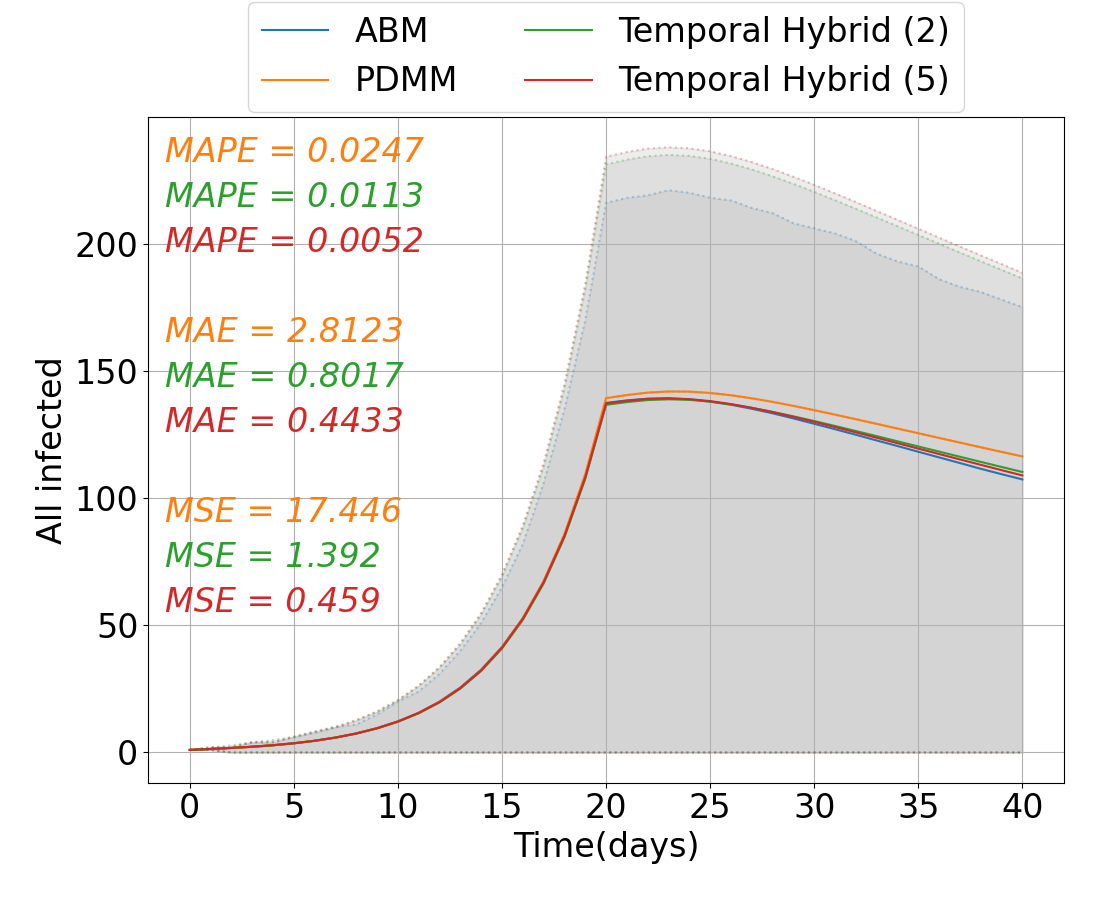"}\label{fig:temporal_hybrid_mean_combined}}
    \sidesubfloat[]{\includegraphics[width=0.33\textwidth]{"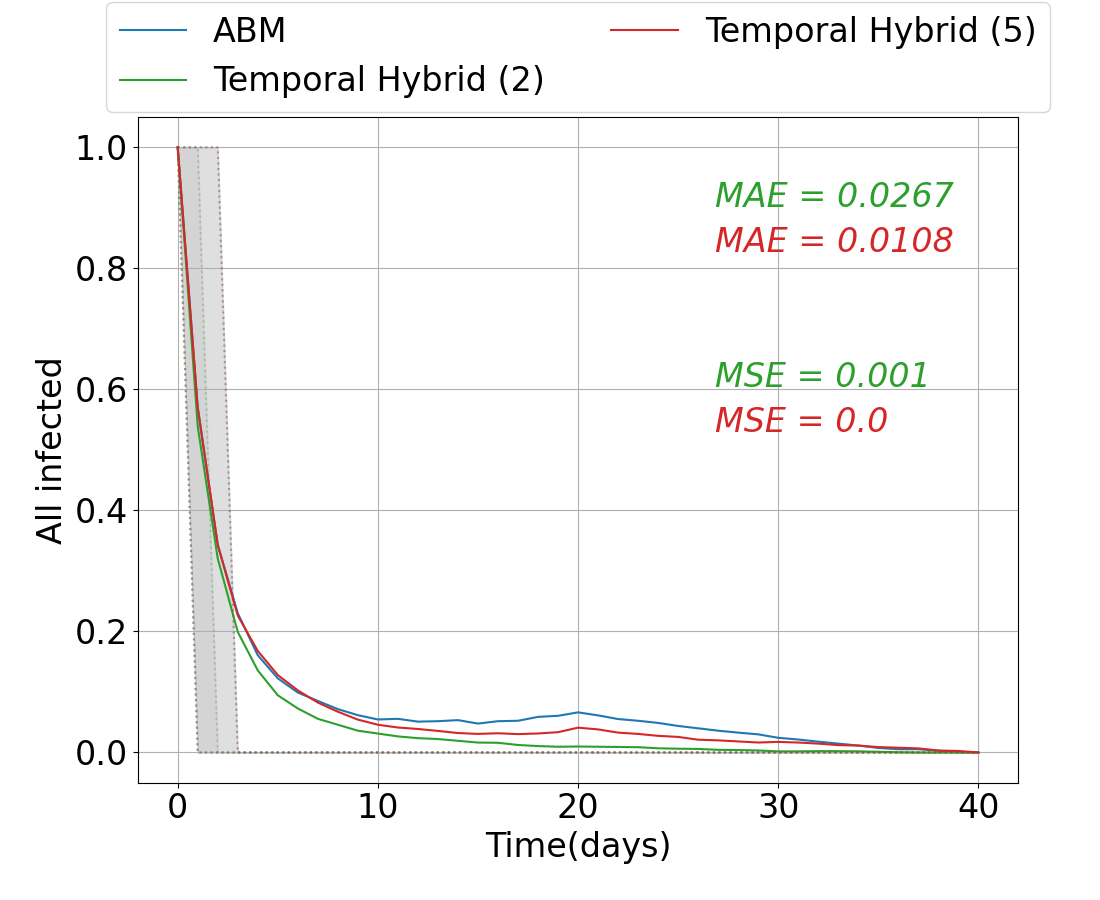"}\label{fig:temporal_hybrid_mean_extinction}}
    \sidesubfloat[]{\includegraphics[width=0.33\textwidth]{"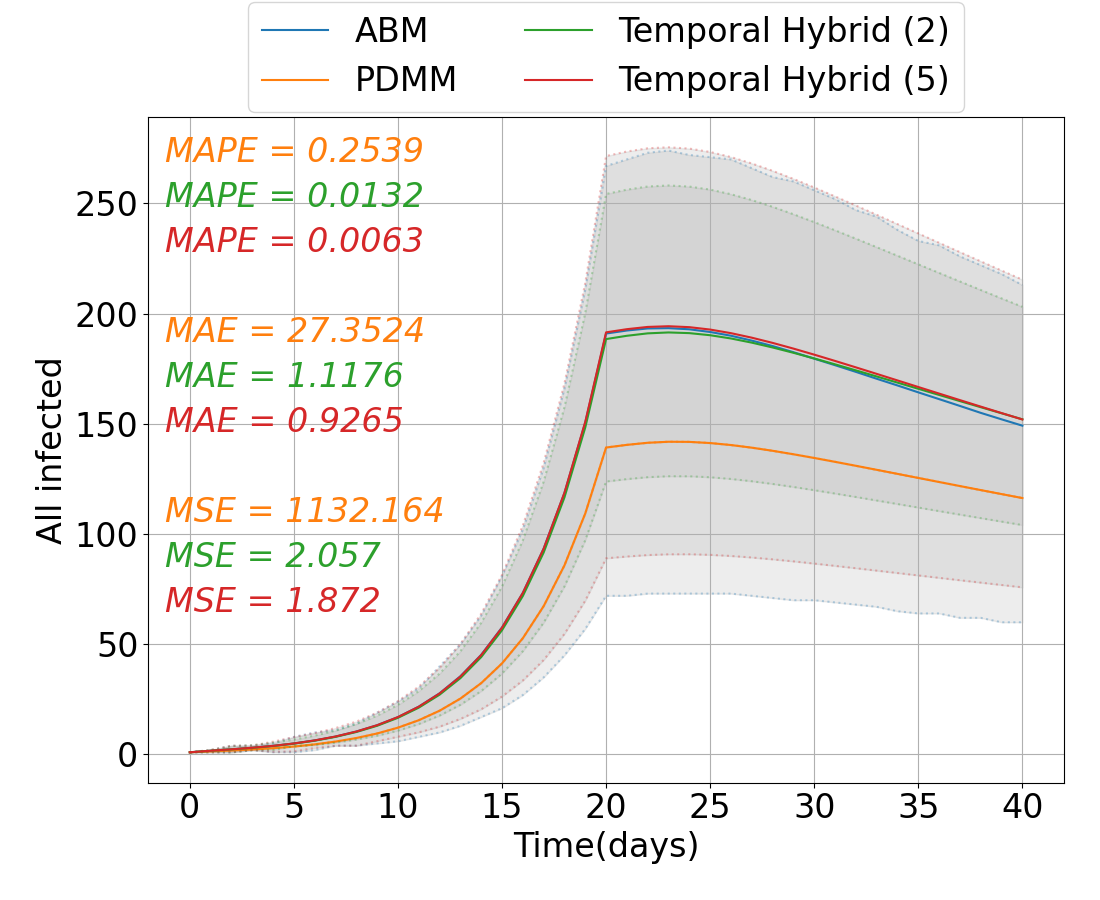"}\label{fig:temporal_hybrid_mean_survival}}
    \caption{\textbf{Temporal hybridization for the single well example.} Shown is the sum of compartments $E$, $C$ and $I$ for (a) all runs, (b) extinction runs with virus extinction, and (c) survival runs without virus extinction for ABM, PDMM and temporal-hybrid models with $s=2$ and $s=5$. The figures show the mean outcomes in solid lines with a partially transparent face between the p25 and p75 percentiles from 10,000 runs with 10,000 agents.}
    \label{fig:temporal_hybrid_mean}
\end{figure}

Here, the model selection function $\Gamma$ from~\cref{alg:Temporal_hybridization} returns the PDMM if the number of infected agents exceeds $s$, and the ABM otherwise. We chose $n_a=$10,000, $t_0 = 0$, $t_{max}=40$, and a fixed step size of $\Delta t = 0.2$ for this Algorithm. The single initially infected agent is chosen randomly from the entire population. \cref{fig:temporal_hybrid_mean} shows the results of 10,000 simulations of ABM, PDMM, and the temporal-hybrid model with $s=2$ and $s=5$.

The reduction of $\rho^{(1)}$ at $t_{NPI}=20$ leads to a kink in the number of infected agents (compartments $E$, $C$ and, $I$),~see~\cref{fig:temporal_hybrid_mean_combined}. Since in $30\%$ of all ABM simulations the virus dies out, the 25th-percentile is almost entirely zero, except for right after the initialization. We can also see that the percentiles as well as the means of the temporal-hybrid models with $s=2$ and $s=5$ are \MK{close together and the errors of both hybrid models are lower than the errors of the PDMM. As expected, the temporal-hybrid model with threshold $s=5$ has even lower errors than the one with threshold $s=2$}. As the PDMM is deterministic, the percentiles and means of it are all the same and \MK{have a MAPE of $0.0247$ for all runs.} While~\cref{fig:temporal_hybrid_mean_combined} shows the results for all simulations, \cref{fig:temporal_hybrid_mean_extinction} shows the results for all simulations where the virus becomes extinct and \cref{fig:temporal_hybrid_mean_survival} the corresponding curves for all simulations where the virus survives, hence, an outbreak occurs. The PDMM is not shown in~\cref{fig:temporal_hybrid_mean_extinction} as it is not able to capture the extinction scenario. In~\cref{fig:temporal_hybrid_mean_survival}, both temporal-hybrid models and ABM curves have shifted upwards compared to the results of all simulations in~\cref{fig:temporal_hybrid_mean_combined}. The PDMM result remains trivially the same and therefore underestimates the mean number of infected in the survival scenarios. For completeness, we also provided the results for the later compartment $I$, see~\cref{fig:temporal_hybrid_mean_comp_I}, where we see a rather smooth transition compared to the kink in~\cref{fig:temporal_hybrid_mean}.

Both switching thresholds seem to be a good choice as the resulting temporal-hybrid models are very close to the ABM results, \MK{i.e., they have low errors} - the mean as well as the percentiles. Note, however, that the low thresholds are due to a large transmission rate $\rho^{(1)}$ and dense contact network realized through the diffusion-drift process. The temporal-hybrid model with $s=2$ does not capture the increase of the ABM mean in the extinction runs until $t_{NPI}$, see~\cref{fig:temporal_hybrid_mean_extinction} while the hybrid model with $s=5$ does.

The ABM runs significantly faster in extinction runs than in survival runs (see~\cref{tab:temporal_hybrid_timings_extinction_survival}), as there are no pairwise interactions to consider once there are no infected agents anymore. Furthermore, the PDMM is - as expected - multiple orders of magnitude faster than the ABM, making it desirable to use it whenever possible. On average, the temporal-hybrid model \MK{for $s=5$ is 20-times and the one for $s=2$ even 40-times} faster than the ABM in the results for all simulations (see~\cref{tab:temporal_hybrid_timings_all}), \MK{8-to 9-}times for the extinction simulations (see~\cref{tab:temporal_hybrid_timings_extinction_survival}), and \MK{22- ($s=5$) and 49-times ($s=2$)} for the survival simulations (see~\cref{tab:temporal_hybrid_timings_extinction_survival}). When comparing the total simulation time for the different switching thresholds, all simulations show a time gain when using the lower threshold of $s=2$ \MK{while for the survival simulations the time gain is the greatest}. For all stochastic models (ABM and temporal-hybrid), the difference in minimum and maximum time to the mean time correlates with the large variation in simulation outcomes; see the percentiles in~\cref{fig:temporal_hybrid_mean}. For this application, we do not further examine the runtime scaling behavior, as it does not behave significantly different to the results for the spatial-hybrid model, see~\cref{sec::spatial_hybrid_results} and~\cref{fig:runtime}. The runtime of the temporal-hybrid model is dominated by the proportion of time (instead of population) in which the ABM is used for the simulation.

\begin{table}[h]
    \centering
    \begin{tabular}{l||r|r|r|}
        & \multicolumn{3}{c|}{\textbf{All simulations}} \\
        \hline
        \textbf{Model} & \textbf{min} & \textbf{mean} & \textbf{max}  \\
        \hline\hline
        \textit{ABM} & 41.80 & 331.76 & 1835.93 \\
        \textit{PDMM} & 0.00040 & 0.00047 & 0.01140 \\
        \textit{Temporal-hybrid} $s=2$ & 0.5587 & 8.4082 & 92.6604\\
        \textit{Temporal-hybrid} $s=5$ & 0.5591 & 16.6405 & 104.497 \\
    \end{tabular}
    \caption{\textbf{Summarized simulation (in sec) timings for ABM, PDMM and temporal-hybrid models for all 10,000 simulations.}}
    \label{tab:temporal_hybrid_timings_all}
\end{table}

\begin{table}[h]
    \centering
    \begin{tabular}{l||r|r|r|r|r|r|}
        & \multicolumn{3}{c|}{\textbf{Extinction simulations}} & \multicolumn{3}{c|}{\textbf{Survival simulations}} \\
        \hline
        \textbf{Model} & \textbf{min} & \textbf{mean} & \textbf{max} & \textbf{min} & \textbf{mean} & \textbf{max} \\
        \hline\hline 
        \textit{ABM} & 41.80 & 48.77 & 109.147 & 91.07 & 442.20 & 1835.93 \\
        \textit{PDMM} & - & - & - & 0.00040 & 0.00047 & 0.01140 \\
        \textit{Temporal-hybrid} $s=2$ & 0.5587 & 5.4414 & 85.2773 & 0.8115 & 9.5341 & 92.6604 \\
        \textit{Temporal-hybrid} $s=5$ & 0.5591 & 6.2384 & 99.2354 & 2.9843 & 20.7604 & 104.497 \\  
    \end{tabular}
    \caption{\textbf{Summarized simulation timings (in sec) for ABM, PDMM and temporal-hybrid models for simulations with and without virus extinction.}}
    \label{tab:temporal_hybrid_timings_extinction_survival}
\end{table}

\MK{
\subsection{Energy consumption}
\label{sec: energy_comsumption}

To quantify the effects of the proposed hybrid models on the estimated $\text{CO}_2$ emissions, we measured the energy consumption of the ABMs and the hybrid models for the chosen setups in the quadwell application~\cref{sec:qw_example},~\cref{fig:results_qw}, the Munich application~\cref{sec:Munich_example},~\cref{fig:results_munich} and the single well application~\cref{sec: results_temporal_hybrid},~\cref{fig:temporal_hybrid_mean}. We used \textit{LIKWID}~\cite{likwid-paper-2010} to measure the energy consumption in Joules for each model using 56 cores in parallel, see~\cref{tab:energy_consumption}. According to extrapolations from the German Federal Environment Agency~\cite{icha_entwicklung_2024}, 1 kWh power produced 380g $\text{CO}_2$ emissions in 2023. Using this extrapolation, we calculated the $\text{CO}_2$ savings achieved for all applications by replacing a full ABM through a hybrid ABM-ODE model. For the quadwell application, the spatial-hybrid model emitted 1180g $\text{CO}_2$ less than the full ABM which is a saving of 97.75\%. For the Munich potential, the savings are - like the runtime saving - less than for the quadwell potential, but we still reduced the $\text{CO}_2$ emissions by a factor of roughly 3.4. The reductions of the temporal-hybrid models for the single well application are nearly as high as of the spatial-hybrid model for the quadwell application with 94.67\% ($s=5$) and 97.28\% ($s=2$) corresponding to a saving of 2524g $\text{CO}_2$ and 2594g $\text{CO}_2$, respectively.}

\begin{table}[h]
    \centering
    \begin{tabular}{l||r|r|r}
        \textbf{Model} & %\textbf{Energy[J]} & 
        \textbf{Power[kWh]} & \textbf{$CO_2$ emissions[g]} & \textbf{$CO_2$ reduction} \\
        \hline\hline
        ABM quadwell & %11,444,013 & 
        3.17889 &  1207.97915 & -\\
        Spatial-hybrid quadwell & %257,819 & 
        0.07162 & 27.21423 & 97.75\%\\
        ABM Munich & %14,954,394 & 
        4.15400 & 1578.51937 & -\\
        Spatial-hybrid Munich & %4,352,027 & 
        1.20890 & 459.38063 & 70.90\%\\
        ABM single well & 7.01801 & 2666.84380 & -\\
        Temporal-hybrid (5) single well &  %1,345,676 & 
        0.37380 & 142.04358 & 94.67\%\\
        Temporal-hybrid (2) single well & %687,412 & 
        0.19095 & 72.56005 & 97.28\%
    \end{tabular}
    \caption{\MK{\textbf{Energy consumption of all models}. The measurements for the quadwell potential represent $500$ runs as in~\cref{sec:qw_example}, the measurements for the Munich potential represent $500$ runs as in~\cref{sec:Munich_example} and the measurements for the single well potential represent $10,000$ runs as in~\cref{sec: results_temporal_hybrid}. Power in kilo-Watt-hours (kWh) computed from measured energy and {$CO_2$ emissions[g]} computed from~\cite{icha_entwicklung_2024}. Reduction (in percent) computed for hybrid model compared to the full, reference ABM.}}
    \label{tab:energy_consumption}
\end{table}

\section{Discussion and Conclusion}
\label{sec::Discussion_and_Conclusion}
While the number of research articles on hybrid models \MK{for infectious disease dynamics} in the sense of our understanding is very small, several authors have published pioneering works in~\cite{bobashev_hybrid_2007,yoneyama_hybrid_2012,kasereka_hybrid_2014,bradhurst_hybrid_2015,hunter_hybrid_2020} over the last two decades. The approaches coming closest to our works are given in~\cite{bobashev_hybrid_2007,hunter_hybrid_2020}. However, runtime and computational cost have, e.g., not been considered in~\cite{bobashev_hybrid_2007}. In~\cite{hunter_hybrid_2020} various interesting results have been published for a model running on a laptop computer.

In this work, we formalized two hybridization approaches and introduced a general framework of spatial- and temporal-hybrid models, combining agent-based and ODE- or metapopulation-based models. In particular, the temporal-hybrid model \MK{with its dynamic model choice} is also an \textit{adaptive} model and future works may also consider adaptivity in time and space through spatio-temporal-hybrid models.

We presented two suitable models recently introduced, which were adapted for our examples and combined in the suggested manner. \MK{The examples provided show the application of the proposed hybridization framework in different theoretical settings, thus demonstrating its runtime benefit while maintaining a high resolution in a specific area or time frame. Although the parameters used in the examples are motivated by COVID-19, the application is not limited to a particular pathogen. Moreover, since this paper is intended to serve as a proof of concept, we did not consider a true real-world application.} The critical aspect of combining the two model types is the definition of exchange rules, an issue that has not been discussed a lot yet, but which has an essential impact on the accuracy and the performance of the hybridization. Models that are derived from one another in a theoretical way, like in~\cite{winkelmann_mathematical_2021,schmidtchen_multiscale_2018} offer a more direct way to implement exchanges. Nevertheless, nuances in projections or mappings can become important as it has been observed in our simulations. With the presented random walk movement of agents, the same agent may cross region boundaries multiple times in a short time, which might drastically change the hybrid model's outcome compared to the outcome of the fine model if this effect is not accounted for in the coarse model.

The movement pattern of the ABM, given by a potential and a Brownian motion is a rather theoretical movement pattern that delivered us two suitable models to be combined, but, \MK{especially when wanting to develop an actual real-world application, a more realistic human mobility pattern or an ABM using contact networks should be considered. Additionally, the transmission model of the ABM could be enhanced by incorporating more features such as an individual viral load.} In future work, we will investigate more complex ABMs realizing discrete locations such as households, schools, and workplaces. For these, more sophisticated methods for projection and, also, reconstruction of information will be needed.

As shown, hybrid models can combine the best of both worlds and deliver fine-granular insights through the use of ABMs, using substantially reduced computational resources. Both hybridization approaches were able to capture disease dynamics of the ABM better than the PDMM.

For the spatial hybridization, the runtime gain depends on the total number of agents in the focus region, relative to all other regions modeled by \MK{metapopulations}. Similarly, the temporal hybridization runtime gain depends mostly on the proportion of simulation time that the ABM is used. From our simulations, we have seen that up to 98~\% of the simulation time can be saved by replacing an ABM by a spatial-hybrid model. Although this number depends highly on the chosen models and the individual setting, a reduction of 90~\% of the simulation time or a speedup of 10 can be achieved easily.

\section*{Acknowledgments}
The authors would like to thank Nata\v{s}a Djurdjevac Conrad and Johannes Zonker for discussions on the models initially developed in~\cite{winkelmann_mathematical_2021}.
The authors JB and MJK have received funding by the German Federal Ministry of Education and Research under grant agreement 031L0297B (Project INSIDe). RS and MJK have received funding by the German Federal Ministry for Digital and Transport under grant agreement FKZ19F2211A (Project PANDEMOS), RS has received funding by the German Federal Ministry for Digital and Transport under grant agreement FKZ19F2211B (Project PANDEMOS). MJK has received funding from the Initiative and Networking Fund of the Helmholtz Association (grant agreement number KA1-Co-08, Project LOKI-Pandemics).

\section*{Competing interests}
The authors declare to not have any competing interests.

\section*{Data availability}
The MEmilio code repository is publicly available on Github under \url{https://github.com/SciCompMod/memilio}, the developed models are to be found on the fork release \url{https://github.com/reneSchm/memilio/releases/tag/hybrid-paper-v2} (based on the branch hybrid-model-omp); The models can be found in the directory /cpp/hybrid\_paper, with simulation algorithms under /cpp/models/mpm. All input data can be generated with or is in the code repository. Visualized simulation data can be shared upon request.

\section*{Author Contributions}

    \noindent\textbf{Conceptualization:} Julia Bicker, René Schmieding, Martin Kühn\\
    \noindent\textbf{Data Curation:} Julia Bicker, René Schmieding\\
    \noindent\textbf{Formal Analysis:} Julia Bicker, René Schmieding, Martin Kühn\\
    \noindent\textbf{Funding Acquisition:} Martin Kühn, Michael Meyer-Hermann\\
    \noindent\textbf{Investigation:} Julia Bicker, René Schmieding, Martin Kühn\\
    \noindent\textbf{Methodology:} Julia Bicker, René Schmieding, Martin Kühn\\
    \noindent\textbf{Project Administration:} Martin Kühn\\
    \noindent\textbf{Resources:} Martin Kühn, Michael Meyer-Hermann\\
    \noindent\textbf{Software:} Julia Bicker, René Schmieding\\
    \noindent\textbf{Supervision:} Martin Kühn, Michael Meyer-Hermann \\
    \noindent\textbf{Validation:} All authors \\
    \noindent\textbf{Visualization:} Julia Bicker, René Schmieding\\
    \noindent\textbf{Writing – Original Draft:} Julia Bicker, René Schmieding, Martin Kühn\\
    \noindent\textbf{Writing – Review \& Editing:} All authors

\newpage
\appendix

\begin{figure}[H]
    \centering
    \sidesubfloat[]{\includegraphics[width=0.35\textwidth]{"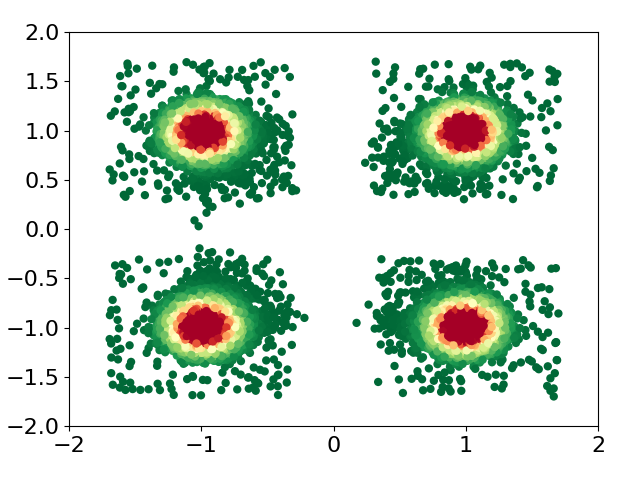"}\label{fig:sigma_0.3}}%
    \sidesubfloat[]{\includegraphics[width=0.35\textwidth]{"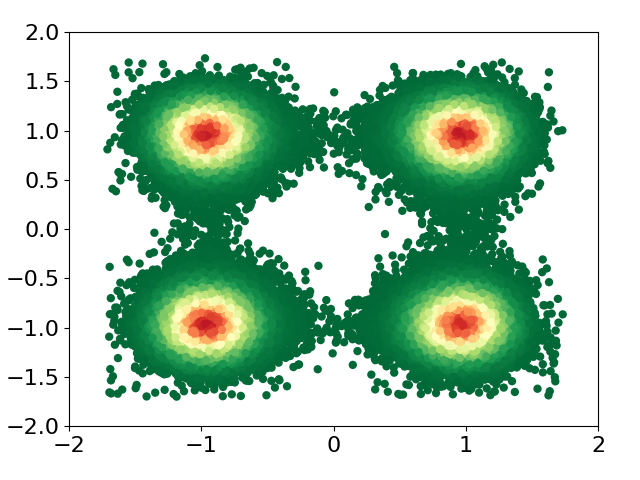"}\label{fig:sigma_0.5}}
    \sidesubfloat[]{\includegraphics[width=0.35\textwidth]{"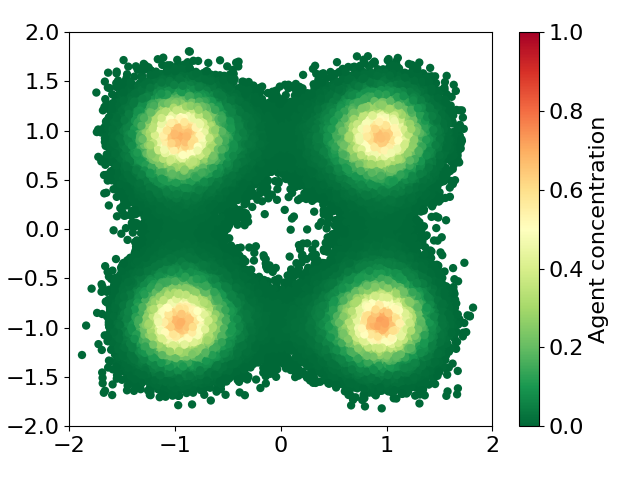"}\label{fig:sigma_0.6}}%
    \caption{\textbf{Position distribution for a simulation of $800$ agents for $50$ days in the quadwell scenario.} Results for noise term (a) $\sigma=0.3$, (b) $\sigma=0.5$ and (c) $\sigma=0.6$. Agents are initialized with positions having $0.3$ distance from the axes with $x=0$ and $y=0$.}
    \label{fig:sigma_movement}
\end{figure}

\begin{figure}[H]
    \centering
    \includegraphics[width=0.6\textwidth]{"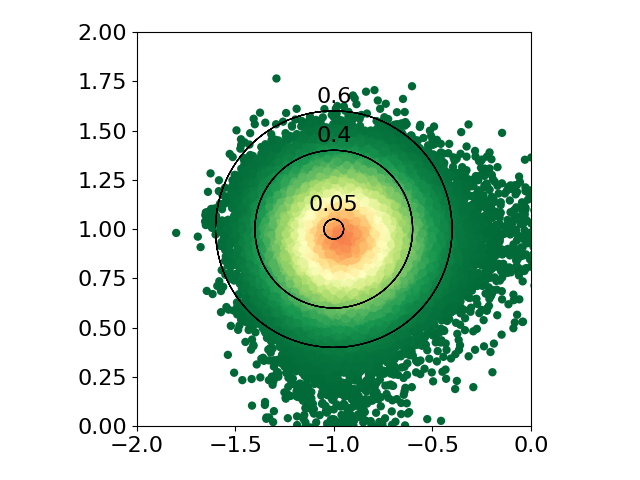"}
    \caption{\MK{\textbf{Different interaction radii and position distribution in $\Omega_1$ for $\sigma=0.55$.} The positions of $800$ agents for $50$ days are shown zoomed in to well $\Omega_1$ together with interaction radius $r=0.05$, $r=0.4$ and $r=0.6$.}}
    \label{fig:radii}
\end{figure}

\begin{figure}[H]
    \centering
    \sidesubfloat[]{\includegraphics[width=0.34\textwidth]{"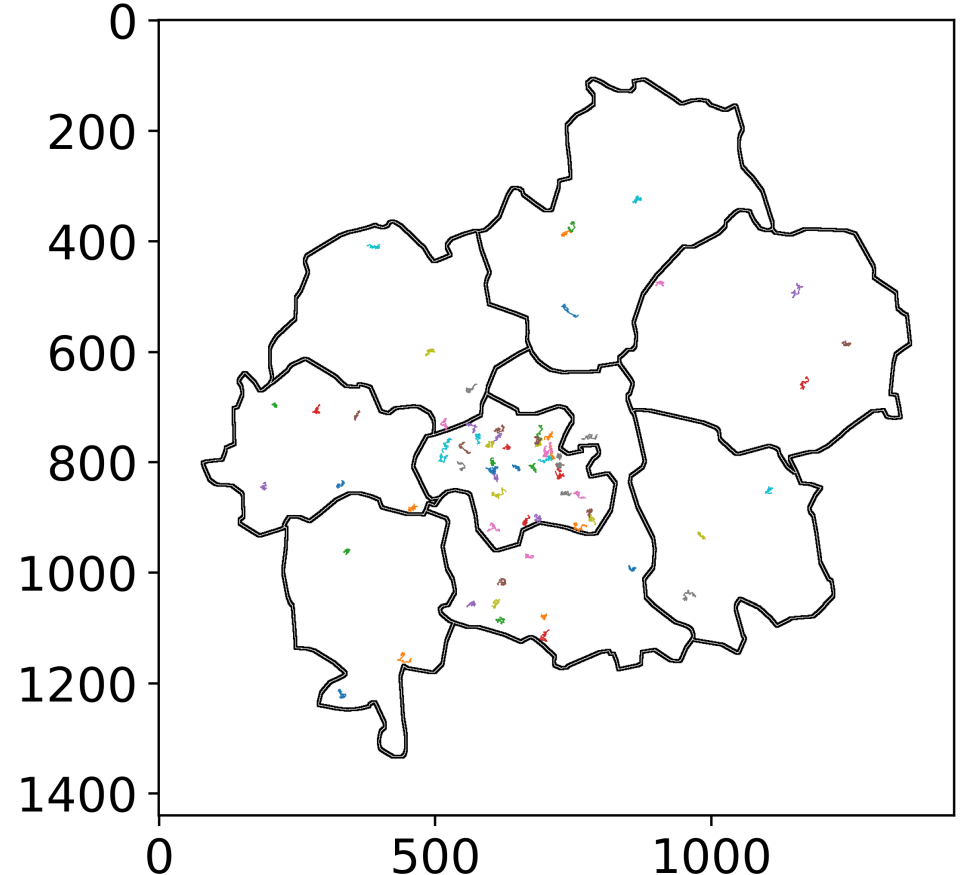"}\label{fig:sigma_5}}%
    \sidesubfloat[]{\includegraphics[width=0.34\textwidth]{"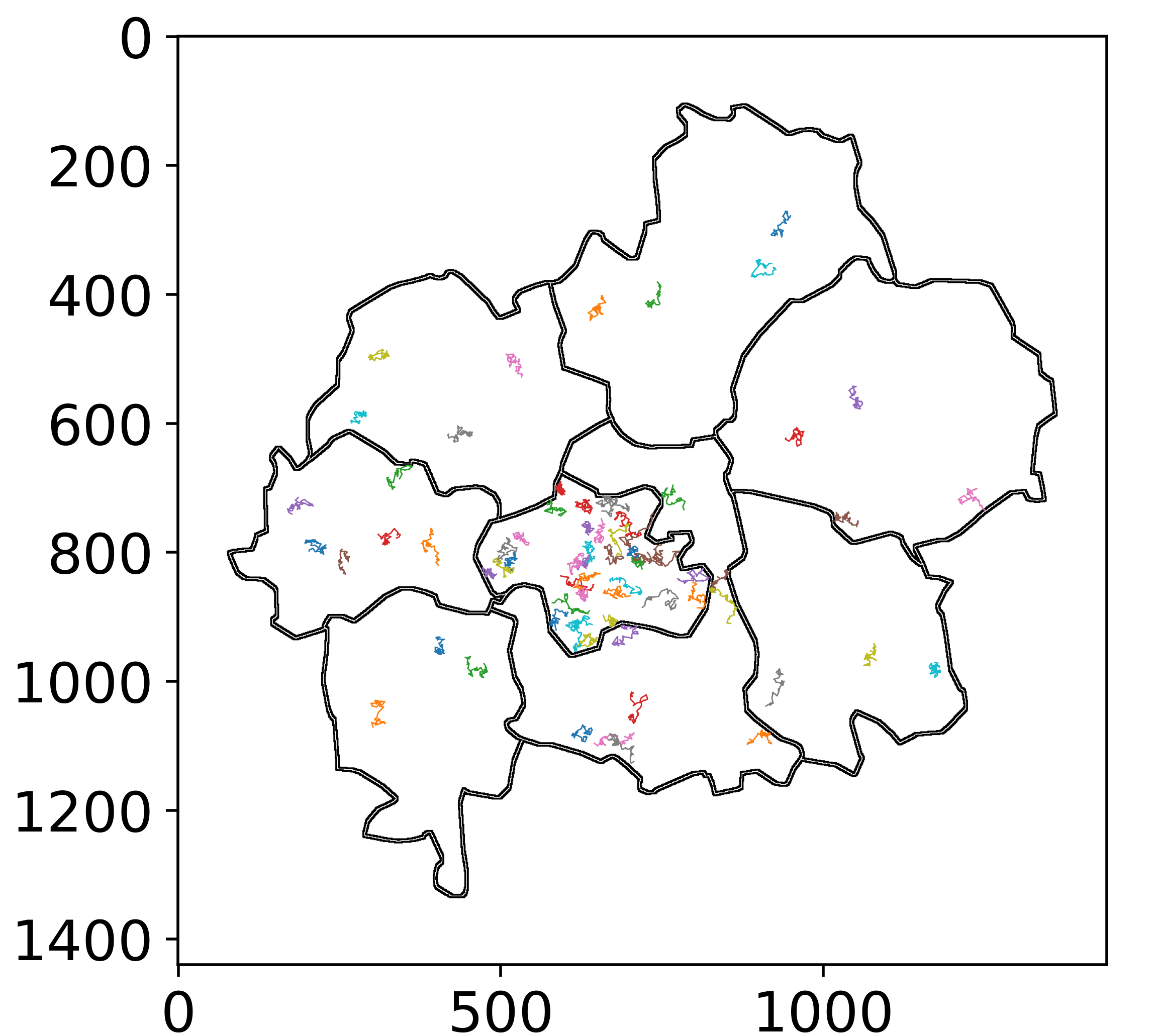"}\label{fig:sigma_10}}
    \sidesubfloat[]{\includegraphics[width=0.34\textwidth]{"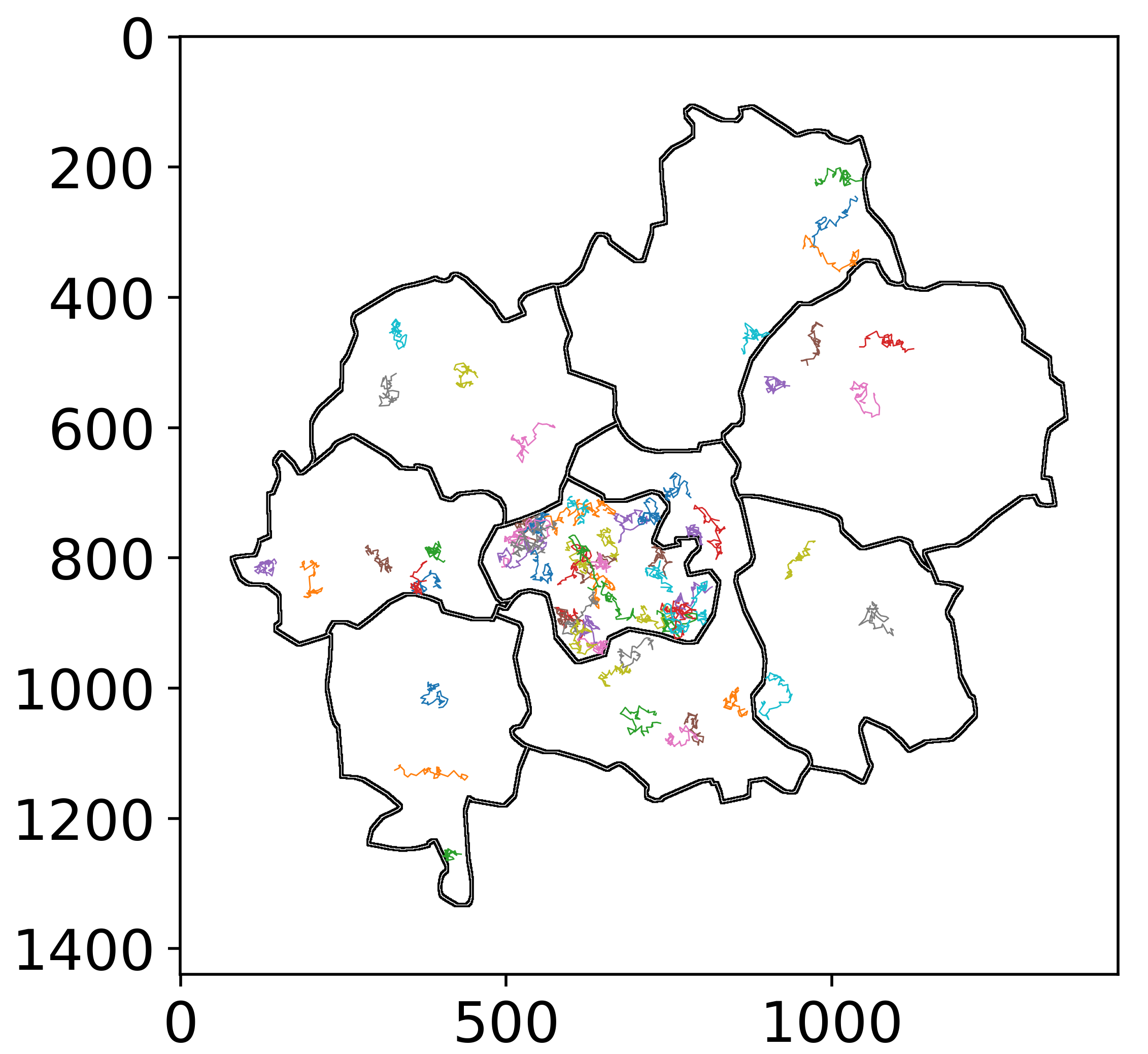"}\label{fig:sigma_15}}%
    \caption{\MK{\textbf{Intracounty movement of $35$ agents for $50$ days in the Munich scenario.} Results for noise term (a) $\sigma=5$, (b) $\sigma=10$ and (c) $\sigma=15$. The plots show only the movement given by the diffusion process, without commuting.}}
    \label{fig:sigma_movement_munich}
\end{figure}

\begin{figure}[H]
    \centering
    \includegraphics[width=0.6\textwidth]{"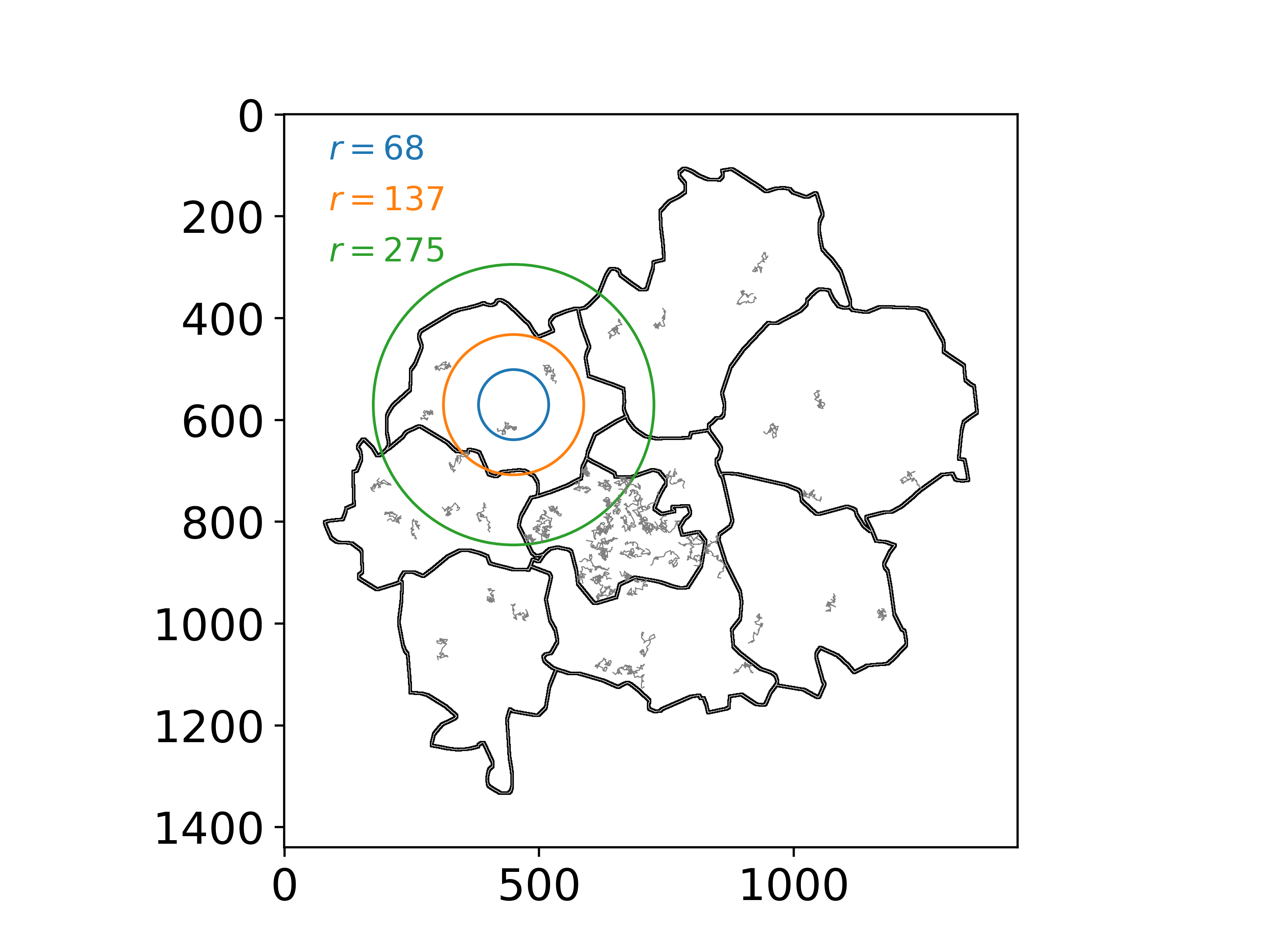"}
    \caption{\MK{\textbf{Different interaction radii for the Munich potential.} Shown are interaction radius $r=68$, $r=137$ and $r=275$. Cross interaction of agents in different subregions is not possible, hence $r=275$ corresponds to the current subregion being the interaction area.}}
    \label{fig:radii_munich}
\end{figure}

\MK{\section{Sensitivity analysis}
\label{sec:sensitivity_analysis}
A sensitivity analysis was conducted for the ABM, the PDMM, and the spatial-hybrid model for both, the quadwell application and the Munich application. In the temporal hybridization only ABM or PDMM is used at a time. We did not perform another sensitivity analysis for the temporal setting as the most influential parameters for the temporal-hybrid model are defined by the individual models used and, hence, align with the results for the Munich and quadwell applications. We used the Morris screening method \cite{morris_factorial_1991} to perform the sensitivity analysis. In this method the elementary effect for parameter $i$ is calculated as
\begin{align}
\label{eq:elem_effect}
    d_i(\textbf{x}) = \frac{y(x_1,\ldots,x_{i-1}, (1+\Delta_i)x_i,x_{i+1},\ldots,x_n)-y(\textbf{x})}{\Delta_i}
\end{align}
with $\textbf{x}=(x_1,\ldots,x_n)$ an input vector, $\Delta_i$ the relative change of input $x_i$ and $y$ the model output. To weaken the effect of the models' inherent stochasticity, we performed $112$ runs per model and used the mean to calculate the outputs for the elementary effects. The number 112 was chosen to completely fill two compute nodes with 56 cores each. We considered the elementary effects of the following outputs:
\begin{itemize}
    \item Maximum number of agents simultaneously in infection state $I$:\\ $\max_{t\in\{0,\ldots,t_{max}\}}N_I(t)$,
    \item Total number of transmissions: $\sum_{t=1}^{t_{max}} \left(N_S(t-1) - N_S(t)\right)$,
    \item Total number of deaths: $N_D(t_{max})$.
\end{itemize}
The number $N_{j}(t)$ is given by
\begin{align*}
N_{j}(t) =
\begin{cases}
    \sum_{k=1}^{n_R} N_{j}^{(k)}(t), & \text{for the PDMM,}\\[10pt]
    \sum_{\alpha = 1}^{n_a} \delta_j(z_\alpha), &\text{for the ABM.}
\end{cases}
\end{align*}
For each parameter, we calculated $40$ elementary effects resulting in $40\cdot 112 = 4480$ runs per parameter plus $40 \cdot 112$ runs for the base values $y(\textbf{x})$. For all models we included two dummy variables which do not influence the model outputs in the sensitivity analysis and serve as reference values. These dummy variables are used to quantify the effect of output variation due to the models' stochastic nature. The parameters shared between all models and their ranges are given in~\cref{tab:global_param_ranges}, while the parameters that vary between the applications are given in~\cref{tab:qw_param_ranges} for the quadwell and in~\cref{tab:munich_param_ranges} for the Munich application. For parameters related to infection dynamics, the ranges are motivated by wild-type COVID-19 and based on \cite{kuhn_assessment_2021}, apart from $\xi_C^{(k)}$, $\xi_I^{(k)}$ which are set to $1.0$ to reduce model complexity. The commute and spatial transition rates for the Munich application are based on mobility matrices used in~\cite{kuhn_assessment_2021} which were derived from mobility data from~\cite{bmas_pendlerverflechtungen_2020} and~\cite{twitter_twitter_2020}. For the ranges of ABM parameters related to movement, we made an educated guess based on preliminary investigations of the model behavior, see~\cref{fig:sigma_movement},~\cref{fig:radii},~\cref{fig:sigma_movement_munich},~\cref{fig:radii_munich}. The ranges for the PDMM spatial transition rates are chosen such that they match the ABM movement.}

\begin{figure}[H]
    \centering
    \includegraphics[width=\textwidth]{"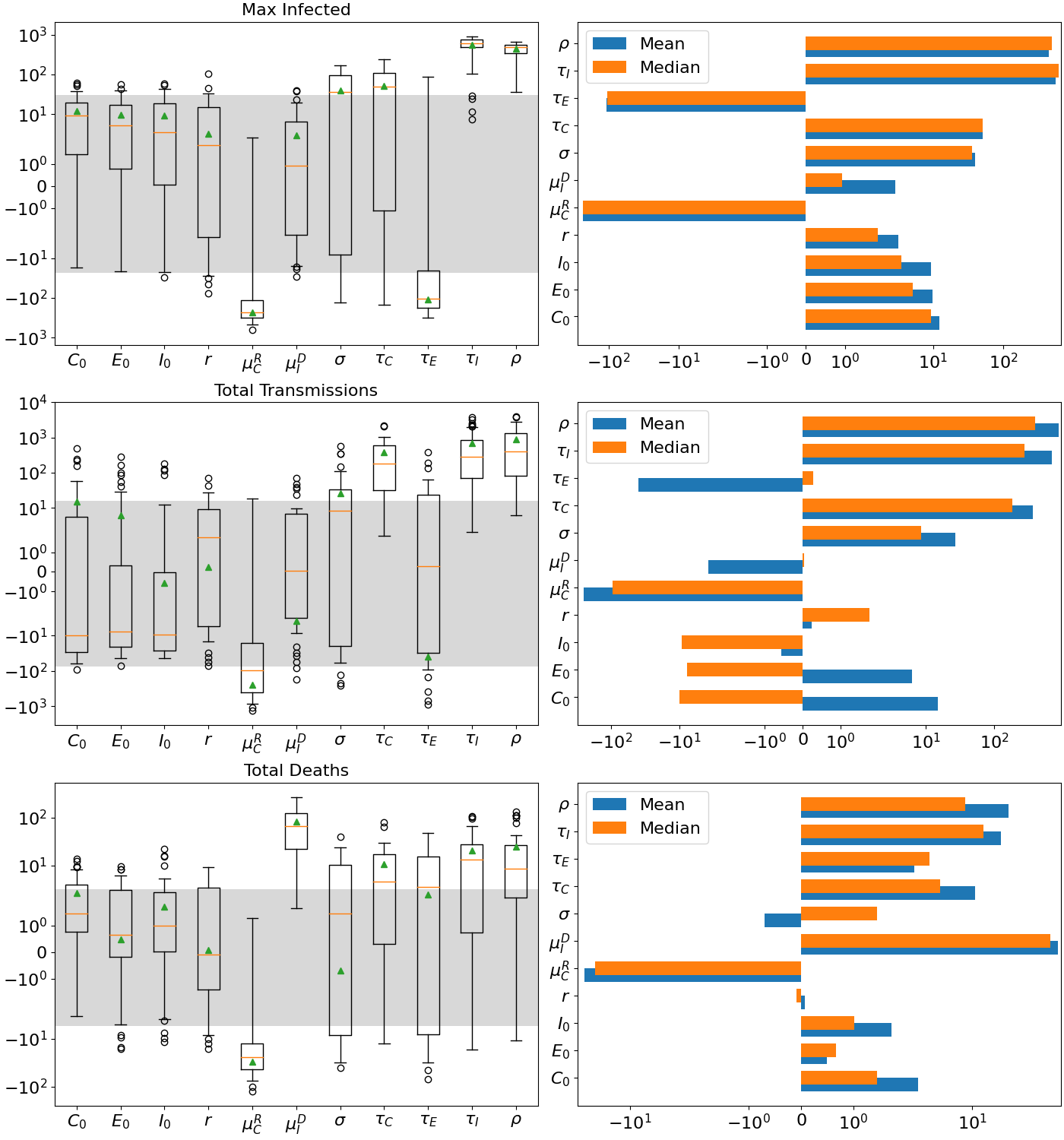"}
    \caption{\MK{\textbf{Sensitivity analysis of the ABM for the quadwell potential.} Shown are the elementary effects of the maximum number of agents simultaneously in infection state $I$ (top), the total number of transmissions (middle) and the total number of deaths (bottom). Model stochasticity is displayed by the grey bars that show the p25 and p75 percentiles of $112$ runs without parameter variation.}}
    \label{fig:sensitivity_ABM_qw}
\end{figure}

\begin{figure}[H]
    \centering
    \includegraphics[width=\textwidth]{"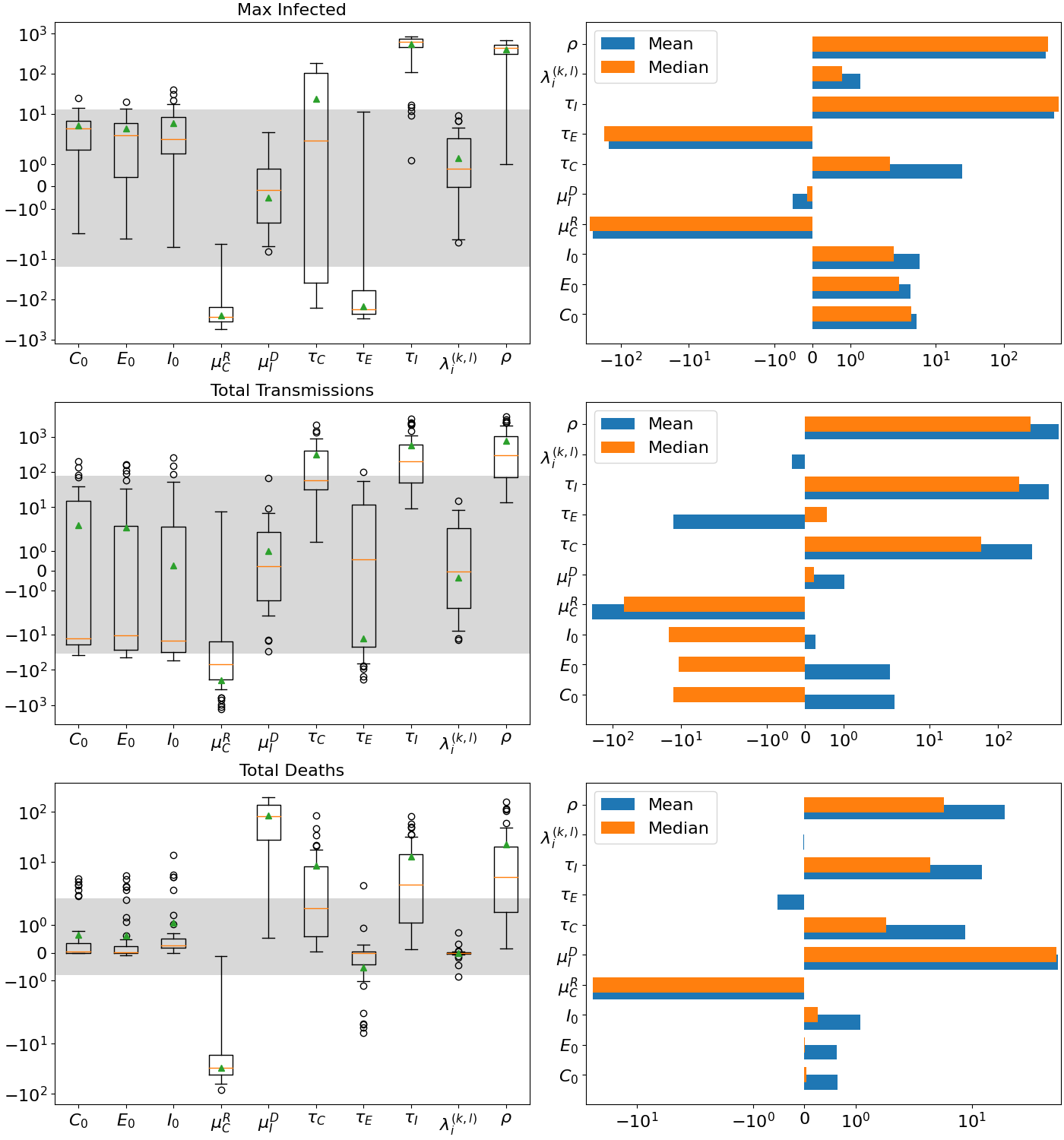"}
    \caption{\MK{\textbf{Sensitivity analysis of the PDMM for the quadwell potential.} Shown are the elementary effects of the maximum number of agents simultaneously in infection state $I$ (top), the total number of transmissions (middle) and the total number of deaths (bottom). Model stochasticity is displayed by the grey bars that show the p25 and p75 percentiles of $112$ runs without parameter variation.}}
    \label{fig:sensitivity_PDMM_qw}
\end{figure}

\begin{figure}[H]
    \centering
    \includegraphics[width=\textwidth]{"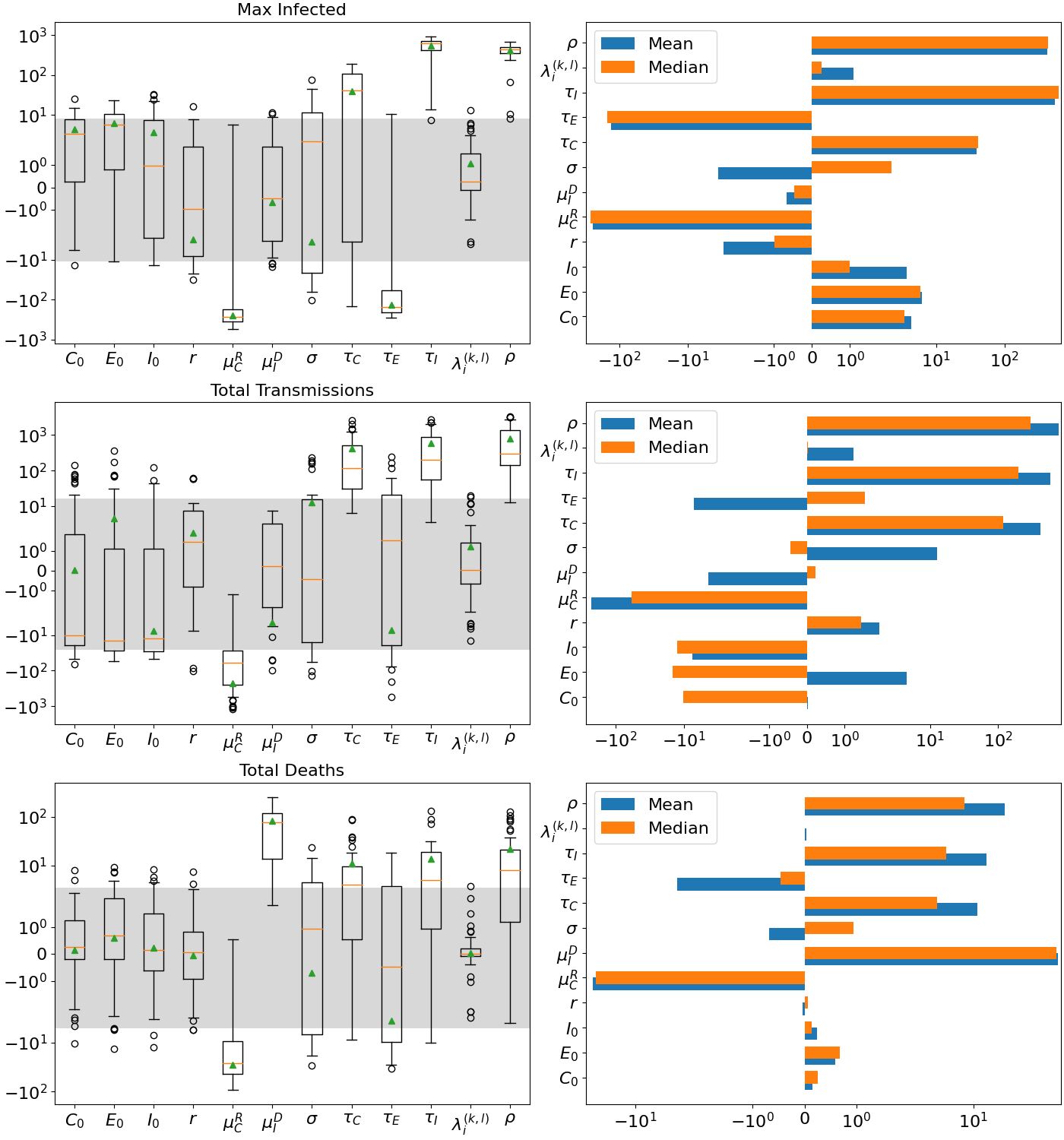"}
    \caption{\MK{\textbf{Sensitivity analysis of the spatial-hybrid model for the quadwell potential.} Shown are the elementary effects of the maximum number of agents simultaneously in infection state $I$ (top), the total number of transmissions (middle) and the total number of deaths (bottom). Model stochasticity is displayed by the grey bars that show the p25 and p75 percentiles of $112$ runs without parameter variation.}}
    \label{fig:sensitivity_Hybrid_qw}
\end{figure}

\MK{\subsection{Quadwell Potential}
The distribution of elementary effects gives information about how sensitive the considered output is to the parameter. As all models are stochastic, the elementary effects are not only influenced by the changed parameter value, but also by the models' inherent stochasticity as displayed by the grey bars in~\cref{fig:sensitivity_ABM_qw},~\cref{fig:sensitivity_PDMM_qw} and~\cref{fig:sensitivity_Hybrid_qw}. We first see that the remaining stochasticity is small (absolute values of 1-50, i.e., without the denominator in~\eqref{eq:elem_effect}) compared to the overall size of the system (4000 agents).  A high variance in the calculated elementary effects of a parameter indicates that the parameter's influence is highly dependent on the value of other parameters, while a small variance and a high absolute mean indicates parameters with high impact. For all models, the transmission rate $\rho$ and the average time in the Infected state $\tau_I$ have the highest positive influence on the maximum number of agents simultaneously in infection state $I$ with $\tau_I$ having a slightly higher mean than $\rho$. The probability of being asymptomatic $\mu_C^R$ followed by the average time in the Exposed state $\tau_E$ have the highest negative influence, i.e., an increase of $\mu_C^R$ and $\tau_E$ leads to a lower number of agents simultaneously in infection state $I$. This is in line with our expectations as the number of agents simultaneously in $I$ grows if agents stay on average longer in that infection state, which is caused by a large value of $\tau_I$. Additionally, more transmissions - caused by a large value of $\rho$ - also lead to more agents being in infection state $I$. In contrast, a larger proportion of agents recovering before developing symptoms lead to less agents in infection state $I$. Looking at the total number of transmissions, $\rho$ is - as expected - the most influential parameter, followed by $\tau_I$ and $\tau_C$ the average time in the Infected and Carrier state, respectively. The only parameter having a relevant negative influence on the total number of transmissions is $\mu_C^R$. For the number of total deaths, the probability of dying $\mu_I^D$ is the most influential parameter having a positive influence, followed by $\rho$, $\tau_I$ and $\tau_C$. The significant impact of $\mu_I^D$ on the number of total deaths is trivial. Moreover, large values of $\rho$, $\tau_I$, and $\tau_C$ lead to more agents in infection state $I$ which again has a positive influence on the number of deaths -- as agents can only die from infection state $I$; see~\cref{fig:transmission_model}. As for the maximum number of agents simultaneously in infection state $I$ and the total number of transmissions, the probability of being asymptomatic $\mu_C^R$ has again the highest negative influence on the total number of deaths.}

\begin{figure}[H]
    \centering
    \includegraphics[width=\textwidth]{"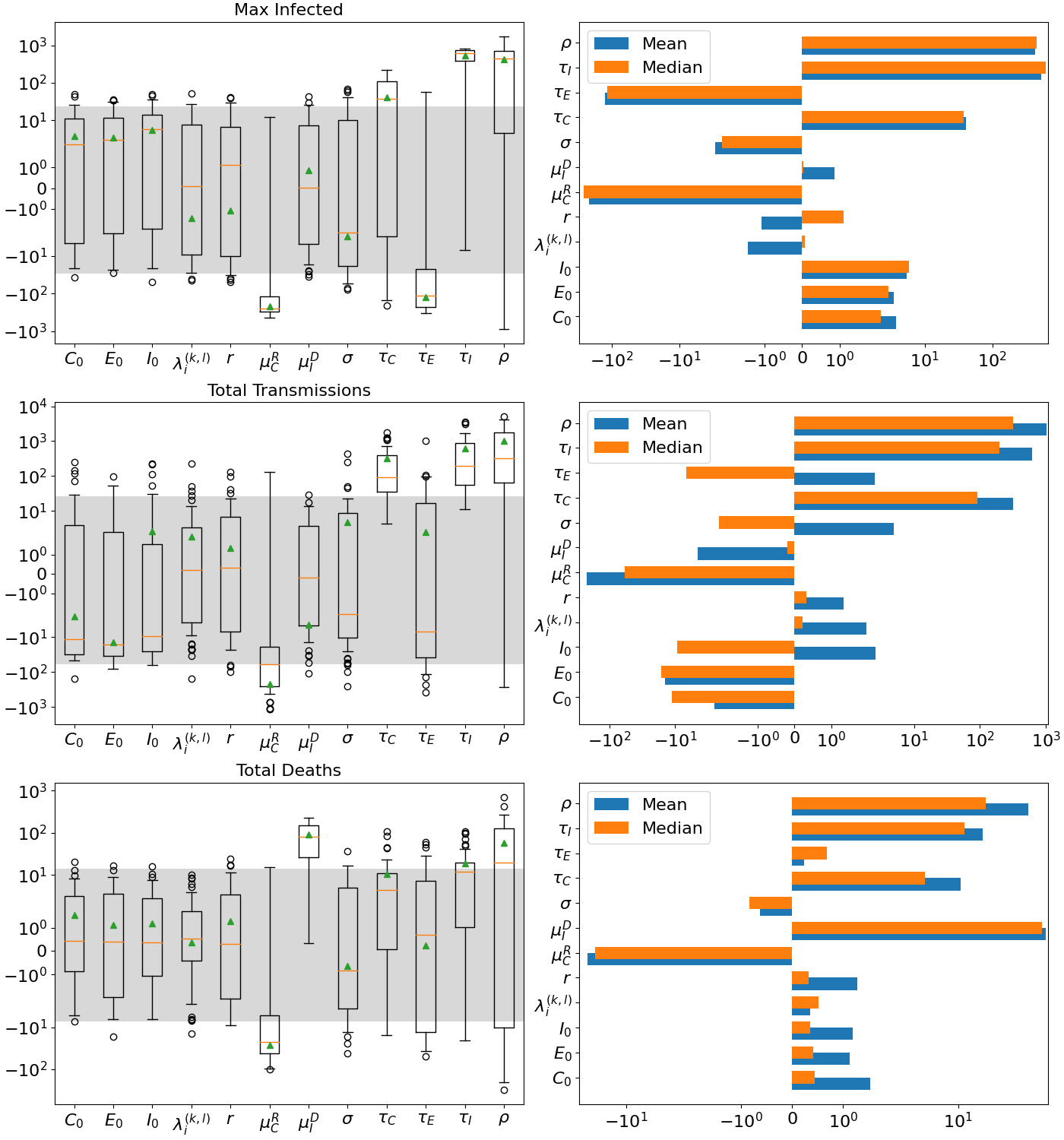"}
    \caption{\MK{\textbf{Sensitivity analysis of the ABM for the Munich potential.} Shown are the elementary effects of the maximum number of agents simultaneously in infection state $I$ (top), the total number of transmissions (middle) and the total number of deaths (bottom). Model stochasticity is displayed by the grey bars that show the p25 and p75 percentiles of $112$ runs without parameter variation.}}
    \label{fig:sensitivity_ABM_munich}
\end{figure}

\begin{figure}[H]
    \centering
    \includegraphics[width=\textwidth]{"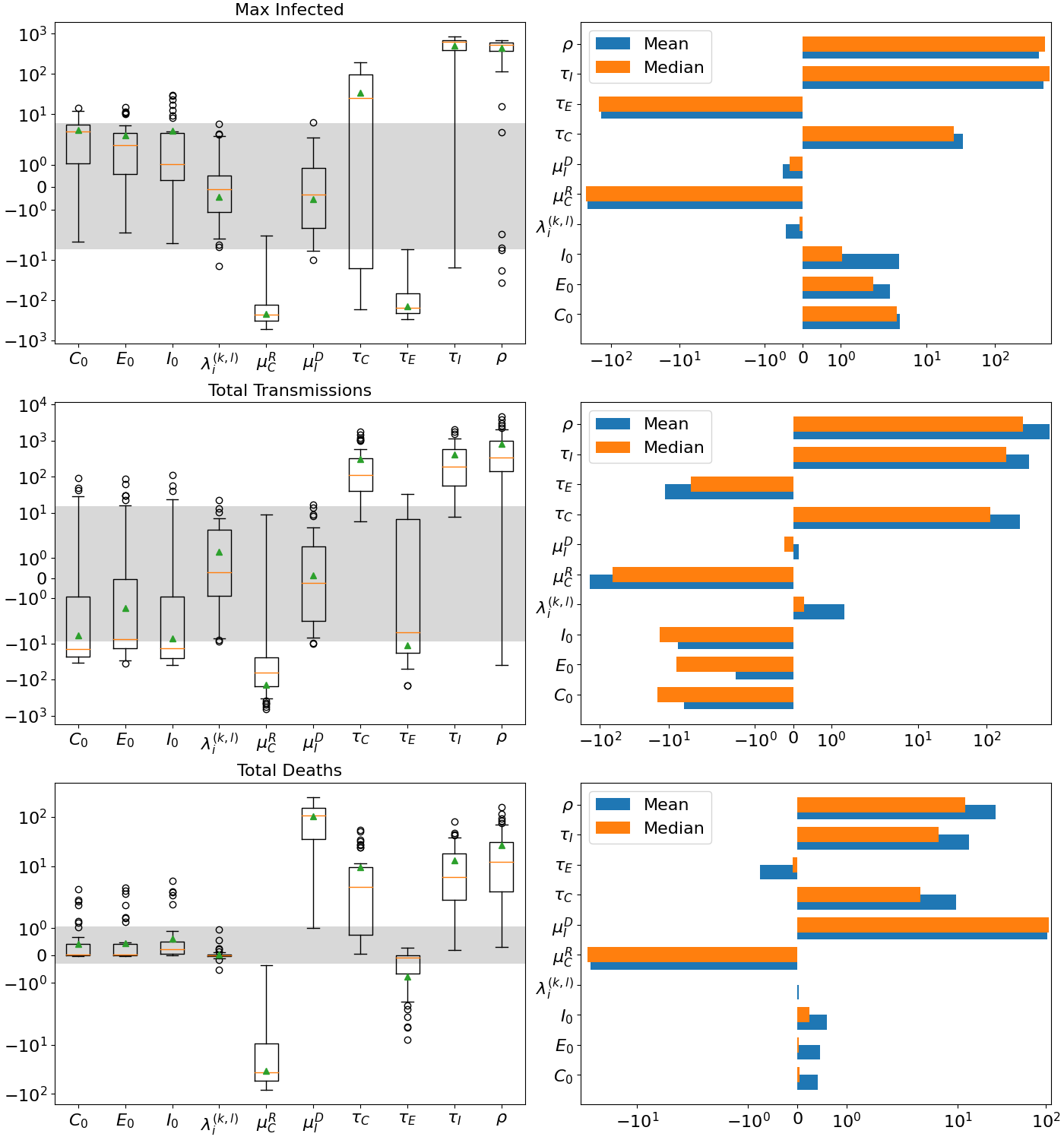"}
    \caption{\MK{\textbf{Sensitivity analysis of the PDMM for the Munich potential.} Shown are the elementary effects of the maximum number of agents simultaneously in infection state $I$ (top), the total number of transmissions (middle) and the total number of deaths (bottom). Model stochasticity is displayed by the grey bars that show the p25 and p75 percentiles of $112$ runs without parameter variation.}}
    \label{fig:sensitivity_PDMM_munich}
\end{figure}

\begin{figure}[H]
    \centering
    \includegraphics[width=\textwidth]{"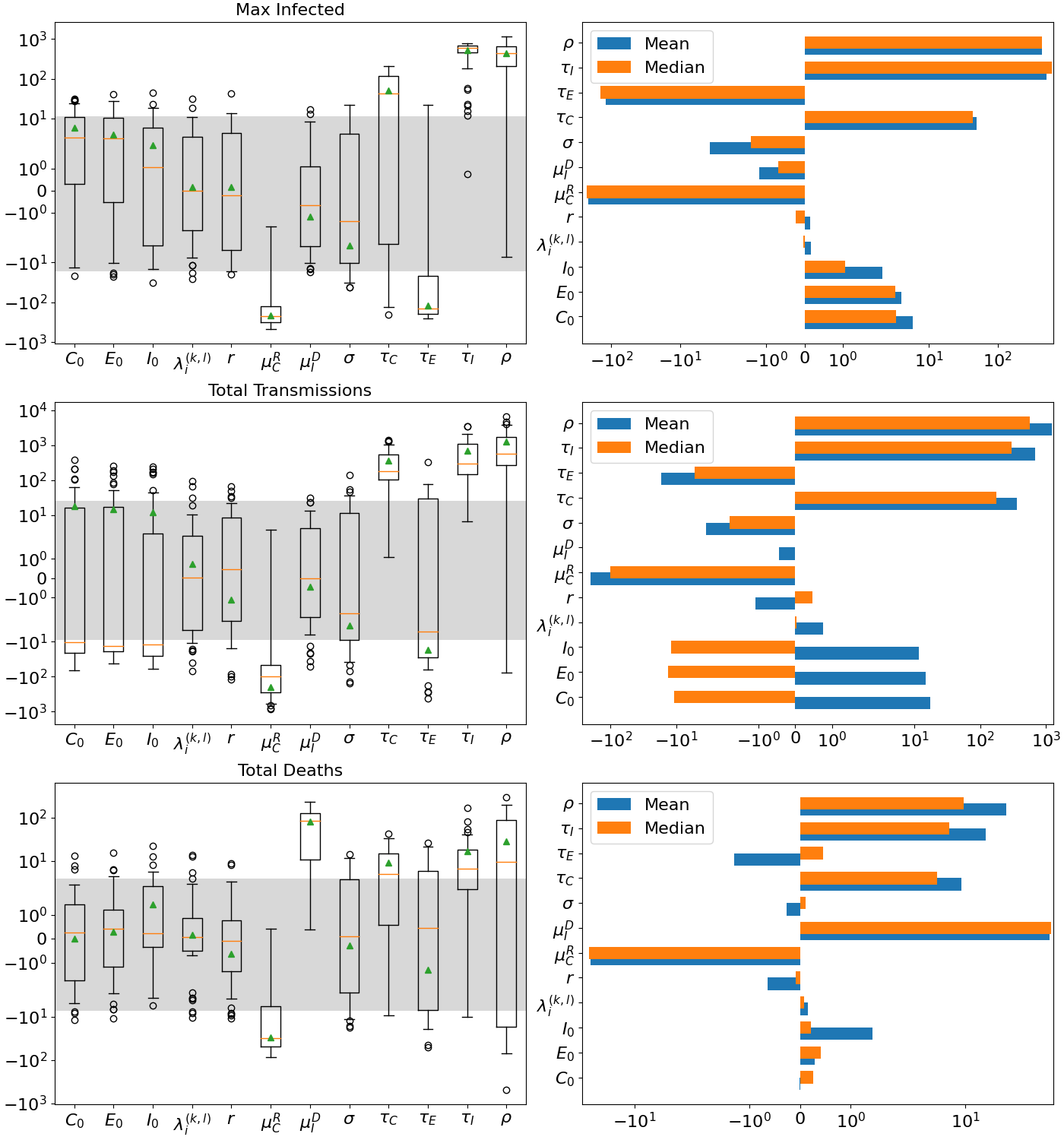"}
    \caption{\MK{\textbf{Sensitivity analysis of the spatial-hybrid model for the Munich potential.} Shown are the elementary effects of the maximum number of agents simultaneously in infection state $I$ (top), the total number of transmissions (middle) and the total number of deaths (bottom). Model stochasticity is displayed by the grey bars that show the p25 and p75 percentiles of $112$ runs without parameter variation.}}
    \label{fig:sensitivity_Hybrid_munich}
\end{figure}

\MK{\subsection{Munich Potential}
By adding the commuting term to the ABM movement, we get one additional parameter for the Munich application ABM and consequently also one additional parameter for the spatial-hybrid model. However, the results for the Munich potential align with the results for the quadwell potential, see~\cref{fig:sensitivity_ABM_munich},~\cref{fig:sensitivity_PDMM_munich} and~\cref{fig:sensitivity_Hybrid_munich} for more details.}

\begin{figure}[H]
    \centering
    \sidesubfloat[]{\includegraphics[width=\textwidth]{"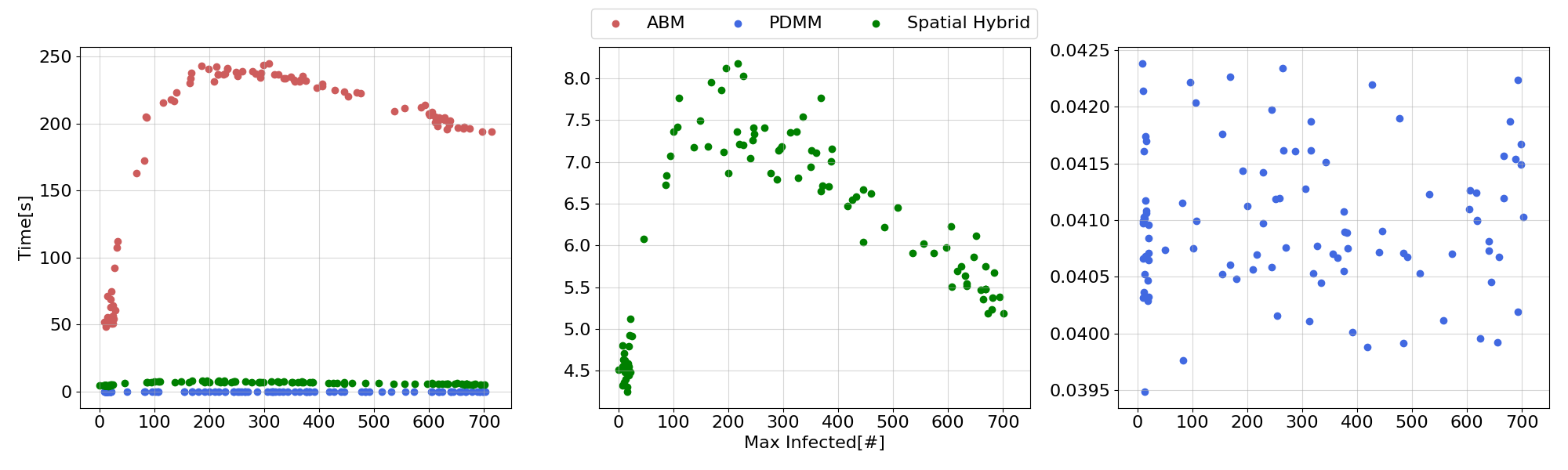"}\label{fig:qw_max_infected}}\\
    \sidesubfloat[]{\includegraphics[width=\textwidth]{"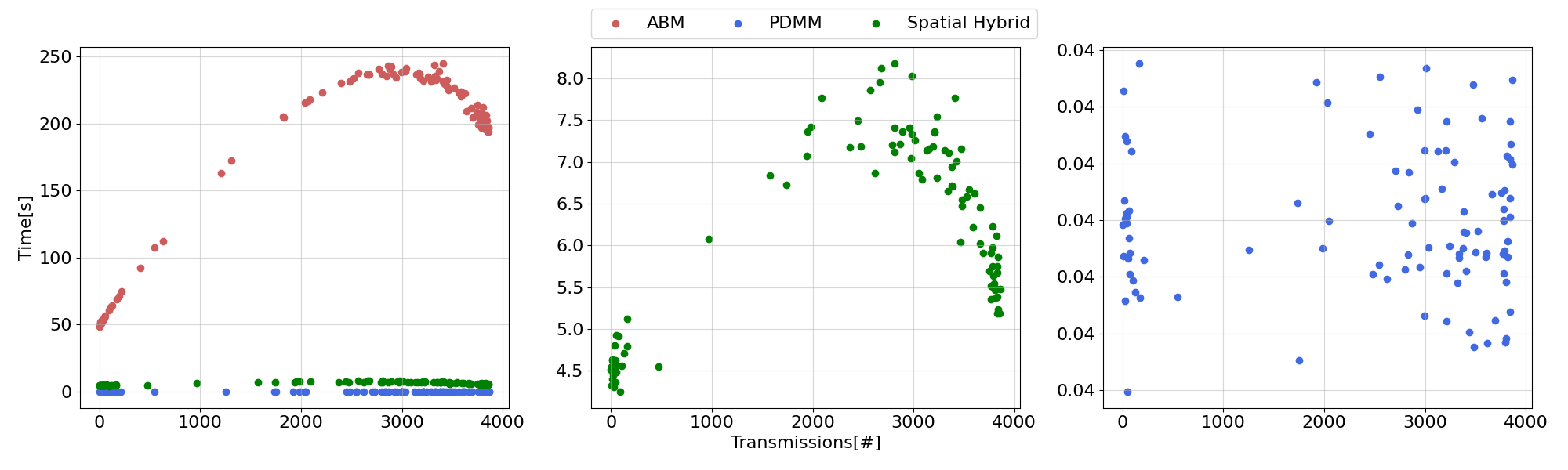"}\label{fig:qw_transmissions}}\\
    \sidesubfloat[]{\includegraphics[width=\textwidth]{"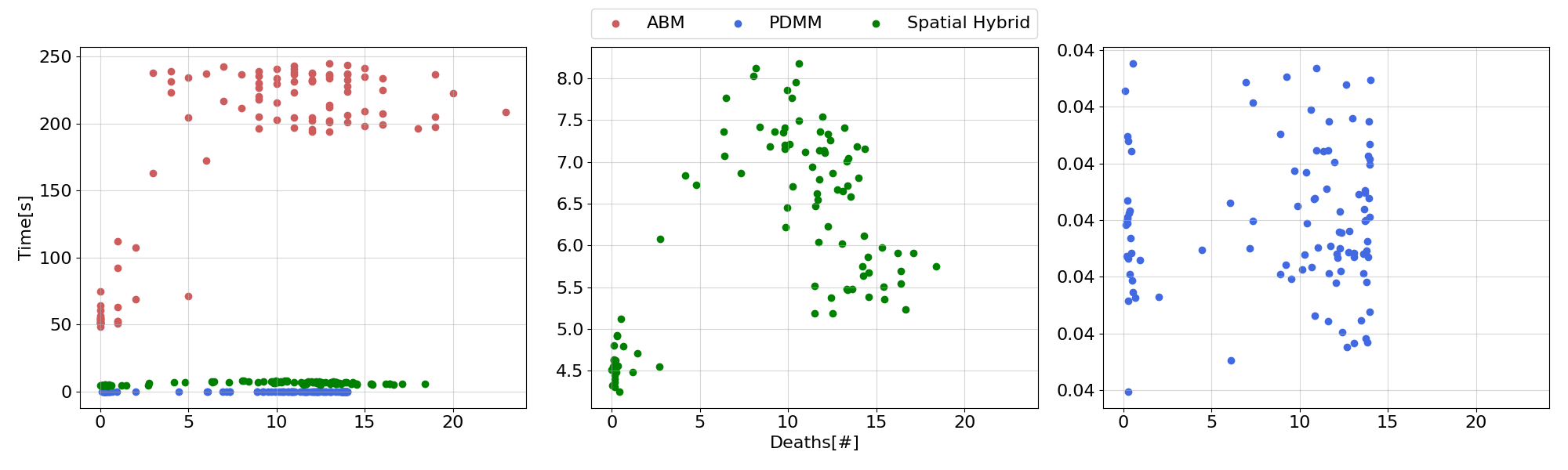"}\label{fig:qw_deaths}}
    \caption{\MK{\textbf{Runtime (in seconds) dependent on various model outputs for ABM, PDMM and spatial-hybrid model for the quadwell potential.} Shown is the runtime dependent on (a) the maximum number of agents simultaneously in infection state $I$, (b) the total number of transmissions and (c) and the total number of deaths for $100$ simulations with $4000$ agents per simulation.}}
    \label{fig:qw_runtime_sensi}
\end{figure}

\begin{figure}[H]
    \centering
    \sidesubfloat[]{\includegraphics[width=0.5\textwidth]{"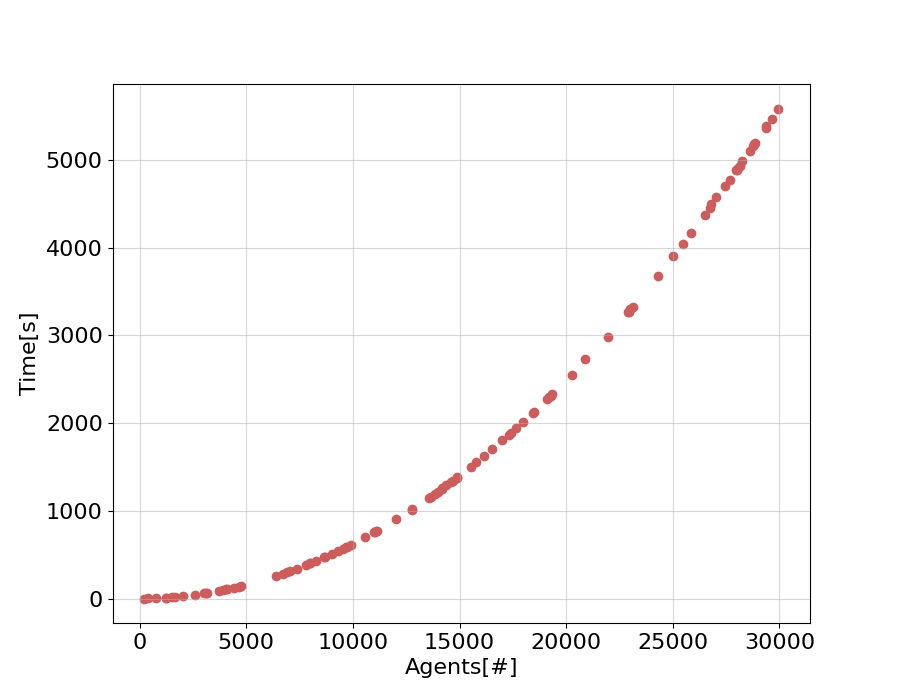"}\label{fig:qw_sus_ABM}}
    \sidesubfloat[]{\includegraphics[width=0.5\textwidth]{"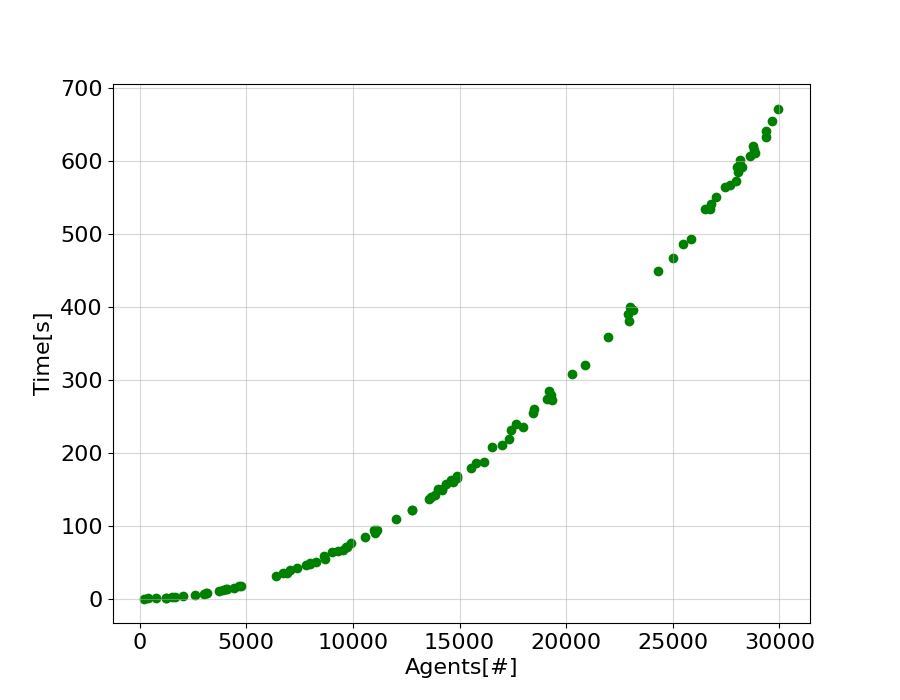"}\label{fig:qw_sus_Hybrid}}\\
    \sidesubfloat[]{\includegraphics[width=0.5\textwidth]{"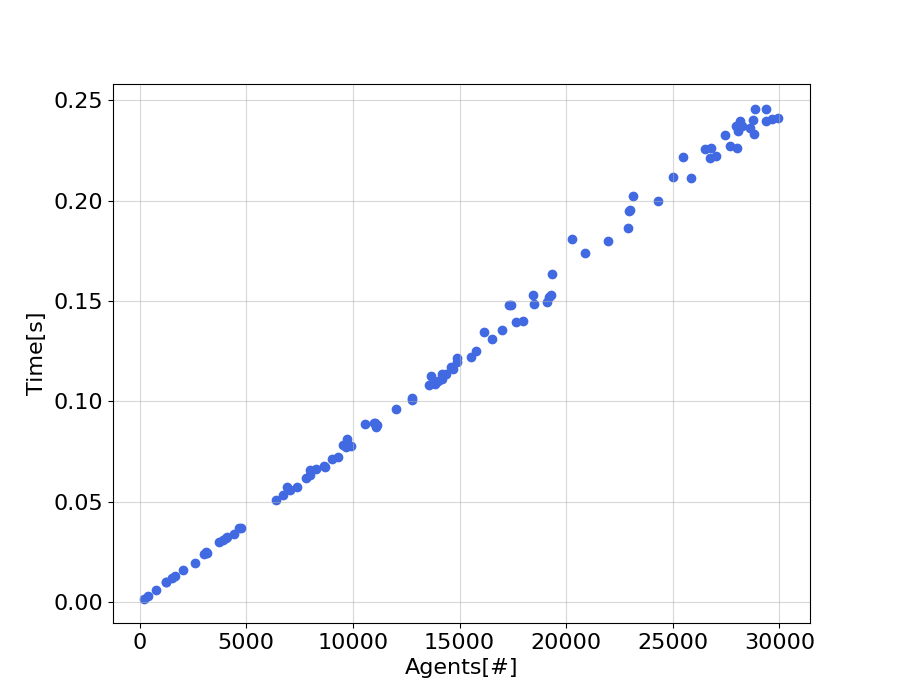"}\label{fig:qw_sus_PDMM}}
    \sidesubfloat[]{\includegraphics[width=0.5\textwidth]{"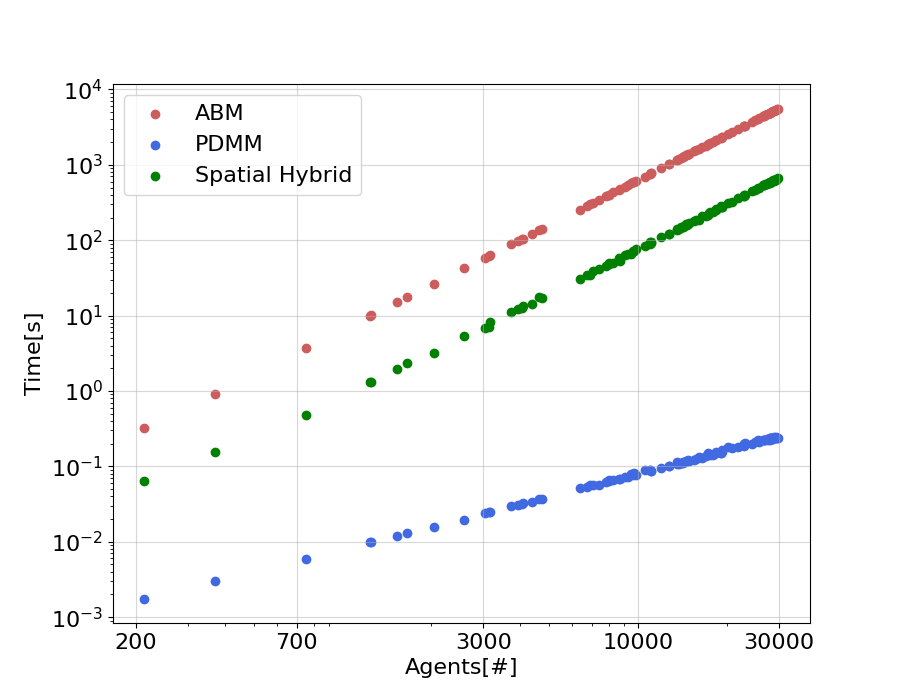"}\label{fig:qw_sus}}
    \caption{\MK{\textbf{Runtime (in seconds) for a completely susceptible population dependent on the number of agents for ABM, PDMM and spatial-hybrid model for the quadwell potential.} Shown is the runtime dependent on the number of agents for (a) the ABM, (b) the spatial-hybrid model, (c) the PDMM and (d) all models with log-scaled axes for $100$ simulations where the whole population is susceptible, i.e., there are no infection state adoptions.}}
    \label{fig:qw_runtime_sus}
\end{figure}

\begin{figure}[H]
    \centering
    \sidesubfloat[]{\includegraphics[width=\textwidth]{"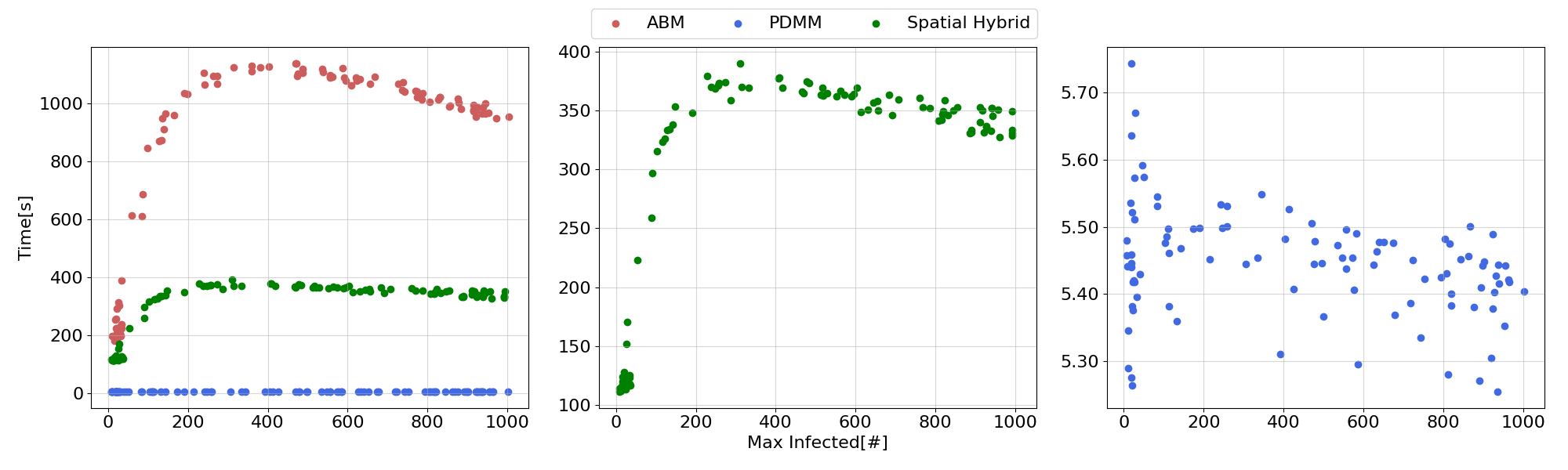"}\label{fig:munich_max_infected}}\\
    \sidesubfloat[]{\includegraphics[width=\textwidth]{"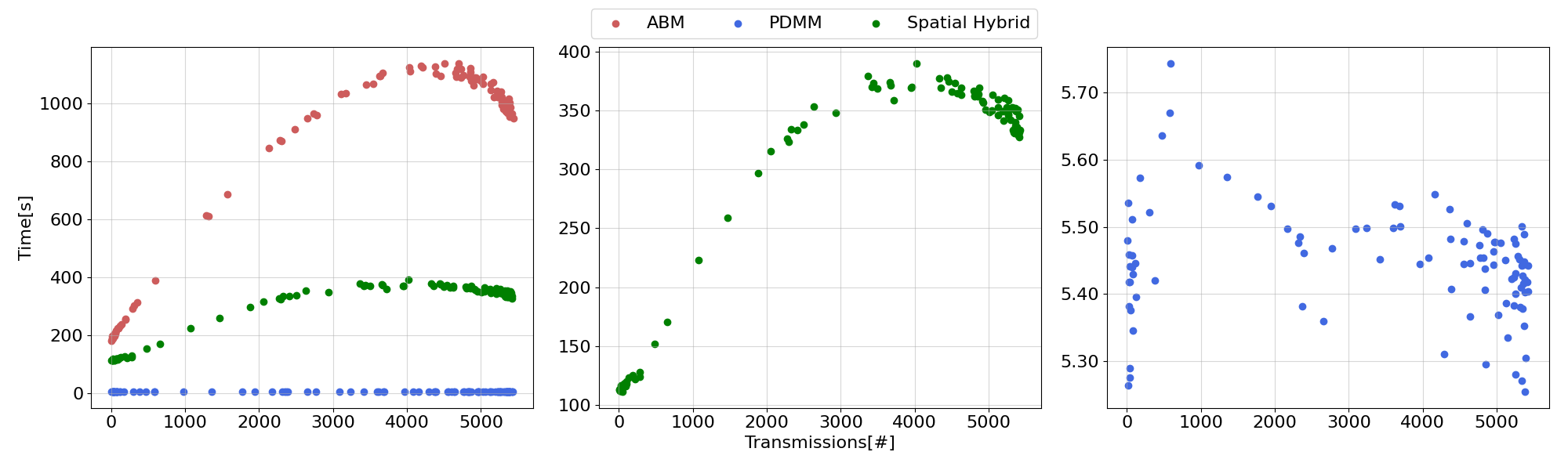"}\label{fig:munich_transmissions}}\\
    \sidesubfloat[]{\includegraphics[width=\textwidth]{"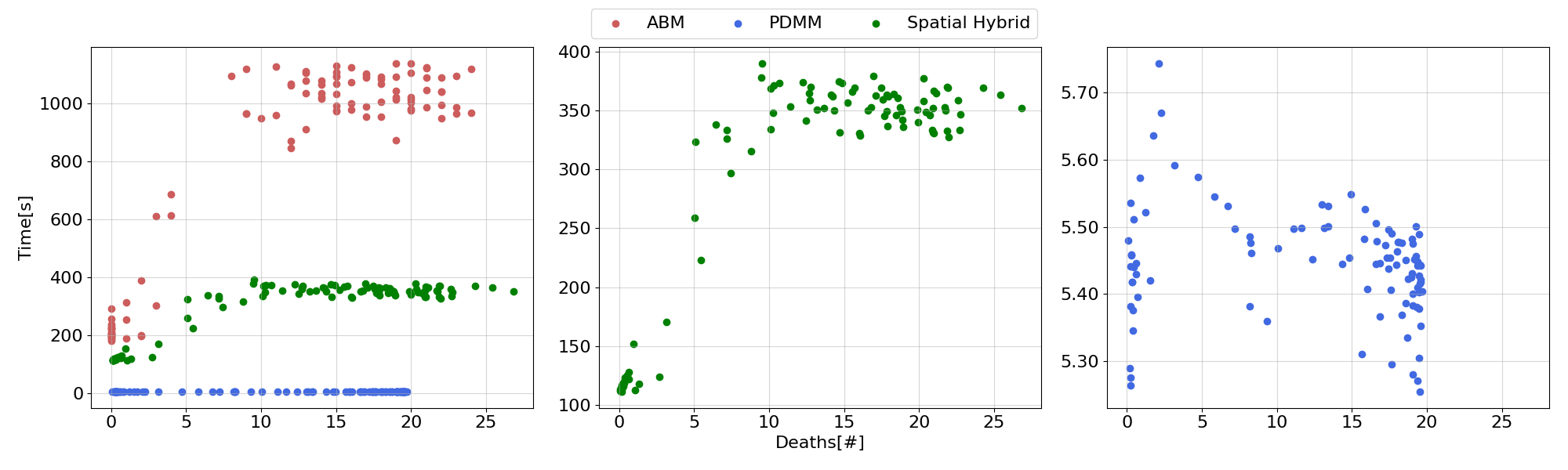"}\label{fig:munich_deaths}}
    \caption{\MK{\textbf{Runtime (in seconds) dependent on various model outputs for ABM, PDMM and spatial-hybrid model for the Munich potential.} Shown is the runtime dependent on (a) the maximum number of agents simultaneously in infection state $I$, (b) the total number of transmissions and (c) and the total number of deaths for $100$ simulations with $5600$ agents per simulation.}}
    \label{fig:munich_runtime_sensi}
\end{figure}

\begin{figure}[H]
    \centering
    \sidesubfloat[]{\includegraphics[width=0.5\textwidth]{"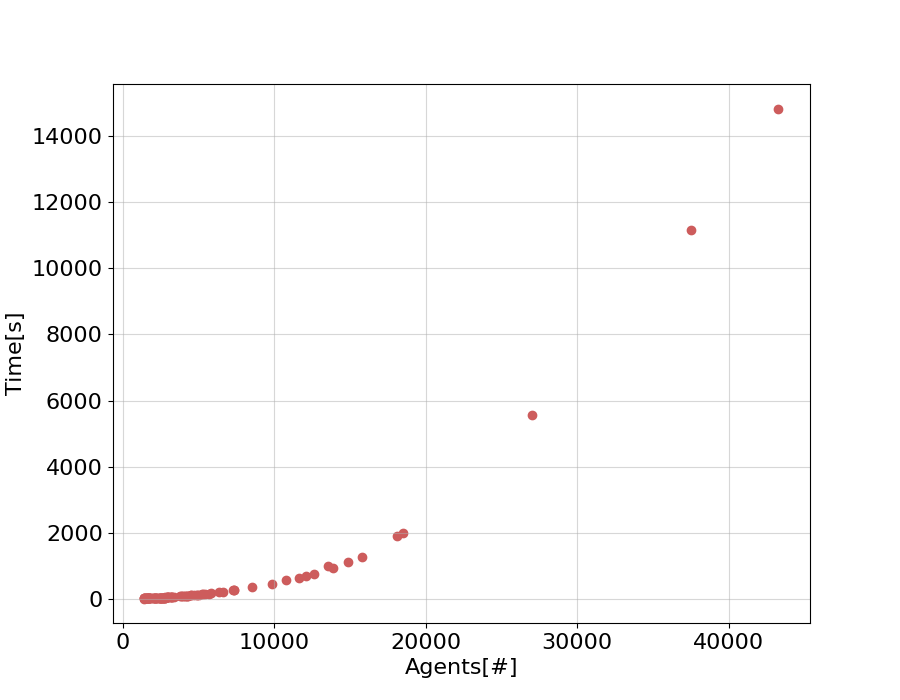"}\label{fig:munich_sus_ABM}}
    \sidesubfloat[]{\includegraphics[width=0.5\textwidth]{"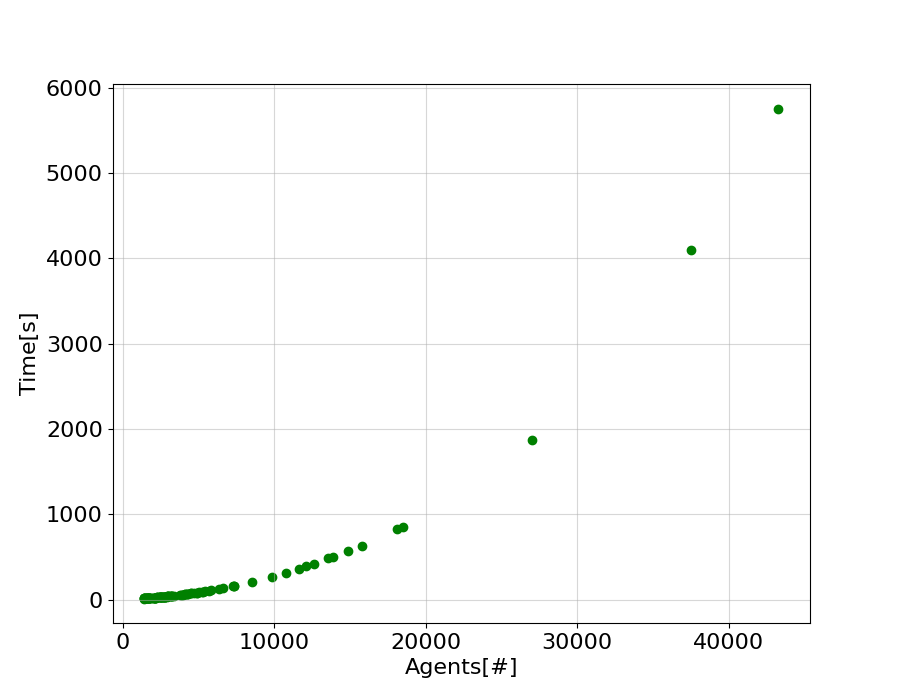"}\label{fig:munich_sus_Hybrid}}\\
    \sidesubfloat[]{\includegraphics[width=0.5\textwidth]{"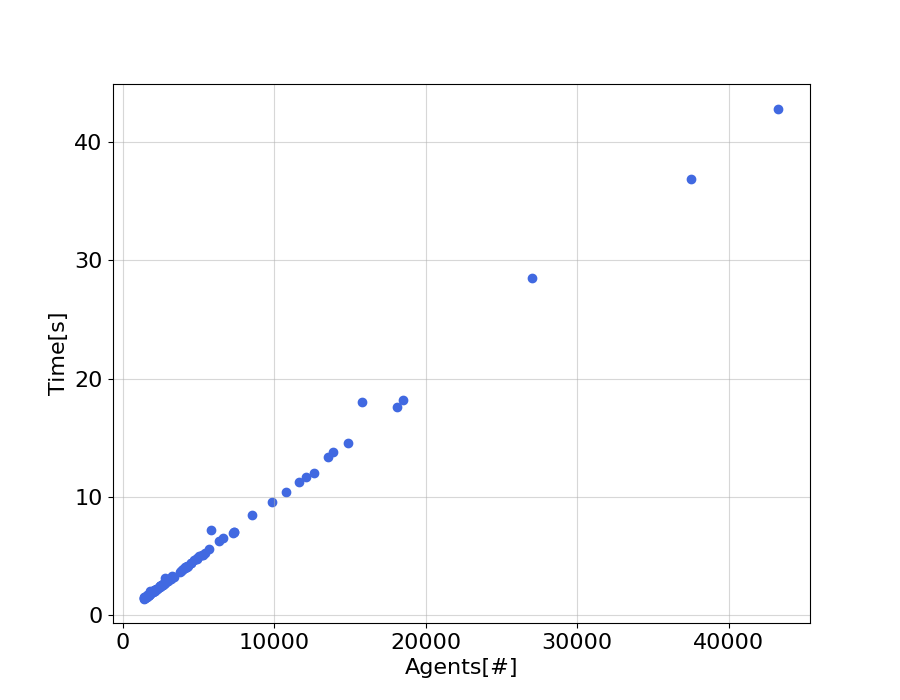"}\label{fig:munich_sus_PDMM}}
    \sidesubfloat[]{\includegraphics[width=0.5\textwidth]{"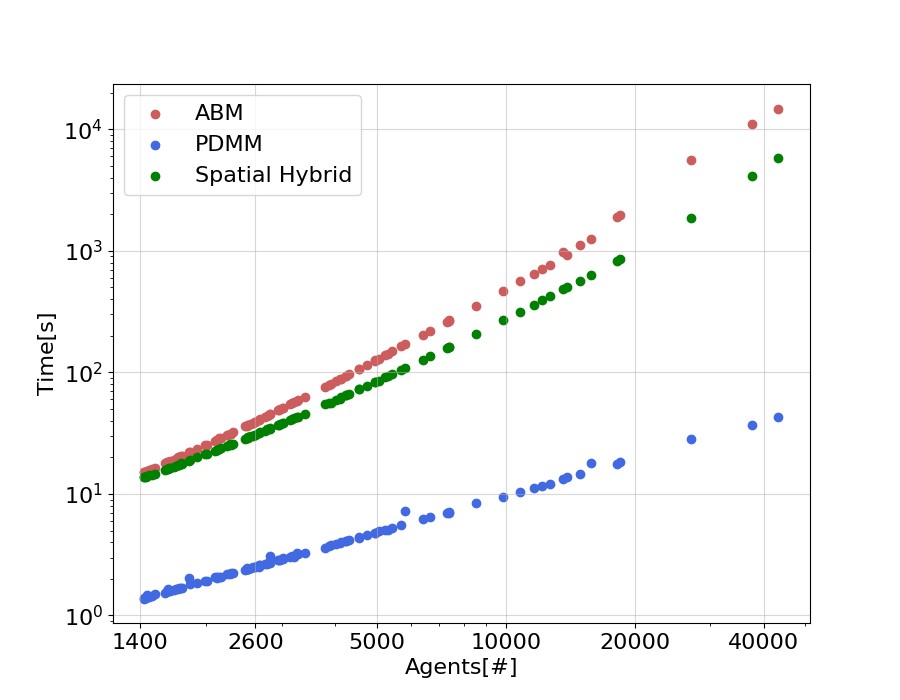"}\label{fig:munich_sus}}
    \caption{\MK{\textbf{Runtime (in seconds) for a totally susceptible population dependent on the number of agents for ABM, PDMM and spatial-hybrid model for the Munich potential.} Shown is the runtime dependent on the number of agents for (a) the ABM, (b) the spatial-hybrid model, (c) the PDMM and (d) all models with log-scaled axes for $100$ simulations where the whole population is susceptible, i.e., there are no infection state adoptions.}}
    \label{fig:munich_runtime_sus}
\end{figure}

\MK{\section{Runtime analysis}
\label{sec:Runtime analysis}
We used the results from the sensitivity analysis (see~\ref{sec:sensitivity_analysis}) to assess the influence of model parameters on the runtime. For that we performed $100$ simulations of ABM, PDMM, and spatial-hybrid model with $4000$ agents per simulation and considered the runtime dependence on the model outputs that we already considered in the sensitivity analysis, i.e., the maximum number of agents simultaneously in infection state $I$, the total number of transmissions, and the total number of deaths. As we identified with the sensitivity analysis the most influential parameters on the considered model outputs, we can, after having analyzed the influence of the outputs considered in the sensitivity analysis on the runtime, deduce which parameters have an influence on model runtime. For instance, if increasing the value of a specific parameter leads to an increase in the total number of transmissions, and an increase of the total number of transmissions leads to an increase in runtime, we can - using transitivity - deduce that an increase of the parameter value leads to an increase in runtime.  The results of the corresponding runtime analysis can be found in~\cref{fig:qw_runtime_sensi} for the quadwell potential and in~\cref{fig:munich_runtime_sensi} for the Munich potential. First, the figures indicate that both potentials have a very similar runtime behavior. Secondly, considering the left column of~\cref{fig:qw_runtime_sensi} and~\cref{fig:munich_runtime_sensi}, it can be seen that the ABM is several orders slower than the PDMM and the spatial-hybrid model whereby the runtime difference between the spatial-hybrid model and the ABM for the Munich potential is smaller than for the quadwell potential. As already mentioned this is due to high proportion of agents modeled with an ABM in the Munich spatial-hybrid model -- which is with $75\%$ three times higher compared to the quadwell where only $25\%$ of agents were modeled with an ABM. The runtime of ABM and spatial-hybrid model first increases with the maximum number of agents simultaneously in infection state $I$ (\cref{fig:qw_max_infected},~\cref{fig:munich_max_infected}) and the number of total transmissions (\cref{fig:munich_transmissions},~\cref{fig:qw_transmissions}) until it saturates and even slightly decreases again at the end. This effect stems from more transmissions and more infected individuals, which lead to more computations for infection state adoptions - first- and second-order. However, if the majority or even the whole population has been infected, no more computations for infection state adoptions are necessary, as there are no transitions from Recovered or Dead to another infection state. In contrast, there is no correlation between PDMM runtime and the number of infected agents, i.e., the runtime was not observed to increase with increasing transmissions or maximum number of agents simultaneously in infection state $I$. The runtime scaling for the total number of deaths looks similar (\cref{fig:qw_deaths},~\cref{fig:munich_deaths}), which is because more deaths arise from more infections, meaning more computations for infection state adoptions for the ABM and the spatial-hybrid model.~\cref{fig:qw_runtime_sus} and~\cref{fig:munich_runtime_sus} show the runtime scaling of all models for a total susceptible population, i.e., computations are only made for movement. The plots display the linear runtime scaling of the PDMM (see~\cref{fig:qw_sus_PDMM}, ~\cref{fig:munich_sus_PDMM}) and the superlinear scaling for ABM (see~\cref{fig:qw_sus_ABM},~\cref{fig:munich_sus_ABM}) and spatial-hybrid model (see~\cref{fig:qw_sus_Hybrid},~\cref{fig:munich_sus_Hybrid}).} 

\begin{figure}[H]
    \centering
    \includegraphics[width=0.6\textwidth]{"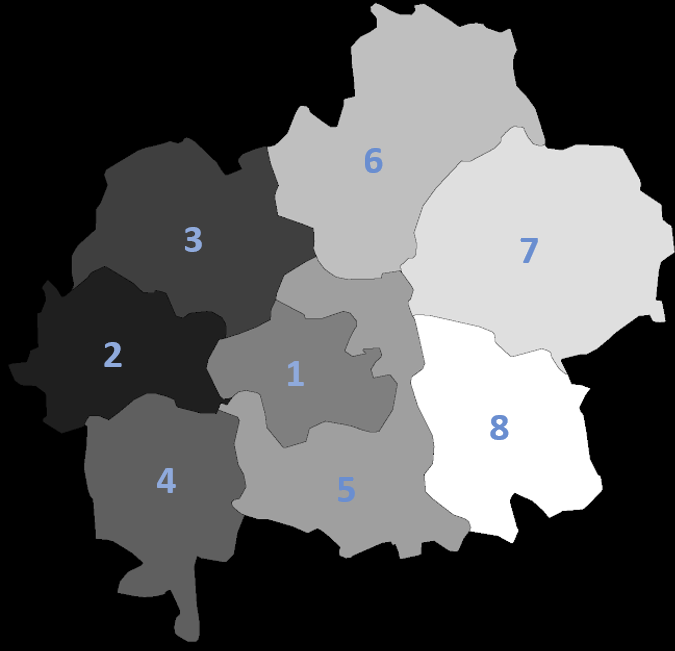"}
    \caption{\MK{\textbf{Regions and corresponding indices for the Munich potential.} The regions, i.e., potential wells are 1 - Munich City, 2 - Fürstenfeldbruck, 3 - Dachau, 4 - Starnberg, 5 - München Land, 6 - Freising, 7 - Erding, 8 - Ebersberg.}}
    \label{fig:metaregions_munich}
\end{figure}

\MK{\section{Commute weights for the Munich potential}
The commute weights used in~\cref{eq:prob_K} for the commuting term $K$ (\cref{eq:K}) in the Munich potential are given by the matrix $\Lambda \in \mathbb{R}^{8\times 8}$ with
\begin{align}
\label{eq:comm_matrix}
    \Lambda = \begin{pmatrix}
        0 & 0.03212 & 0.02734 & 0.02059 & 0.11452 & 0.02289 & 0.02989 & 0.02053\\
        0.21861 & 0 & 0.02286 & 0.03115 & 0.0388 & 0.00586 & 0.00236 & 0.0027\\
        0.26164 & 0.03214 & 0 & 0.00623 & 0.04413 & 0.01992 & 0.0063 & 0.00455\\
        0.22401 & 0.04979 & 0.00709 & 0 & 0.04785 & 0.00572 & 0.004 & 0.00305\\
        0.48702 & 0.02424 & 0.01961 & 0.01871 & 0 & 0.03798 & 0.02593 & 0.04008\\
        0.188 & 0.00707 & 0.0171 & 0.00432 & 0.07334 & 0 & 0.07431 & 0.00901\\
        0.31851 & 0.00369 & 0.00701 & 0.00392 & 0.06496 & 0.09641 & 0 & 0.04246\\
        0.21125 & 0.00408 & 0.00489 & 0.00289 & 0.09699 & 0.01129 & 0.04101 & 0
    \end{pmatrix}.
\end{align}
Entry $(k,l)$, $k,l=1,\ldots,8$ of $\Lambda$ is the commute rate $\lambda^{(k,l)}$ as well as the PDMM spatial transition rate between region $\Omega_k$ and $\Omega_l$, see~\cref{fig:metaregions_munich} for a mapping of the regions. The matrix $\Lambda$ corresponds to a submatrix of the commuter matrix used in~\cite{kuhn_assessment_2021} with entry $(k,l)$ divided by the population size of county $k$ such that we have relative instead of absolute numbers.} 

% \newpage
\section{Supplementary Tables}
\label{supp:tables}

\begin{table}[H]
    \centering
    \renewcommand{\arraystretch}{1.5}
    \begin{tabular}{p{0.15\textwidth}|p{0.5\textwidth}|p{0.14\textwidth}|p{0.15\textwidth}}
        \hline
         \textbf{Parameter} & \textbf{Description} & \textbf{Range} & \textbf{Source}  \\
         \hline
         $n_a$ & Number of agents & &\\
        $\rho^{(k)}$ & Transmission rate & $\left[ 0.025, 0.6\right]$ & \cite{kuhn_assessment_2021}\\
        $\xi_C^{(k)}$ & Increase or decrease factor on transmission probability for Carrier agents & $1.0$ & (A)\\
        $\xi_I^{(k)}$ & Increase or decrease factor on transmission probability for Infected agents & $1.0$ & (A)\\
        $\tau_E$ & Average time in Exposed state & $\left[2.67, 4.0\right]$ & \cite{kuhn_assessment_2021}\\
        $\tau_C$ & Average time in Carrier state & $\left[1.2, 6.73\right]$ & \cite{kuhn_assessment_2021}\\
        $\tau_I$ & Average time in Infected state & $\left[5.0, 12.0\right]$ & \cite{kuhn_assessment_2021}\\
        $\mu_C^R$ & Probability of transitioning from Carrier to Recovered & $\left[0.15, 0.3\right]$ & \cite{kuhn_assessment_2021}\\
        $\mu_I^D$ & Probability of transitioning fron Infected to Dead & $\left[0.0, 0.08\right]$ & \cite{kuhn_assessment_2021}\\
        $\gamma_{E, C}^{(k)}$ & Rate from Exposed to Carrier in region $k$ & $\frac{1}{\tau_E}$ & see $\tau_E$\\
        $\gamma_{C, R}^{(k)}$ & Rate from Carrier to Recovered in region $k$ & $\frac{\mu_C^R}{\tau_C}$& see $\mu_C^R$, $\tau_C$\\
        $\gamma_{C, I}^{(k)}$ & Rate from Carrier to Infected in region $k$ & $\frac{1-\mu_C^R}{\tau_C}$ & see $\mu_C^R$, $\tau_C$\\
        $\gamma_{I, D}^{(k)}$ & Rate from Infected to Dead in region $k$ & $\frac{\mu_I^D}{\tau_I}$ & see $\mu_I^D$, $\tau_I$\\
        $\gamma_{I, R}^{(k)}$ & Rate from Infected to Recovered in region $k$ & $\frac{1-\mu_I^D}{\tau_I}$ & see $\mu_I^D$, $\tau_I$\\
        $E_0$ & Proportion of initially Exposed agents & $\left[ 0.0, 0.01\right]$\\
        $C_0$ & Proportion of initial Carrier agents & $\left[ 0.0, 0.01\right]$\\
        $I_0$ & Proportion of initial Infected agents & $\left[ 0.0, 0.01\right]$\\
    \end{tabular}
    \caption{\textbf{Parameters shared between all models and their ranges used for all sensitivity analyses}. Parameter ranges are either based on \cite{kuhn_assessment_2021} corresponding to wild-type COVID-19 or are assumed to be fixed (A) to reduce model complexity.}
    \label{tab:global_param_ranges}
\end{table}

\begin{table}[H]
    \centering
    \renewcommand{\arraystretch}{1.5}
    \begin{tabular}{p{0.15\textwidth}|p{0.5\textwidth}|p{0.14\textwidth}|p{0.15\textwidth}}
        \hline
         \textbf{Parameter} & \textbf{Description} & \textbf{Range} & \textbf{Source}  \\
         \hline
        $r$ & Interaction radius$^*$ & $\left[0.05,0.6\right]$ & (B), see~\cref{fig:radii}\\
        $\sigma$ & Noise term of the diffusion process$^*$ & $\left[0.4,0.6\right]$ & (B), see~\cref{fig:sigma_movement}\\
        $\lambda_i^{(k,l)}$ & Spatial transition rate from region $k$ to region $l$ being in infection state $i$$^{**}$ & $\left[2e^{-5},0.0125\right]$ & (B), match range for $\sigma$
    \end{tabular}
    \caption{\textbf{Parameters and their ranges used for the sensitivity analysis of the quadwell application.} The asterisk $^*$ denotes exclusive ABM parameters and $^{**}$ denotes exclusive PDMM parameters. (B) denotes ranges that are based on preliminary investigations done to determine a reasonable range for the parameter.}
    \label{tab:qw_param_ranges}
\end{table}

\begin{table}[H]
    \centering
    \renewcommand{\arraystretch}{1.5}
    \begin{tabular}{p{0.15\textwidth}|p{0.5\textwidth}|p{0.14\textwidth}|p{0.15\textwidth}}
        \hline
         \textbf{Parameter} & \textbf{Description} & \textbf{Range} & \textbf{Source}  \\
         \hline
        $r$ & Interaction radius$^*$ & $\left[65.0,260.0\right]$ & (B), see~\cref{fig:radii_munich}\\
        $\sigma$ & Noise term of the diffusion process$^*$ & $\left[5.0,15.0\right]$ & (B), see~\cref{fig:sigma_movement_munich}\\
        $\lambda^{(k,l)}$ & Commute rate$^*$ and spatial transition rate$^{**}$ from region $k$ to region $l$& $\left[1e^{-3},0.22\right]$ & \cite{kuhn_assessment_2021}, \cite{bmas_pendlerverflechtungen_2020}, \cite{twitter_twitter_2020}
    \end{tabular}
    \caption{\textbf{Parameters and their ranges used for the sensitivity analysis of the Munich application.} The asterisk $^*$ denotes exclusive ABM parameters and $^{**}$ denotes exclusive PDMM parameters. (B) denotes ranges that are based on preliminary investigations done to determine a reasonable range for the parameter.}
    \label{tab:munich_param_ranges}
\end{table}

\begin{table}[H]
    \centering
    \renewcommand{\arraystretch}{1.5}
    \begin{tabular}{p{0.2\textwidth}|p{0.2\textwidth}}
        \hline
         \textbf{Parameter} & \textbf{Value}  \\
         \hline
        $\xi_C^{(k)}$ & 1.0\\
        $\xi_I^{(k)}$ & 1.0\\
        $\tau_E$ & 3\\
        $\tau_C$ & 3\\
        $\tau_I$ & 5\\
        $\mu_C^R$ & 0.1\\
        $\mu_I^D$ & 0.004
    \end{tabular}
    \caption{\textbf{Parameter values shared for all applications.}}
    \label{tab:joint_transmission_params}
\end{table}

\begin{table}[H]
    \centering
    \renewcommand{\arraystretch}{1.5}
    \begin{tabular}{p{0.6\textwidth}|p{0.2\textwidth}}
        \hline
         \textbf{Parameter} & \textbf{Value}  \\
         \hline
        $\rho^{(k)}$, $k=1,3,4$ & 0.1\\
        $\rho^{(2)}$ & 0.3\\
        $\sigma$  & 0.55\\
        $r$ & 0.4\\
        $\lambda_D^{(k,l)}$ & 0.0\\
        $\lambda_i^{(k,l)}$, $(k,l)\in \mathcal{D} :=\{(1,4),(4,1),(2,3),(3,2)\}$ & $1e^{-7}$\\
        $\lambda_i^{(k,l)}$, $(k,l)\in\{1,2,3,4\}^2 \setminus \mathcal{D}$, $i=S,R$ & 0.0048\\
        $\lambda_i^{(k,l)}$, $(k,l)\in\{1,2,3,4\}^2 \setminus \mathcal{D}$, $i=E,C,I$ & 0.0033
    \end{tabular}
    \caption{\textbf{Parameter values for the quadwell potential.}}
    \label{tab:params_qw}
\end{table}

\begin{table}[H]
    \centering
    \renewcommand{\arraystretch}{1.5}
    \begin{tabular}{p{0.4\textwidth}|p{0.2\textwidth}}
        \hline
         \textbf{Parameter} & \textbf{Value}  \\
         \hline
        $\rho^{(k)}$, $k=1,\ldots,8$ & 0.2\\
        $\sigma$  & 10\\
        $r$ & 280
    \end{tabular}
    \caption{\textbf{Parameter values for the Munich potential.}}
    \label{tab:params_munich}
\end{table}

\begin{table}[H]
    \centering
    \renewcommand{\arraystretch}{1.5}
    \begin{tabular}{p{0.2\textwidth}|p{0.2\textwidth}}
        \hline
         \textbf{Parameter} & \textbf{Value}  \\
         \hline
        $\rho^{(1)}$ & 0.6\\
        $\sigma$  & 0.5\\
        $r$ & 0.1\\
        $\kappa$ & 0.8
    \end{tabular}
    \caption{\textbf{Parameter values for the single well potential.}}
    \label{tab:params_sw}
\end{table}

% \newpage
\section{Supplementary Figures}
\label{supp:figures}

\begin{figure}[H]
    \centering
    \includegraphics[width=1.05\textwidth]{"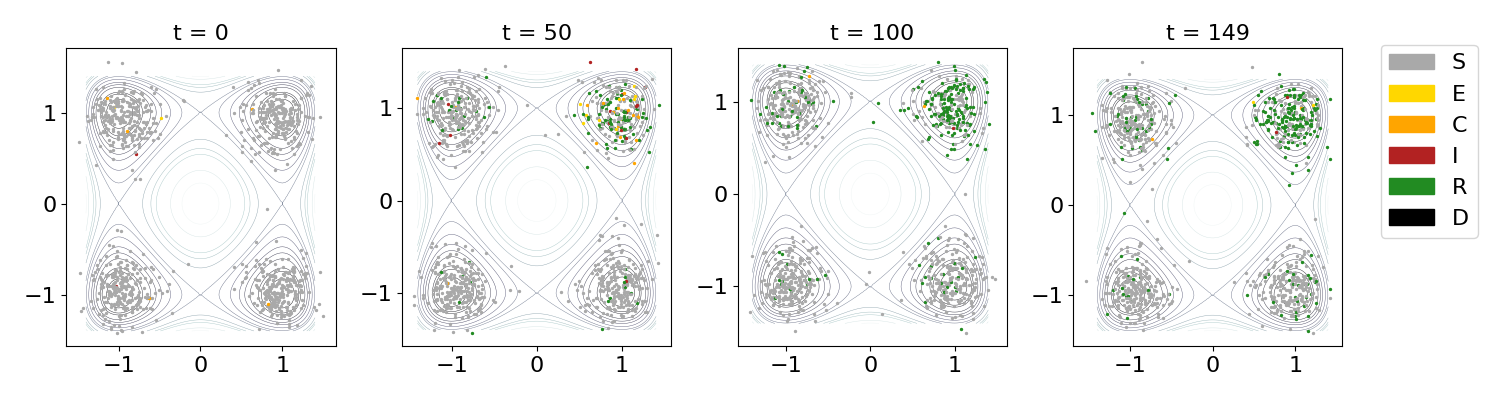"}
    \caption{\MK{\textbf{System state for one ABM realization for the quadwell potential at four different time points.} Shown are the time points $t=0, 50, 100$ and $149$ for a simulation with $1000$ agents for the quadwell application.}}
    \label{fig:qw_trajectory}
\end{figure}

\begin{figure}[H]
    \centering
    \includegraphics[width=0.6\textwidth]{"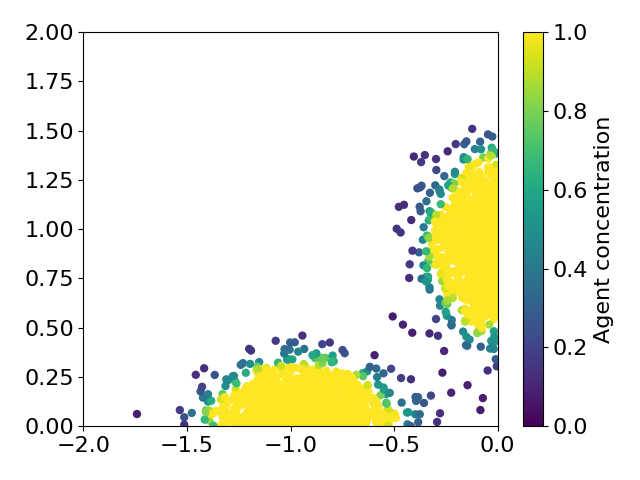"}
    \caption{\MK{\textbf{Position distribution of $2850$ agents transitioning to $\Omega_1$ (focus region) in the quadwell potential.}}}
    \label{fig:qw_trans_dist}
\end{figure}

\begin{figure}[H]
    \centering
    \sidesubfloat[]{\includegraphics[width=0.5\textwidth]{"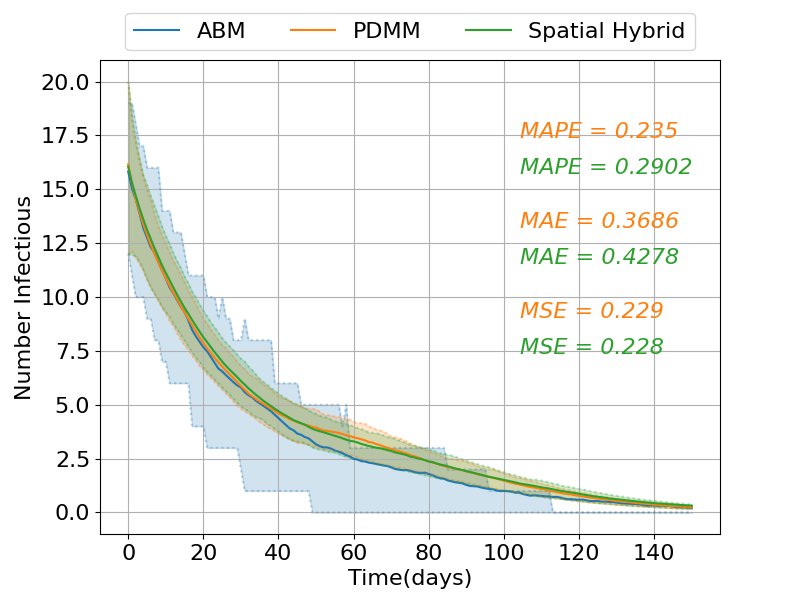"}\label{fig:qw_r3}}
    \sidesubfloat[]{\includegraphics[width=0.5\textwidth]{"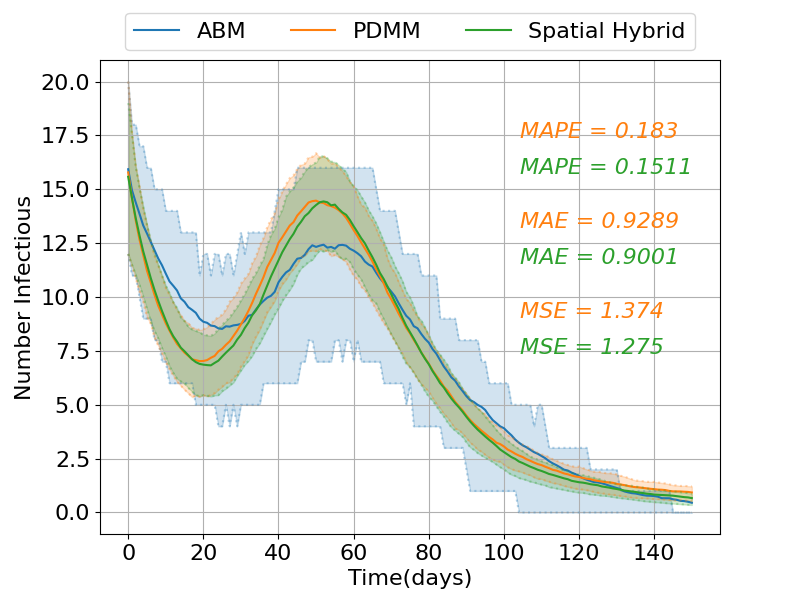"}\label{fig:qw_r4}}%
    \caption{\MK{\textbf{Spatial hybridization for the quadwell potential: Region 3 and 4.} Number of infectious agents (compartments $C$ and $I$) for (a) $\Omega_3$ and (b) $\Omega_4$. The figures show the mean outcomes in solid lines with a partially transparent face between the p25 and p75 percentiles from 500 runs.}}
    \label{fig:results_qw_r3_r4}
\end{figure}

\begin{figure}[H]
    \centering
    \includegraphics[width=1.05\textwidth]{"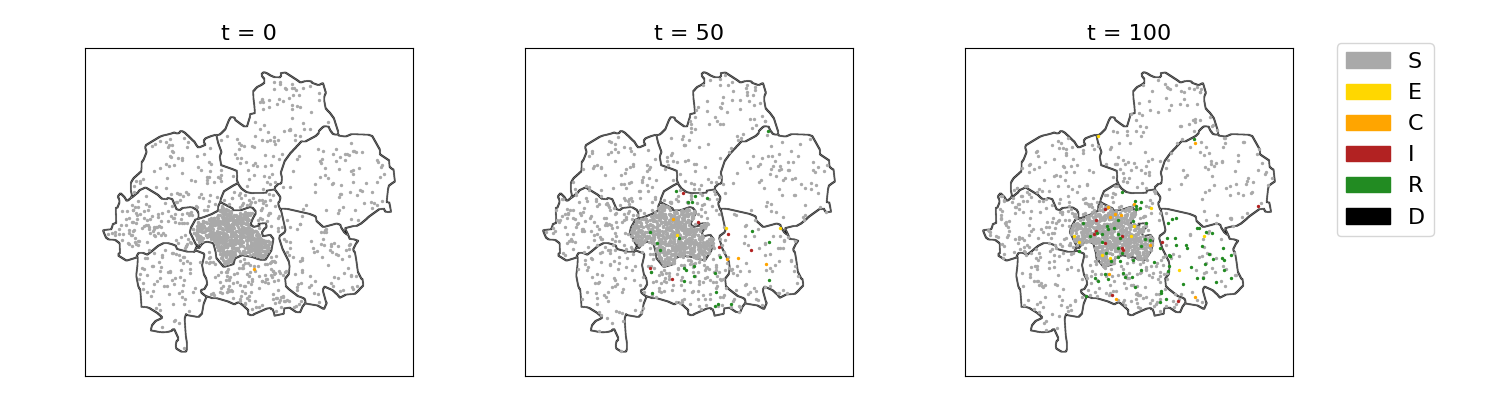"}
    \caption{\MK{\textbf{System state for one ABM realization for the Munich potential at three different time points.} Shown are the time points $t=0, 50$ and $100$ for a simulation with $1400$ agents for the Munich application.}}
    \label{fig:munich_trajectory}
\end{figure}

\begin{figure}
    \centering
    \sidesubfloat[]{\includegraphics[width=\textwidth]{"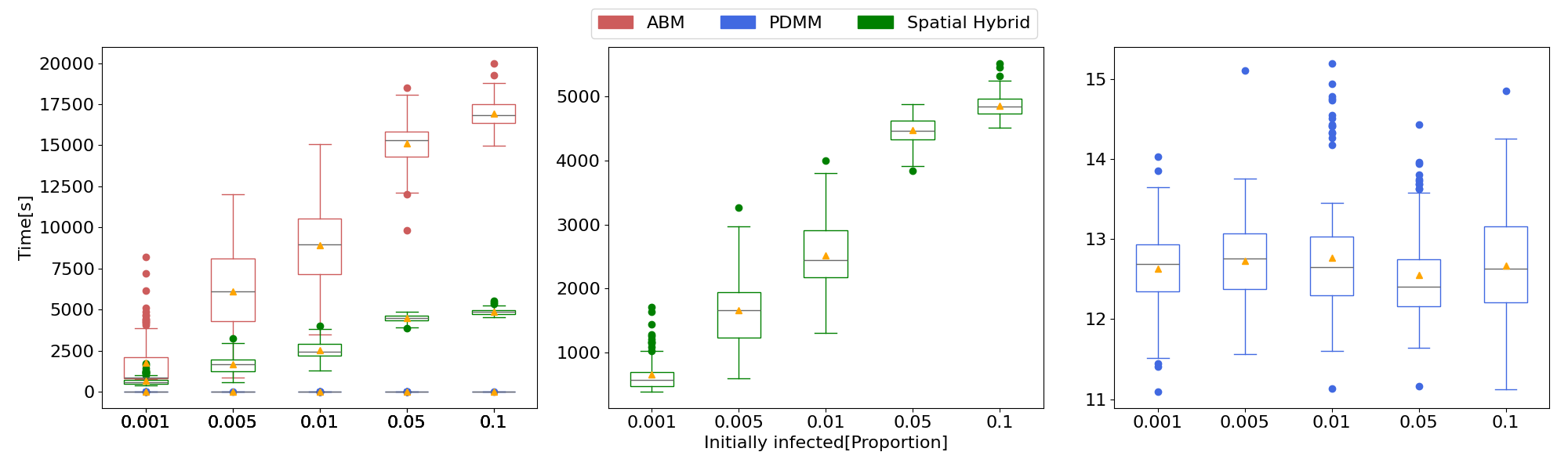"}\label{fig:munich_initially_infected}}\\
    \sidesubfloat[]{\includegraphics[width=\textwidth]{"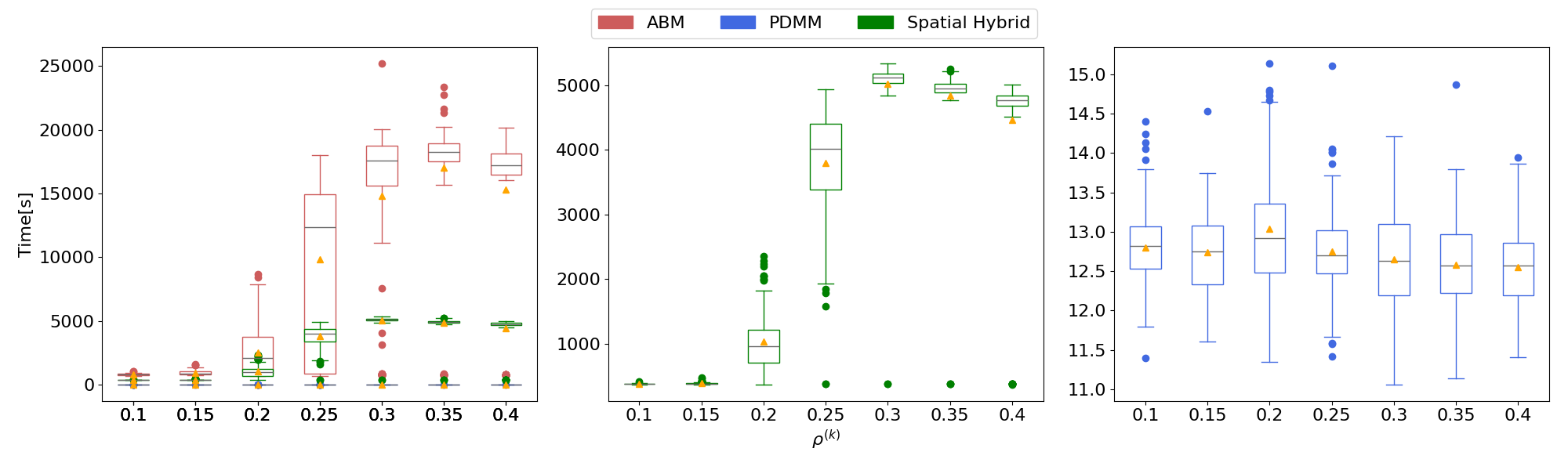"}\label{fig:munich_rho}}%
    \caption{\MK{\textbf{Runtime distribution for ABM, PDMM and spatial-hybrid model for a varying proportion of initially infected in $\Omega_5$ and varying values of $\rho^{(k)}$.}  Shown are the runtimes of $112$ runs for (a) five different values for the proportion of initially infected with the corresponding proportion distributed equally to compartments $E$, $C$ and $I$ and (b) seven different values for the transmission rate $\rho^{(k)}$ in all regions. In (a) $\rho^{(k)}=0.2$ for all runs and in (b) $0.05\%$ is Exposed, $0.05\%$ is Carrier and $0.1\%$ is Infected in the population of $\Omega_5$ in all runs.}}
    \label{fig:runtime_munich}
\end{figure}

\begin{figure}[H]
    \centering
    \includegraphics[width=1.05\textwidth]{"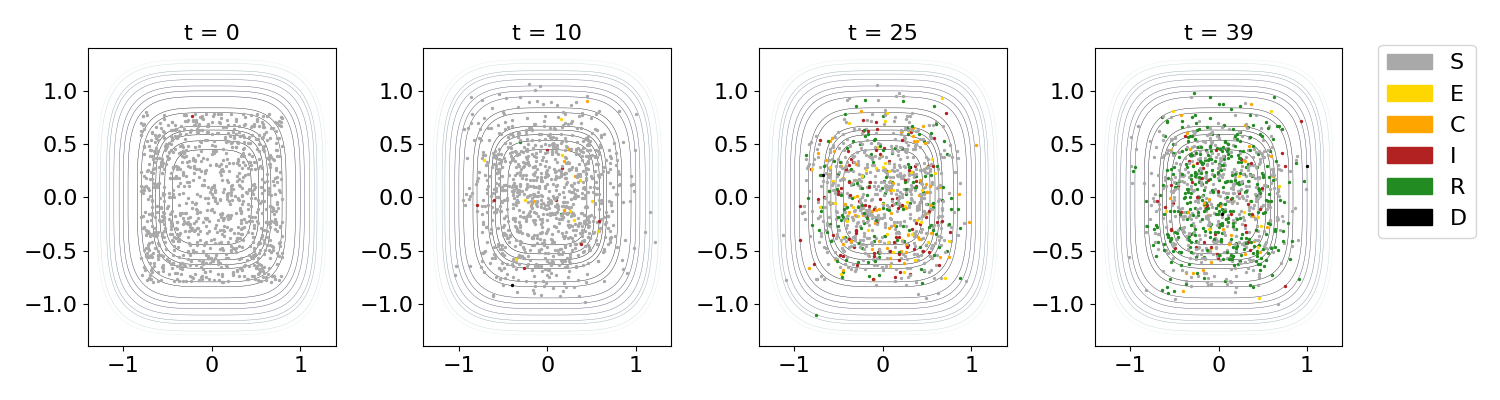"}
    \caption{\MK{\textbf{System state for one ABM realization for the single well potential at four different time points.} Shown are the time points $t=0, 10, 25$ and $39$ for a simulation with $1000$ agents for the single well application.}}
    \label{fig:sw_trajectory}
\end{figure}

\begin{figure}[H]
    \centering
    \sidesubfloat[]{\includegraphics[width=0.33\textwidth]{"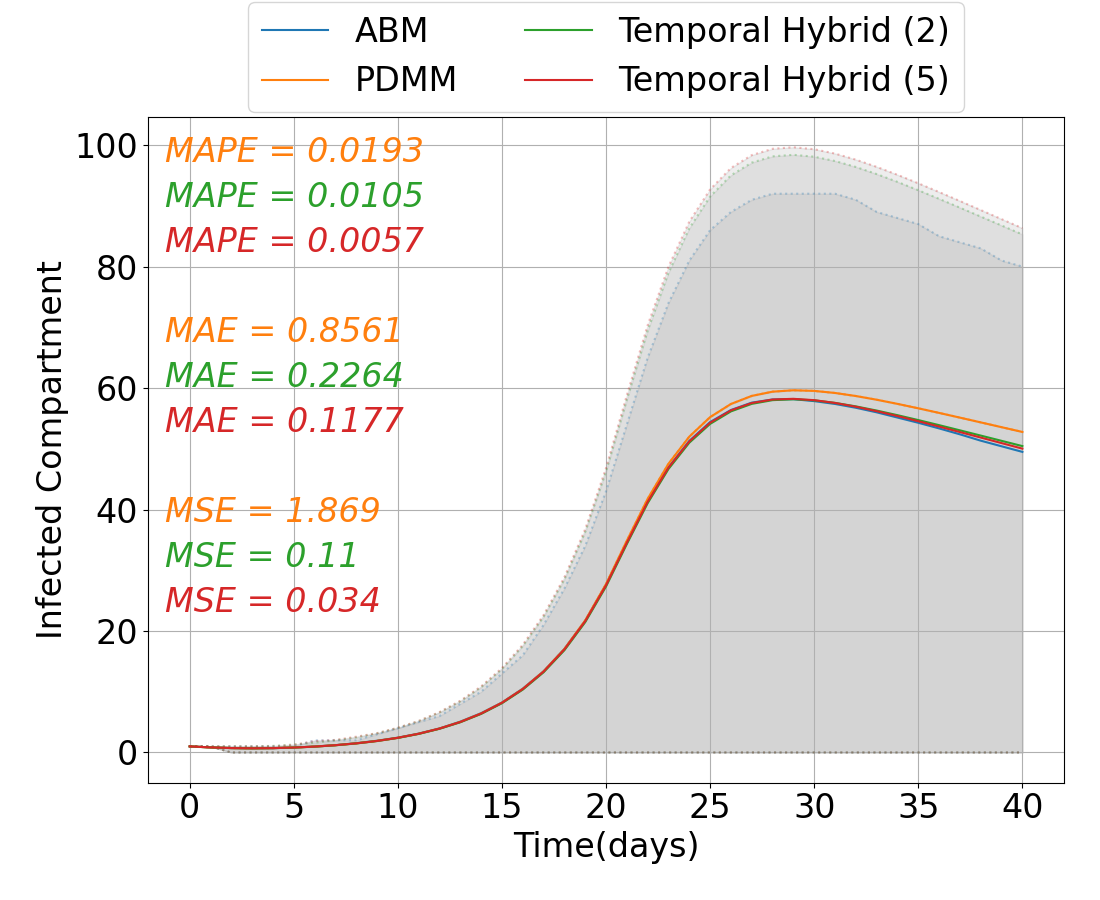"}}
    \sidesubfloat[]{\includegraphics[width=0.33\textwidth]{"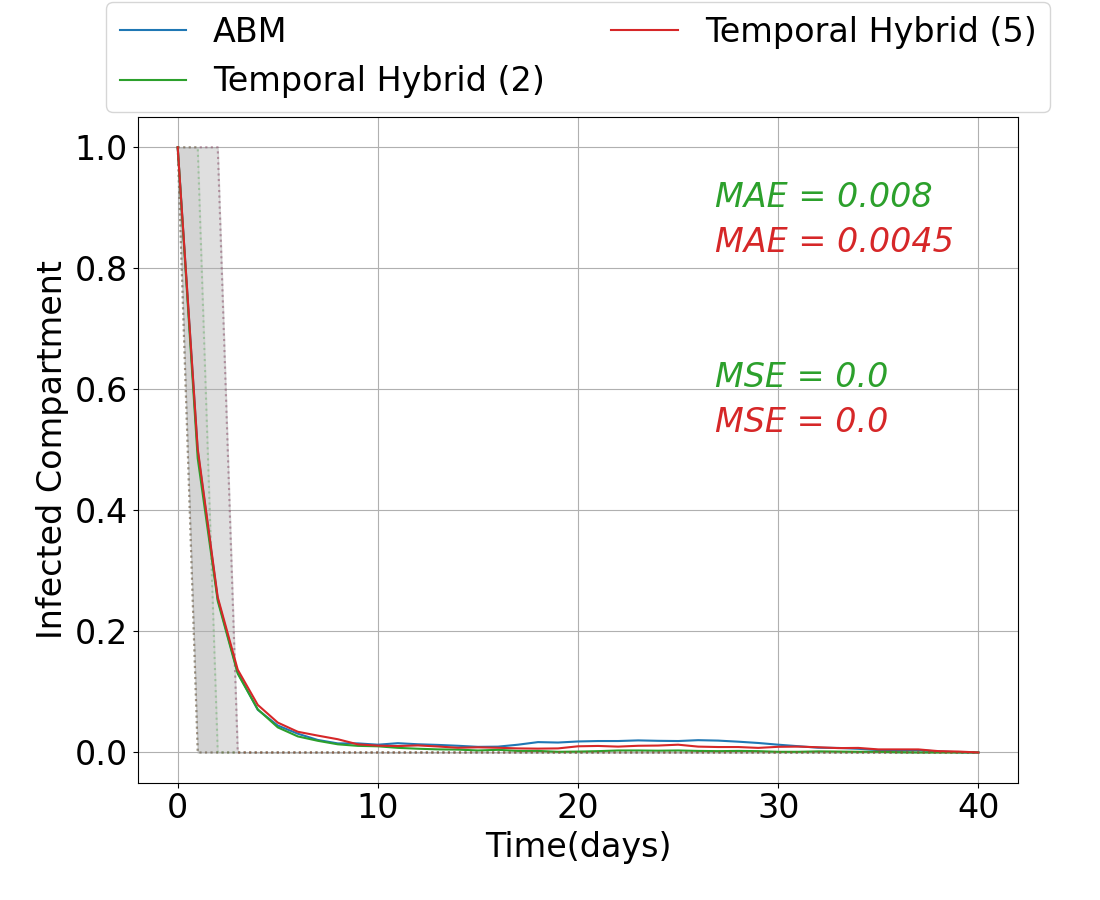"}}
    \sidesubfloat[]{\includegraphics[width=0.33\textwidth]{"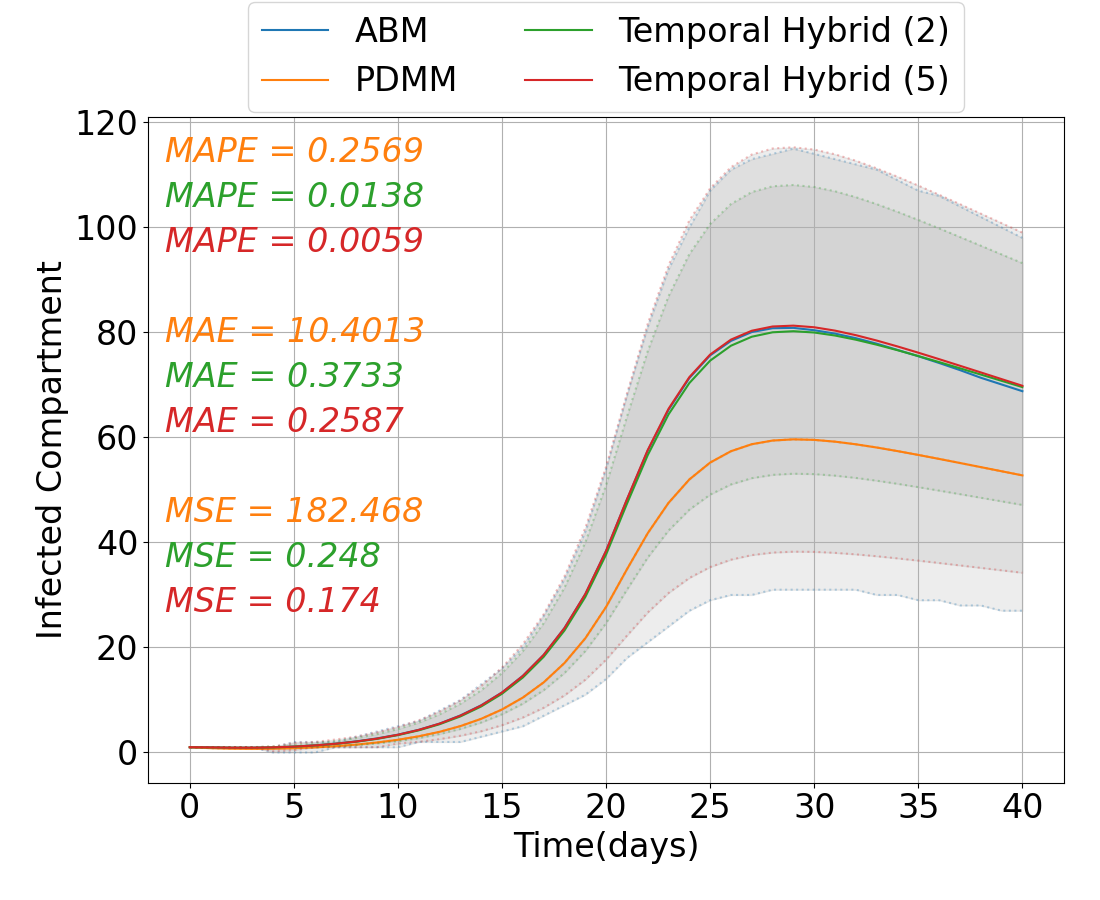"}}
    \caption{\textbf{Compartment $I$ for the single well potential for ABM, PDMM and temporal hybridization.} (a) All runs and selected runs (b) with and (c) without virus extinction for ABM, PDMM and temporal-hybrid models with $s=2$ and $s=5$. The figures show the mean outcomes in solid lines with a partially transparent face between the p25 and p75 percentiles from 10,000 runs with 10,000 agents.}
    \label{fig:temporal_hybrid_mean_comp_I}
\end{figure}

\newpage
\bibliographystyle{elsarticle-num} 
\bibliography{references.bib}


%\begin{thebibliography}{00}



%\end{thebibliography}
\end{document}